\documentclass[twocolumn]{aastex631}
\usepackage{amsmath}
\usepackage{graphicx}
\usepackage{placeins}
\usepackage{float}
\usepackage{txfonts}
\usepackage{hyperref}

\usepackage{lipsum}
\newcommand{\sw}[1]{\texttt{#1}}

\clearpage

\begin{document}
\title{Optical and Radio Analysis of Systematically Classified Broad-lined Type Ic Supernovae from the Zwicky Transient Facility}

\correspondingauthor{Gokul P. Srinivasaragavan}\email{gsriniv2@umd.edu}
\author[0000-0002-6428-2700]{Gokul P. Srinivasaragavan}
\affiliation{Department of Astronomy, University of Maryland, College Park, MD 20742, USA}
\affiliation{Joint Space-Science Institute, University of Maryland, College Park, MD 20742, USA}
 \affiliation{Astrophysics Science Division, NASA Goddard Space Flight Center, 8800 Greenbelt Rd, Greenbelt, MD 20771, USA}

\author[0000-0002-2898-6532]{Sheng Yang}\thanks{Please contact for HAFFET questions: \href{sheng.yang@astro.su.se}{sheng.yang@astro.su.se}}
\affiliation{Henan Academy of Sciences, Zhengzhou 450046, Henan, China}

\author[0000-0003-3768-7515]{Shreya Anand}
\affiliation{Division of Physics, Mathematics and Astronomy, California Institute of Technology, Pasadena, CA 91125, USA}

\author[0000-0003-1546-6615]{Jesper Sollerman}
\affiliation{Department of Astronomy, The Oskar Klein Center, Stockholm University, AlbaNova, 10691 Stockholm, Sweden}

\author[0000-0002-9017-3567]{Anna Y. Q.~Ho}
\affiliation{Department of Astronomy, Cornell University, Ithaca, NY 14853, USA}

\author[0000-0001-8104-3536]{Alessandra Corsi}
\affiliation{William H. Miller III Department of Physics and Astronomy, Johns Hopkins University, Baltimore, MD 21218, USA
}

\author[0000-0003-1673-970X]{S. Bradley Cenko}
\affiliation{Astrophysics Science Division, NASA Goddard Space Flight Center, 8800 Greenbelt Rd, Greenbelt, MD 20771, USA}
\affiliation{Joint Space-Science Institute, University of Maryland, College Park, MD 20742, USA}

\author[0000-0001-8472-1996]{Daniel Perley}
\affiliation{Astrophysics Research Institute, Liverpool John Moores University, Liverpool Science Park, 146 Brownlow Hill, Liverpool L3 5RF, UK}

\author[0000-0001-6797-1889]{Steve Schulze}
\affiliation{Center for Interdisciplinary Exploration and Research in Astrophysics (CIERA), Northwestern University, 1800 Sherman Ave., Evanston, IL 60201, USA}

\author{Marquice Sanchez-Fleming}
\affiliation{Department of Astronomy, Cornell University, Ithaca, NY 14853, USA}

\author{Jack Pope}
\affiliation{Department of Astronomy, Cornell University, Ithaca, NY 14853, USA}

 \author[0000-0003-2700-1030]{Nikhil Sarin}
\affiliation{Oskar Klein Centre for Cosmoparticle Physics, Department of Physics,
Stockholm University, AlbaNova, Stockholm SE-106 91, Sweden}
\affiliation{Nordita, Stockholm University and KTH Royal Institute of Technology,
Hannes Alfvéns väg 12, SE-106 91 Stockholm, Sweden}

\author[0000-0002-9646-8710]{Conor Omand}
\affiliation{Astrophysics Research Institute, Liverpool John Moores University, Liverpool Science Park, 146 Brownlow Hill, Liverpool L3 5RF, UK}
\affiliation{Department of Astronomy, The Oskar Klein Center, Stockholm University, AlbaNova, 10691 Stockholm, Sweden}

\author[0000-0001-8372-997X]{Kaustav K. Das}
\affiliation{Division of Physics, Mathematics and Astronomy, California Institute of Technology, Pasadena, CA 91125, USA}

\author[0000-0002-4223-103X]{Christoffer Fremling}
\affiliation{Caltech Optical Observatories, California Institute of Technology, Pasadena, CA 91125, USA}
\affiliation{Division of Physics, Mathematics and Astronomy, California Institute of Technology, Pasadena, CA 91125, USA}

\author[0000-0002-8977-1498]{Igor Andreoni}
\affil{Joint Space-Science Institute, University of Maryland, College Park, MD 20742, USA}
\affil{Department of Astronomy, University of Maryland, College Park, MD 20742, USA}
\affil{Astrophysics Science Division, NASA Goddard Space Flight Center, Mail Code 661, Greenbelt, MD 20771, USA}

\author{Rachel Bruch}
\affiliation{Department of Particle Physics and Astrophysics Weizmann Institute of Science 234 Herzl St. 76100 Rehovot, Israel}

\author[0000-0002-7226-836X]{Kevin B. Burdge}
\affiliation{Department of Physics, Massachusetts Institute of Technology, Cambridge, MA 02139, USA}
\affiliation{Kavli Institute for Astrophysics and Space Research, Massachusetts Institute of Technology,
Cambridge, MA 02139, USA}

\author[0000-0002-8989-0542]{Kishalay De}
\affiliation{MIT-Kavli Institute for Astrophysics and Space Research, 77 Massachusetts Ave., Cambridge, MA 02139, USA}
\affiliation{NASA Einstein Fellow}

\author[0000-0002-3653-5598]{Avishay Gal-Yam}
    \affiliation{Department of Particle Physics and Astrophysics, Weizmann Institute of Science, 76100 Rehovot, Israel}

\author[0000-0002-3884-5637]{Anjasha Gangopadhyay}
\affiliation{Department of Astronomy, The Oskar Klein Center, Stockholm University, AlbaNova, 10691 Stockholm, Sweden}

\author[0000-0002-3168-0139]{Matthew J. Graham}
\affiliation{California Institute of Technology, 1200 E. California Blvd, Pasadena, CA 91125, USA}

\author[0000-0001-5754-4007]{Jacob E. Jencson}
\affiliation{Department of Physics and Astronomy, Johns Hopkins University, 3400 North Charles Street, Baltimore, MD 21218, USA}
\affiliation{Space Telescope Science Institute, 3700 San Martin Drive, Baltimore, MD 21218, USA}

\author[0000-0003-2758-159X]{Viraj Karambelkar}
\affiliation{Division of Physics, Mathematics and Astronomy, California Institute of Technology, Pasadena, CA 91125, USA}

\author[0000-0002-5619-4938]{Mansi M. Kasliwal}
\affiliation{Division of Physics, Mathematics and Astronomy, California Institute of Technology, Pasadena, CA 91125, USA}

\author[0000-0001-5390-8563]{S. R. Kulkarni}
\affiliation{Division of Physics, Mathematics and Astronomy, California Institute of Technology, Pasadena, CA 91125, USA}

\author{Julia Martikainen}
\affiliation{Instituto de Astrofísica de Andalucía, CSIC, Granada, Spain}

\author[0000-0003-4531-1745]{Yashvi S. Sharma}
\affiliation{Division of Physics, Mathematics and Astronomy, California Institute of Technology, Pasadena, CA 91125, USA}

\author[0000-0003-0484-3331]{Anastasios Tzanidakis}
\affiliation{DIRAC Institute, Department of Astronomy, University of Washington, 3910 15th Avenue NE, Seattle, WA 98195, USA}

\author[0000-0003-1710-9339]{Lin Yan}
\affil{Caltech Optical Observatories, California Institute of Technology,
Pasadena, CA 91125, USA}

\author[0000-0001-6747-8509]{Yuhan Yao}
\affiliation{Department of Astronomy, University of California, Berkeley, 501 Campbell Hall, Berkeley, CA 94720, USA}
\affiliation{Miller Institute for Basic Research in Science, 468 Donner Lab, Berkeley, CA 94720, USA}

\author[0000-0001-8018-5348]{Eric C. Bellm}
\affiliation{DIRAC Institute, Department of Astronomy, University of Washington, 3910 15th Avenue NE, Seattle, WA 98195, USA}

\author[0000-0001-5668-3507]{Steven L. Groom}
\affiliation{IPAC, California Institute of Technology, 1200 E. California Blvd, Pasadena, CA 91125, USA}

\author[0000-0002-8532-9395]{Frank J. Masci}
\affiliation{IPAC, California Institute of Technology, 1200 E. California Blvd, Pasadena, CA 91125, USA}

\author[0000-0002-7501-5579]{Guy Nir}
\affiliation{Department of Astronomy, University of California, Berkeley, CA 94720-3411, USA}

\author[0000-0003-1227-3738]{Josiah Purdum}
\affiliation{Caltech Optical Observatories, California Institute of Technology, Pasadena, CA  91125}

\author[0000-0001-7062-9726]{Roger Smith}
\affiliation{Caltech Optical Observatories, California Institute of Technology, Pasadena, CA  91125}

\author{Niharika Sravan}
\affiliation{Department of Physics, Drexel University, Philadelphia, PA 19104, USA}

\begin{abstract}
We study a magnitude-limited sample of 36 Broad-lined Type Ic Supernovae (SNe Ic-BL) from the Zwicky Transient Facility Bright Transient Survey (detected between March 2018 and August 2021), which is the largest systematic study of SNe Ic-BL done in literature thus far. We present the light curves (LCs) for each of the SNe, and analyze the shape of the LCs to derive empirical parameters, along with the explosion epochs for every event. The sample has an average absolute peak magnitude in the $r$ band of $\rm{\overline{M}_{r, max}} = -18.51 \pm 0.15$ mag. Using spectra obtained around peak light, we compute expansion velocities from the Fe II 5169 $\AA$ line for each event with high enough signal-to-noise ratio spectra, and find an average value of $\rm{\overline{v_{ph}}} = 16,100 \pm 1,100 $ km s$^{-1}$. We also compute bolometric LCs, study the blackbody temperature and radii evolution over time, and derive the explosion properties of the SNe. The explosion properties of the sample have average values of $\rm{\overline{M}_{Ni}} = 0.37^{+0.08}_{-0.06} \, \rm{M_{\odot}}$, $\rm{\overline{M}_{ej}} = 2.45^{+0.47}_{-0.41} \, \rm{M_{\odot}}$, and $\rm{\overline{E}_{\rm{K}}} = (4.02^{+1.37}_{-1.00})  \times 10^{51} $ erg. Thirteen events have radio observations from the Very Large Array, with 8 detections and 5 non-detections. We find that the populations that have radio detections and radio non-detections are indistinct from one another with respect to their optically-inferred explosion properties, and there are no statistically significant correlations present between the events' radio luminosities and optically-inferred explosion properties. This provides evidence that the explosion properties derived from optical data alone cannot give inferences about the radio properties of SNe Ic-BL, and likely their relativistic jet formation mechanisms. 

\end{abstract}

\section{Introduction}
\label{sec:intro}
Type Ic supernovae (SNe) represent the final fate of massive stars ($M_{\rm{ZAMS}} \gtrsim 8 \, \rm{M_{\odot}}$) whose hydrogen and helium envelopes have been stripped prior to explosion \citep{Gal-Yam2017}. These SNe are part of a larger sample of stripped-envelope SNe whose progenitors are either very massive stars ($M_{\rm{ZAMS}} \lesssim 30 \, M_{\odot}$) whose outer layers have been stripped due to stellar winds or eruptions \citep{Conti1976}, or less massive stars ($M_{\rm{ZAMS}} \gtrsim 30 \, M_{\odot}$) whose outer hydrogen and/or helium layers have been stripped by binary interactions \citep{Yoon:2010aa, Lyman2016, Taddia2018a}.

Optical spectra of Type Ic SNe usually display photospheric expansion velocities of up to 10,000 km $\rm{s^{-1}}$ at peak from their Fe II lines \citep{Modjaz2016}. A subset of these events display broader Fe II and O I lines in their spectra, corresponding to velocities between 10,000 and 30,000 $\rm{km \, s^{-1}}$. These events are referred to as broad-lined Type Ic (Ic-BL) SNe, and are usually found in lower metallicity environments \citep{Arcavi:2011aa, Schulze2021} than normal stripped-envelope SNe. Their progenitors are also younger and more massive \citep{Sanders2012, Cano:2013aa}. The lightcurves (LCs) of Type Ic-BL events also rise faster than normal stripped-envelope SNe and are brighter at peak magnitude \citep{Drout2011, Cano:2013aa, taddia2015sdss, Lyman2016, prentice2016}. LC modeling of these events has shown that they often have a larger amount of $^{56}\rm{Ni}$ synthesized in the explosion, and they sometimes possess explosion energies ($10^{52}$ erg) an order of magnitude higher than normal stripped-envelope events ($10^{51}$ erg); though peculiarly, their ejecta masses are similar (e.g., \citealt{Cano:2013aa, taddia2015sdss, Taddia2018a, Lyman2016, prentice2016, Barbarino2021}).

SNe Ic-BL challenge the standard picture associated with the standard explosion mechanism of CCSNe, as the extra energy possessed by some of these explosions necessitate a deviation from the traditional picture of neutrino irradiation from a proto-neutron star (NS) reviving the bounce-back shock in the progenitor's core. A major open question is understanding how the same amount of ejecta can lead to higher amounts of $^{56}\rm{Ni}$ and higher kinetic energies. A hypothesized scenario is that a relativistic jet driven from the core of the progenitor's proto-NS transfers the extra energy needed to the surrounding stellar medium such that an explosion can reach the order of $10^{52}$ erg observed in some events \citep{Woosley2003}. In fact, \citet{Rodriguez2024} found evidence of a non-radioactive power source for the majority of stripped-envelope SNe in their sample of 54 events, including 9 SNe Ic-BL.

This scenario is supported observationally by the detection of long gamma-ray bursts (LGRBs) unambiguously associated with a handful of SNe Ic-BL. Nearly all nearby LGRBs have observationally associated SNe Ic-BL (see, e.g., \citealt{Galama1998, Hjorth2003, Pian2006, Melandri2012, D'Elia2015, cano2017,Melandri2019, Hu2021, Kumar2022, Rossi2022, Blanchard2023, Srinivasaragavan2023, Srinivasaragavan2024}). For more distant events, non-detections are not particularly constraining (with the exception of a few events, see \citealt{GalYam2006, Fynbo2006, DelleValle2006, Tanga2018}). This is due to a few reasons -- at high redshifts, detected GRBs usually have very luminous afterglows that wash out their associated SN signature in their LCs, the SN spectral peak shifts to the near-infrared at high redshifts making it difficult to detect with optical telescopes, and Type Ic-BL SNe rarely get brighter than -20 mag, giving a constraint on how far they can be detected regardless of their associated GRB. 

However, it is clear that the majority of SNe Ic-BL do not have GRB counterparts. It has been suggested that some SNe Ic-BL may produce off-axis GRBs, whose jets are initially out of our line-of-sight, and emerge within our viewing angle through their radio emission at later times. Studies have shown that relativistic ejecta are not fully ubiquitous to SNe Ic-BL, and viewing angle effects soley cannot account for the lack of GRBs associated with most Type Ic-BL events \citep{Soderberg+2006, Corsi2016, Corsi2024}. Therefore, there are possible intrinsic differences in the explosion mechanisms between jet-powered SNe Ic-BL and normal SNe Ic-BL, and understanding this dichotomy can provide important insights into the current understanding of the landscape of massive stellar explosions.

In this work, we present a sample of 36 SNe Ic-BL observed with the Zwicky Transient Facility \citep[ZTF;][]{Graham2019,Bellm2019, Dekany2020, Masci2019}, and analyze their optical properties. This sample builds on the sample of ZTF's predecessor, the intermediate Palomar Transient Facility (iPTF), whose 34 SNe Ic-BL were analyzed in \citet{taddia2019iptf}. This work is the largest systematic study done on SNe Ic-BL in literature thus far. In addition to presenting the sample and analyzing its optical characteristics, we perform systematic comparisons of key optical properties between events that have radio detections and non-detections, to see if the optical properties of SNe Ic-BL can provide a link to their radio observations, and therefore any insight into their jet formation mechanisms. 

The structure of the paper is: in \S \ref{SN_Sample} we describe how the sample was created; in \S \ref{Observations} we describe the facilities used to obtain observations of our sample; in \S \ref{LCs} we describe the analysis done on the photometric observations; in \S \ref{spectra} we describe the analysis done on the spectroscopic observations; in \S \ref{Bollabel} we describe the creation of bolometric LCs and their analysis; in \S \ref{explosionlabel} we describe the derivation of explosion properties; in \S \ref{Multiwavelengthlabel} we describe analysis done on a subset of events with radio observations; and in \S \ref{Conclusion} we present a summary and conclusions of the work. We also present an Appendix, where we include discovery paragraphs on every event, their full spectral sequences, as well as efforts to model the LCs that have multiwavelength data.
 
\section{SN Sample Description}
\label{SN_Sample}

The sample was created by compiling all of the events that passed the internal quality cuts in ZTF's Bright Transient Survey (BTS) that were spectroscopically classifed as SNe Ic-BL. BTS is a magnitude-limited survey that spectroscopically classifies all SNe $\lesssim 18.5$\,mag at peak brightness \citep{Perley2020, fremling2020}. The quality cuts ensure that the objects have adequate LC coverage before and after peak. In addition, they ensure that the reference images used for image subtraction are uncontaminated by transient light, are in fields that are still visible one month after peak, and have low Galactic extinction ($A_V < 1$ mag). The final sample has a total of 36 events, selected from the BTS explorer website. Three events in the sample are subjects of single-object studies already published (SN 2018bvw, SN 2018gep, and SN 2020bvc; \citealt{18aaqjovh, 18abukavn,Pritchard2021, Leung2018, 20aalxlis, Rho2021, Izzo2020, Long2023}), and we refer to those works when necessary. Six events in the sample are also presented in \citet{Anand2024}, in the context of near-infrared follow-up to search for $r$-process nucleosynthesis, as well as eight events in \citet{Corsi2024}, in the context of radio follow-up observations searching for relativistic ejecta. 

\begin{figure}
    \centering
    \includegraphics[width = \linewidth]{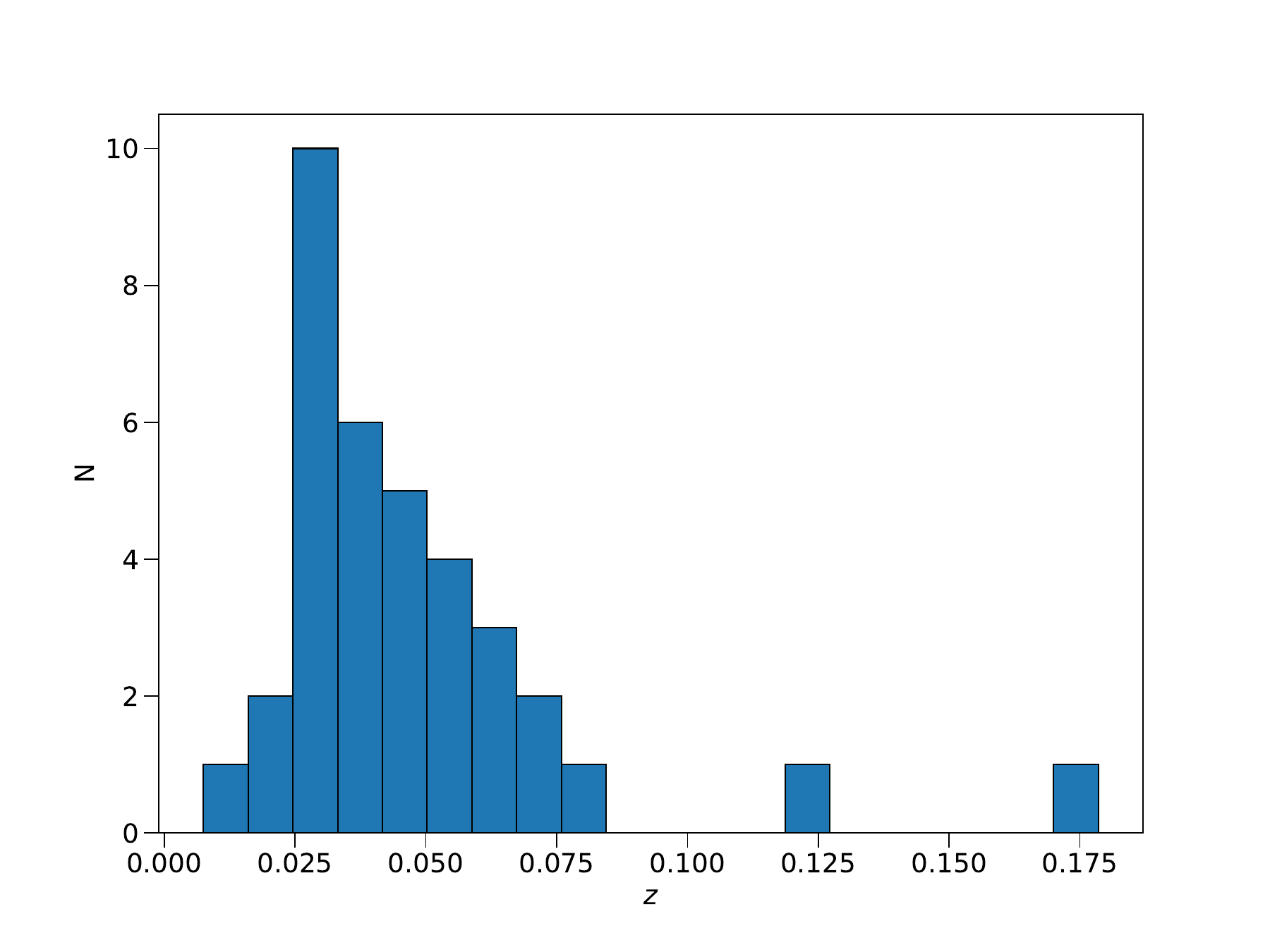}
    \caption{Redshift distribution for the SNe Ic-BL sample.}
    \label{redshift}
\end{figure}

In Figure \ref{redshift}, we show the redshift distribution of events in our sample, which ranges from $z = 0.017$ to $z = 0.1785$. All of the events except for SN 2020wgz and SN 2018hsf have $z < 0.082$, and K-corrections are close to negligible for these events. To quantify this, we generate a SN 1998bw-like LC using \sw{SNCosmo} \citep{SNCosmo}, setting the peak absolute magnitude of the SN equivalent to the average value of the sample found in \S \ref{empiricalparam} of $-18.51$ mag. We find that at peak light, the K-correction at $z = 0.082$ is just $\sim 0.1$ mag. SN 2020wgz is a unique event that may be a superluminous SN (SLSN; see \S \ref{2020wgz}). The event has poor spectral coverage, with only three spectra obtained. Because of its unique evolution, utilizing existing templates to compute its K-corrections will not be sufficient, and more complex methods are necessary. SN 2018hsf is another event at high redshift ($z = 0.119$), that also has poor spectral coverage with only three spectra, and though it passed the quality cut, there are only a few photometry points pre-peak. This event also is likely not powered by radioactive decay (more in \S \ref{explosionlabel}), and therefore also likely exhibits unique spectral evolution that cannot be modeled with existing templates. Therefore, we do not apply K-corrections to our sample, following \citet{Anand2024} and \citet{Corsi2024}.

In the Appendix, we include descriptions of all the events in this paper that have not been presented in previous works. We provide the first ZTF magnitudes along with the discovery and classification details in each of the descriptions. All magnitudes are reported in the AB system, and UT dates are used throughout this paper. We estimate and use the explosion dates throughout the paper, and measure the phases in rest-frame days with respect to the explosion epochs. We also use a flat $\Lambda$CDM cosmology $H_0 = 69.6$\,km\,s$^{-1}$\,Mpc$^{-1}$, $\Omega_{\rm M }= 0.286$, $\Omega_{\rm vac} = 0.714$ \citep{Bennett2014}, to convert redshifts to distances.
In Table~\ref{discoverytable} we provide the SN IAU name, the ZTF name, RA and Dec coordinates to the transient, redshift, distance, and Milky way extinction \citep{Schlafly11}. All average values and errors are calculated through bootstrapping the sample with replacement 10,000 times, and drawing from the 16th, 50th, and 84th percentile means derived in the process. We use the Hybrid Analytic Flux FittEr for Transients (HAFFET;  \citealt{Yang+2023}) code for the analyses presented in \S \ref{LCs}, \ref{Bollabel}, \ref{explosionlabel}. A detailed description of the methodologies used for these analyses is presented in \citet{Yang+2023}, which we summarize in the applicable sections.

\begin{table*}[]
\centering
\caption{Discovery properties of the SNe Ic-BL sample}
\begin{tabular}{l|l|l|l|l|l|l}
\hline
\hline
ZTF name     & SN name     & RA (hms)           & Dec ($\degr$ $`$ $``$)         & $z$        & $A_{\rm{v, MW}}$ (mag) & First Presented In  \\
\hline
ZTF18aaqjovh & SN 2018bvw & 11:52:43.62 & +25:40:30.1 & 0.054 & 0.062 & \citet{18aaqjovh}\\ 
        ZTF18abhhnnv & SN 2018ell & 16:49:57.02 & +27:38:26.7 & 0.0638 & 0.16 & -- \\ 
        ZTF18abukavn & SN 2018gep & 15:17:02.54 & +03:56:38.7 & 0.0442 & 0.124 & e.g.,\footnote{\citet{18abukavn,Pritchard2021, Leung2018}}\\ 
        ZTF18acbvpzj & SN 2018hsf & 02:40:12.79 & -19:58:44.9 & 0.1184 & 0.093 & --\\ 
        ZTF18acxgoki & SN 2018keq & 23:22:41.97 & +21:00:43.2 & 0.0384 & 0.341 & -- \\ 
        ZTF19aawqcgy & SN 2019hsx & 18:12:56.21 & +68:21:45.2 & 0.020652 & 0.129 & \citet{Anand2024} \\ 
        ZTF19aaxfcpq & SN 2019gwc & 16:03:26.88 & +38:11:02.6 & 0.038 & 0.036 & \citet{Anand2024} \\ 
        ZTF19abfsxpw & SN 2019lci & 16:31:01.61 & +08:28:23.7 & 0.0292 & 0.208 & -- \\ 
        ZTF19ablesob & SN 2019moc & 23:55:45.94 & +21:57:19.7 & 0.055 & 0.171 & \citet{Anand2024} \\ 
        ZTF19abqshry & SN 2019oqp & 16:38:33.20 & +45:37:52.2 & 0.03082 & 0.037 & -- \\ 
        ZTF19abupned & SN 2019pgo & 23:53:00.04 & +25:07:16.4 & 0.0500 & 0.156 & --\\
        ZTF19abzwaen & SN 2019qfi & 21:51:07.89 & +12:25:38.4 & 0.028 & 0.19 & \citet{Anand2024} \\ 
        ZTF20aafmdzj & SN 2020zg & 04:02:36.39 & -16:11:54.4 & 0.0557 & 0.087 & --\\ 
        ZTF20aaiqiti & SN 2020ayz & 12:12:04.89 & +32:44:01.7 & 0.025 & 0.038 & --\\ 
        ZTF20aalxlis & SN 2020bvc & 14:33:57.00 & +40:14:37.3 & 0.0252 & 0.031 & e.g.,\footnote{\citet{20aalxlis, Rho2021, Izzo2020, Long2023}} \\ 
        ZTF20aapcbmc & SN 2020dgd & 15:45:35.54 & +29:18:38.4 & 0.032 & 0.071 & \citet{Anand2024}\\ 
        ZTF20aaurexl & SN 2020hes & 17:47:05.71 & +42:46:39.7 & 0.0700 & 0.106 & --\\ 
        ZTF20aavcvrm & SN 2020hyj & 16:23:47.22 & +29:58:58.5 & 0.055 & 0.077 & -- \\
        ZTF20aazkjfv & SN 2020jqm & 13:49:18.57 & -03:46:10.3 & 0.03696 & 0.096 & \citet{Corsi2024} \\ 
        ZTF20abbplei & SN 2020lao & 17:06:54.60 & +30:16:17.3 & 0.030814 & 0.138 & \citet{Anand2024} \\ 
        ZTF20abrmmah & SN 2020rfr & 22:39:49.30 & -06:26:16.0 & 0.0725 & 0.105 & -- \\ 
        ZTF20abswdbg & SN 2020rph & 03:15:17.81 & +37:00:50.6 & 0.042 & 0.65 & \citet{Anand2024}\\
        ZTF20abzoeiw & SN 2020tkx & 18:40:09.00 & +34:06:59.5 & 0.027 & 0.226 & \citet{Anand2024} \\ 
        ZTF20achvlbs & SN 2020wgz & 08:57:33.27 & +62:34:00.1 & 0.1785 & 0.217 & -- \\
        ZTF20acvcxkz & SN 2020abxl & 05:04:22.76 & -14:02:46.4 & 0.0815 & 0.344 & -- \\ 
        ZTF20acvmzfv & SN 2020abxc & 01:00:34.04 & -08:07:00.7 & 0.0600 & 0.255 & -- \\ 
        ZTF20adadrhw & SN 2020adow & 08:33:42.26 & +27:42:43.7 & 0.0075 & 0.124 & -- \\ 
        ZTF21aagtpro & SN 2021bmf & 16:33:29.41 & -06:22:49.4 & 0.017 & 0.85 & \citet{Anand2024} \\
        ZTF21aaocrlm & SN 2021epp & 08:10:55.27 & -06:02:49.3 & 0.0385 & 0.15 & \citet{Corsi2024} \\ 
        ZTF21aapecxb & SN 2021fop & 07:46:42.90 & +07:12:38.6 & 0.077 & 0.089 & --\\ 
        ZTF21aartgiv & SN 2021hyz & 09:27:36.50 & +04:27:11.0 & 0.046 & 0.125 & \citet{Corsi2024}\\ 
        ZTF21aaxxihx & SN 2021ktv & 11:03:03.88 & +08:51:39.7 & 0.0700 & 0.071 & -- \\ 
        ZTF21abchjer & SN 2021ncn & 22:36:32.92 & +25:45:40.5 & 0.02461 & 0.143 & -- \\ 
        ZTF20abcjdwu & SN 2021qjv & 15:10:47.05 & +49:12:18.0 & 0.03803 & 0.04 & -- \\ 
        ZTF21abmjgwf & SN 2021too & 21:40:54.28 & +10:19:30.4 & 0.035 & 0.171 & \citet{Anand2024} \\ 
        ZTF21acbnfos & SN 2021ywf & 05:14:10.99 & +01:52:52.2 & 0.028249 & 0.292 & \citet{Anand2024}\\ 
\end{tabular}
\label{discoverytable}
\end{table*}

\begin{figure*}
    \centering
    \includegraphics[width=.99\linewidth]{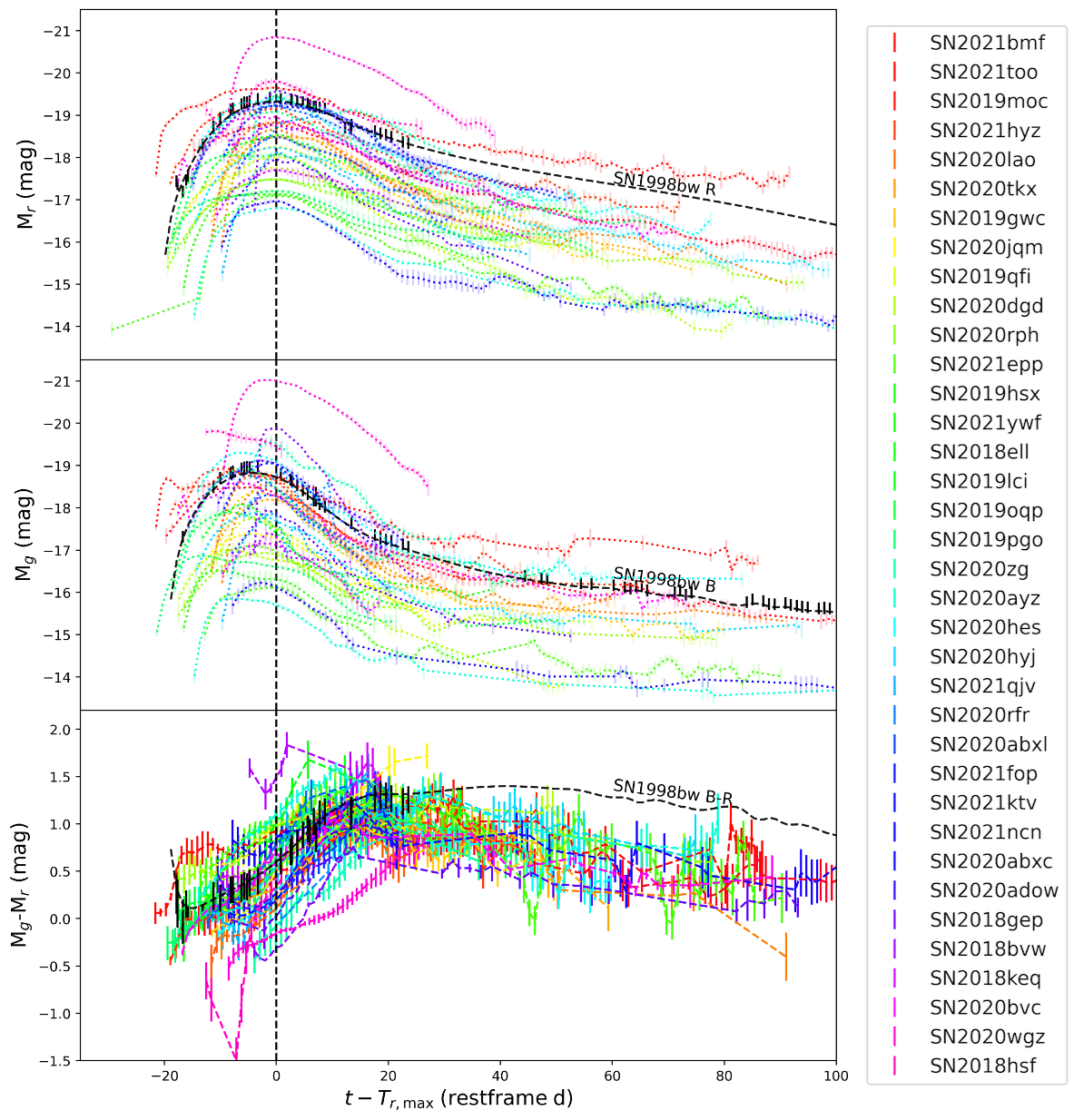}
    \caption{Light curve and color comparison bewteen the SNe Ic-BL sample and SN 1998bw in $r/R$- and $g/B$-bands, with all magnitudes corrected for Galactic extinction, and the GP processed LCs shown. The colors are computed with a combination of the data binned within 1 day, as well as the GP processed LCs. The LC of SN 1998bw \citep{98bwpaper} was interpolated through GP techniques, and is shown as the black dashed lines.}
    \label{ZTFLCs}
\end{figure*}
%%%

\section{Observations}
\label{Observations}
We describe here the facilities used to obtain photometric and spectroscopic observations of the SN sample. We note that for two events (SN 2020abxl and SN 2021epp), we did not obtain any spectroscopic observations, and utilize publicly available spectra from the ESO Spectroscopic Survey of Transient Objects (ePESSTO+; \citealt{epessto}) on the Transient Name Server for clasissification purposes.

\subsection{ZTF}
The ZTF camera \citep{ztfcam} on the Palomar 48-inch telescope was used for supernova discovery and photometric follow-up. ZTF surveys the entire observable northern sky every two to three days in the $r$ and $g$ bands, in addition to the $i$ band for some selected observations, reaching a median 5$\sigma$ detection depth of 20.5\,m$_{\mathrm{AB}}$ in the $g$ and $r$ bands. The default observing mode is 30\,s exposures, and alerts according to 5$\sigma$ changes in brightness relative to the reference image are sent out through an avro format \citep{Patterson2019}. Real-time filtering through machine-learning classifiers \citep{Mahabal2019}, star-galaxy classifiers \citep{Tachibana2018}, and light curve properties is also performed for candidate events. ZTF observations of the SN sample last to $\sim$ 60 days after peak, until the SN is fainter than 20.5\,mag. For more information about the data processing and image subtraction pipelines, see \citet{Masci2019}. We utilized both the GROWTH marshal \citep{Kasliwal2019} and the Fritz SkyPortal \citep{VanderWalt2019,Coughlin2023skyportal} to gather the datasets needed for this work.

\subsection{SEDM}
BTS used the Spectral Energy Distribution Machine's (SEDM, \citealt{Blagorodnova2018, Rigualt2019}) low-dispersion (R$\sim$100) integral field spectrograph (IFU) to obtain classification spectra shortly after discovery for many of the objects in our sample. The spectra obtained with the IFU are reduced through a custom data reduction pipeline \citep{Rigault2019} that utilizes flat-fielding, wavelength calibration, extraction, flux calibration, and telluric corrections.

\subsection{SPRAT}
We used the Spectrograph for the Rapid Acquistion of Transients  (SPRAT; \citealt{Piasik2014}), a low-resolution, (R$\sim 350$), spectrograph mounted on the 2.0 meter Liverpool Telescope (LT; \citealt{Steele2004}) on La Palma, Spain to obtain spectra for some of our events.  Spectra are reduced and flux calibrated using a custom pipeline for the LT \citep{Smith2016}. 

\subsection{LRIS}
We used the Low Resolution Imaging Spectrometer (LRIS; \citealt{Oke1995}) on the 10 meter Keck I telescope to obtain spectra for some of the events in our sample. We utilized typical exposure times of 600 \,s, longslit masks of  1.0$\arcsec$ or 1.5$\arcsec$ width, and the 400/3400 grism on the blue arm and the 400/8500 grating on the red arm, with a central wavelength of 7830 $\rm \AA$. This enabled wavelength coverage from 3,200-10,000 $\rm \AA$. The reduction was done using LPipe \citep{Perley2019}.

\subsection{DBSP}
We used the Double Beam Spectrograph (DBSP; \citealt{Oke1982}) on the Palomar 200-in telescope to obtain low to medium resolution (R$\sim$1000-10000) spectra of many of the events in our sample. DBSP has a pixel scale of 0.293$\arcsec$/pixel (red side) and 0.389 $\arcsec$/pixel (blue side). We utilized a red grating of 316/7500, a blue grating of 600/400, a D55 dichroic, and slitmasks of 1$\arcsec$, 1.5$\arcsec$, and 2$\arcsec$. The data reduction was done using a custom PyRAF DBSP reduction pipeline \citep{Bellm2016} while the rest were reduced using a custom DBSP Data Reduction pipeline relying on Pypeit \citep{Prochaska2019, Roberson2021}.  

\subsection{ALFOSC}
We used the Alhambra Faint Object Spectrograph and Camera (ALFOSC)\footnote{\href{http://www.not.iac.es/instruments/alfosc}{{http://www.not.iac.es/instruments/alfosc}}} on the 2.56\,m Nordic Optical Telescope (NOT) at the Observatorio del Roque de los Muchachos on La Palma (Spain) to obtain low-resolution (R $\sim$ 700) spectra for some events in our sample. The spectra were obtained with a 1.0'' wide slit and grism \#4 with a spectral resolution of 360.
%with a central wavelength of 5800 $\AA$.  
We reduced the data with IRAF and Pypeit.

\subsection{\textit{Swift XRT and UVOT}}
\label{Swiftobs}
The Neils Gehrels \textit{Swift} observatory \citep{Gehrels_2004} observed a handful of events in our sample. \citet{Corsi2024} reports the observations taken with the X-ray telescope (XRT; \citealt{Burrows2005}), with five non-detections (SN 2020lao, SN 2020tkx, SN 2020jqm, SN 2020rph, SN 2021hyz) and 2 detected events (SN 2019hsx, SN 2019ywf). In addition, \citet{18aaqjovh, 18abukavn, 20aalxlis} report \textit{Swift} non-detections for 2 more events in the sample (SN 2018bvw, SN 2018gep) and 1 more detected event (SN 2020bvc). SN 2020adow was also detected by the XRT, but was not included in \citet{Corsi2024}. We utilize the X-ray detections for the subset of events in our sample in \S \ref{Multiwavelengthlabel}. In this work, we also utilize observations taken with \textit{Swift}'s Ultra-Violet/Optical Telescope \citep[UVOT;][]{Roming2005}, to better constrain the blackbody temperature and radii of the events in our sample. UVOT observed SN 2018etk, SN 2018hom, SN 2019hsx, SN 2020jqm, SN 2020lao, SN 2020rph, SN 2020tkx, and SN 2021ywf, across the v, b, u, uvw1, uvm2, and uvw2 filters. In addition to these observations, we also obtained ToOs in 2022 in the same filters as the SN initial measurements, to obtain late-time photometry of the host galaxies of these SNe. We did this in order to correct for any host-galaxy contamination in the SN photometry. We used the \texttt{uvotsource} task to measure the photometry, and utilize the photometry in \S \ref{blackbody} to better constrain the blackbody temperature and radii for part of the sample. We note that SN 2020adow also had measurements from UVOT; however, this event was added to our sample in 2024, and due to \textit{Swift's} pointing constraints, this event will not be visible again until late 2024. Therefore, we omit the \textit{Swift} data.

\subsection{VLA}
\subsubsection{Individual Observations}
\citet{Corsi2024} observed eight events in our sample using the Very Large Array (VLA). Five events in the sample (SN 2020lao, SN 2021hyz, SN 2020rph, SN 2021epp, SN 2019hsx) had radio non-detections or detections that were consistent with emission from the host galaxy. Three events (SN 2020jqm, SN 2020tkx, and SN 2021ywf) have radio detections compatible with point sources along with variability over the timescale of observations. SN 2020adow was also observed by the VLA \citep{2020adowradio}, and has multiple radio point source detections that evolve with time. In addition, \citet{18aaqjovh, 18abukavn, 20aalxlis} and \citet{Izzo2020} report radio point source detections for three more events in our sample (SN 2018gep, SN 2018bvw, SN 2020bvc). SN 2018bvw and SN 2020bvc were clear transient radio sources. 

At the time of publication in \citet{18abukavn}, SN 2018gep's radio detections could not be ruled out from being due to its host galaxy. \citet{18abukavn} reports three detections in the VLA D configuration (of 34, 24.4, and 26.8 $\mu \rm{Jy}$ at 9, 9.7, and 14 GHz) and two non-detections in the VLA C configuration two months after the last detection (of $< 16$ and $<17$ $\mu \rm{Jy}$ at 9 and 14 GHz), which is a comparable level to the declining trend in flux. Because the C configuration has a different resolution than the D configuration, it was unclear whether the detections in the D configuration were from the host galaxy and were simply being resolved out in the C configuration observations. Therefore, in April 2021 more than two years after the initial observations, SN 2018gep's location was observed by VLA program ID 21A-308 (PI Ho) in the D configuration, with the same setup as the initial observations. The source was not detected, to an upper limit of $ < 18 \, \mu \rm{Jy}$, confirming the transient SN 2018gep displays fading radio emission with time. We therefore conclude that the radio detections of SN 2018gep reported in \citet{18abukavn} are from the transient.

\subsubsection{VLASS}
The Karl G. Jansky Very Large Array Sky Survey (VLASS) is conducted across multiple epochs in the S-Band at 2-4 GHZ to monitor and analyze the radio sky. The VLASS is divided into three distinct epochs, each with its own timeline:
\begin{itemize}
\item Epoch 1: Commenced on September 13, 2017, and concluded on July 22, 2019.
\item Epoch 2: Started on May 27, 2020, and ended on March 7, 2022.
\item Epoch 3: Began on February 1, 2023, with observations planned to continue until the end of 2024.
\end{itemize}

The typical RMS noise for an individual epoch detection in the VLASS is around 0.12 mJy, which provided a baseline for assessing the sensitivity of the survey and the significance of detections. 

Using a cross-matching effort between the BTS and the VLASS, we searched for radio emission for the 36 SNe Ic-BL in our sample, and found that SN 2021bmf was the only object that showed transient radio emission across the three epochs. The first VLASS epoch had a non-detection on June 17, 2019, 19 months before the SNe. The second epoch had a detection on September 26, 2021, 8 months after the SN's peak with a peak flux of $5.57 \pm 0.526$ mJy. This corresponds to a peak luminosity of $4.2 \times 10^{28}$ erg s$^{-1}$ Hz${^-1}$.  

Therefore, we consider eight events to have radio counterparts  (SN 2018gep, SN 2018bvw, SN2020bvc, SN 2020jqm, SN 2020tkx, SN 2020adow, SN2021bmf, and SN 2021ywf), and five events to have radio non-detections (SN 2019hsx, SN 2020lao, SN 2020rph, SN 2021epp, SN 2021hyz).

\section{Supernova Lightcurves}
\label{LCs}
%\subsection{Galactic and Host Galaxy Extinction}
%\label{sec:hostext}

%In order to correct for host-galaxy extinction, we use intrinsic SN color-curve templates from \citet{stritzingercolors}, who have shown that most stripped-envelope SNe have similar colors after peak \citep{Drout2011}. We represent the host extinction in terms of the observed colors minus the intrinsic colors obtained from these templates, through 
%\begin{equation}
%\label{eqn:color}
%E(X-Y)_{\rm{host}}=(X-Y)_{\rm obs}-(X-Y)_{\rm int}
%\end{equation}
%where $X$ and $Y$ are the measured magnitudes corrected for the Galactic extinction in two different filters. For ZTF, the filters used to compute these colors are $g$ and $r$. 

%\textbf{Report Results}

\subsection{Light curve interpolation}
\label{interpolation}

Before creating LCs for the events, we correct ZTF photometry for Galactic extinction with the Milky Way color excess $E(B-V)_{MW}$ toward the position of every SN from \citet{Schlafly11}. We use the \citet{1989ApJ...345..245C} extinction law to perform reddening corrections, setting $R_V = 3.1$. To be consistent with \citet{Anand2024} and \citet{Corsi2024}, we also assume zero host galaxy extinction (more in \S \ref{NaI}). 

\begin{figure*}
\centering
    \includegraphics[width=0.8\textwidth]{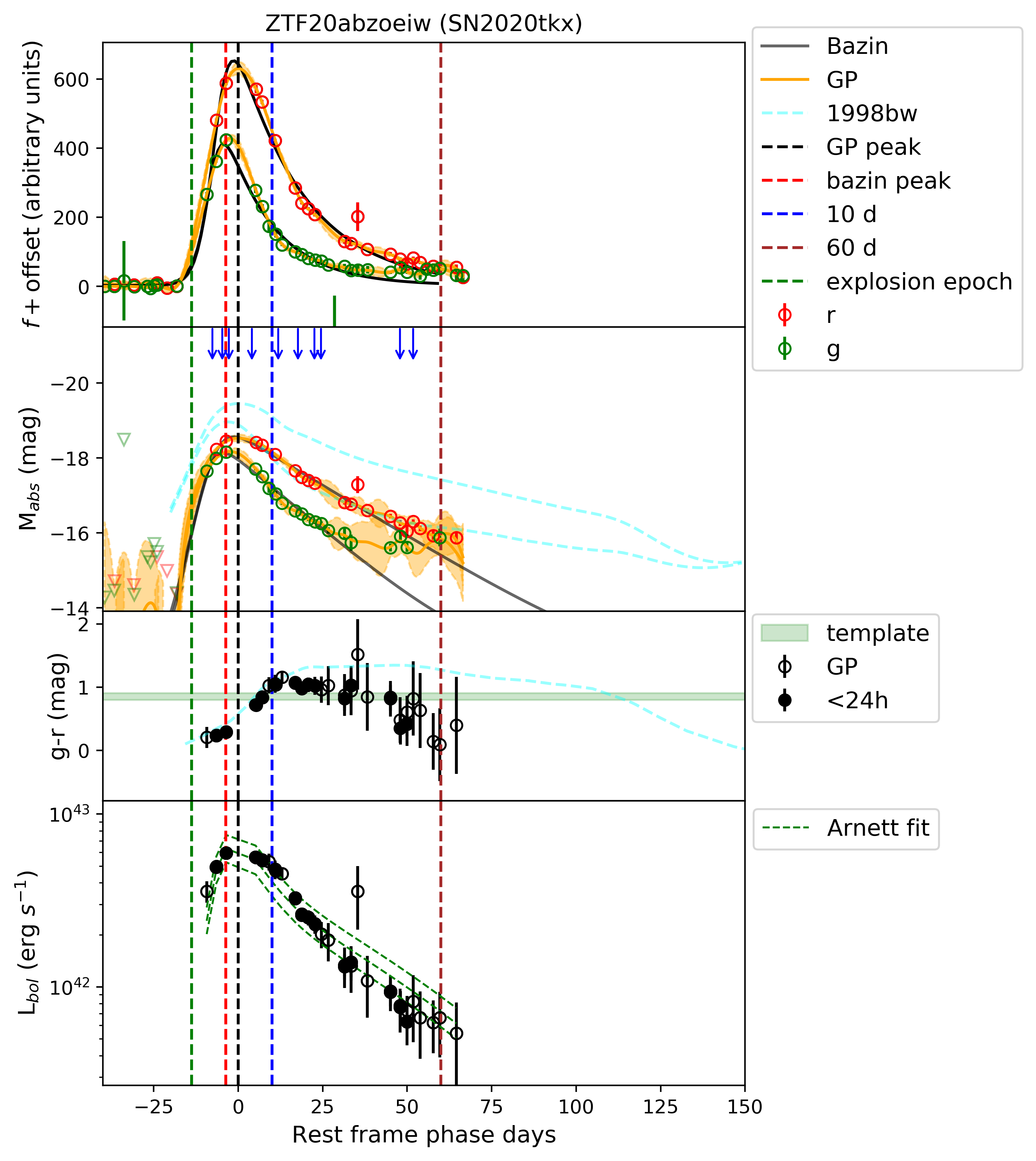}
    \caption{Example of LC fits for SN 2020tkx.  
    \textit{Upper panel}: The photometry in flux space together with the GP interpolation and fit to the Bazin formula. This allows for estimates of peak time and other LC parameters, provided in Table \ref{LCtable}. Peak magnitudes are shown in $r$ band, along wit the derived explosion epoch. \textit{Second panel}: The absolute magnitude LC, along with the GP inteprolation and best-fit Bazin function. SN 1998bw's LCs are plotted as comparison. The blue arrows indicate epochs were spectra were taken. \textit{Third panel}: The color evolution, along with that of SN 1998bw. The template shown is a range of colors for normal stripped envelope SNe from \citet{Sollerman2022}. \textit{Lower panel}: The bolometric luminosity LC, along with the best-fit \citet{arnett1982} model. }
    \label{examplefitting}
\end{figure*}

We use ZTF forced photometry to construct optical LCs for each event in the sample, and show the LCs in Figure \ref{ZTFLCs}, along with the $g-r$ colors. The ZTF forced photometry is unevenly sampled, making it difficult to derive LC parameters. Therefore, we use a non-parametric data-driven interpolation technique, Gaussian Processing (GP) to interpolate the LCs. We focus on the $g$ and $r$ band fluxes and use the \texttt{GEORGE} \citep{GEORGE} package with a stationary Matern 3/2 kernel and a flat mean function for the flux form. When the data either has large uncertainties or becomes more sparse, we fit the interpolated data with the analytic function from \citet{bazin}, where the flux is represented as 
\begin{equation}
\label{eqn:bazin}
F(t) = A \frac{e^{-(t-t_{0, \rm{ Bazin}})/\tau_\mathrm{fall, Bazin}}}{1+e^{-(t-t_{0, \rm{Bazin}})/\tau_\mathrm{rise, 
 Bazin}}} + B
\end{equation}
where $\tau_\mathrm{rise, Bazin}$ and $\tau_\mathrm{fall, Bazin}$ represents the rising (which we note differs from $\tau_\mathrm{rise}$ reported in \S \ref{4.3}, corresponding to the rise time of the SN from the explosion epoch to peak light), and declining time, A is a normalization parameter, $t_{0, \rm{ Bazin}}$ is a characteristic timescale (which we note differs from $t_0$ reported in \S \ref{empiricalparam}), while B is the baseline flux. After running a Monte Carlo simulation on the rest-frame LCs with priors and boundaries listed in Table~\ref{table:bazin}, we find the best-fitting Bazin function for each of our events, and use it in addition to the ZTF forced photometry and GP interpolated data to help derive LC parameters. An example of this fitting procedure is shown in Figure \ref{examplefitting}. This same LC interpolation process was done for the events also presented in \citet{Corsi2024} and \citet{Anand2024}, and for those events we redo the fits and find that they are consistent.  For parameters that were already derived in their works, we re-quote their results.

\begin{table}[]
    \centering
    \begin{tabular}{c|c|c}
      Parameter & Prior & Boundary \\ \hline
    $\tau_\mathrm{rise, Bazin}$ & 10 & [0, 60] days\\
$\tau_\mathrm{fall, Bazin}$ & -15 & [0, 120] days\\
$t_0$ & 0 & [-10, 10] days\\
$A$ & $F_\mathrm{max}$ & [$F_\mathrm{max}$*0.5, $F_\mathrm{max}$*2] \\
$B$ & 5 & [-20, 20] 
    \end{tabular}
    \caption{Priors used for the Bazin fits.}
    \label{table:bazin}
\end{table}

\subsection{Light Curve Empirical Parameters}
\label{empiricalparam}
After interpolating the LCs, we derive empirical LC parameters for every event, and show the results in Table \ref{LCs}. For events shared with \citet{Corsi2024} and \citet{Anand2024}, we report the peak absolute magnitudes in $r$ band ($\rm{M_{r}^{max}}$), the explosion epochs ($t_{\rm{exp,}r}$), and the peak time in $r$ band ($t_0$) derived in their works. 

\begin{figure}
    \centering
    \includegraphics[width=.99\linewidth]{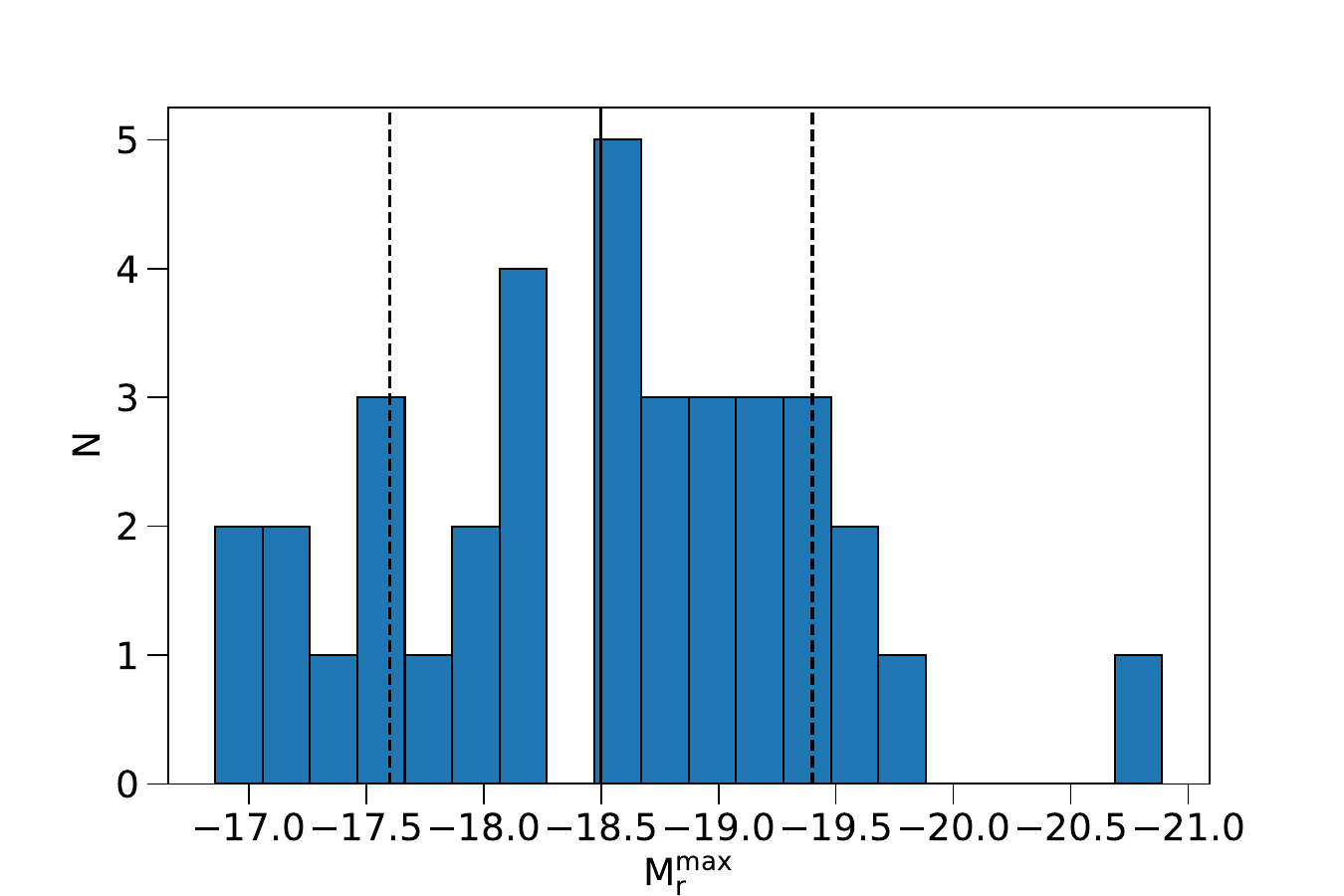}
    \caption{$\rm{M_{r}^{max}}$ distribution of the SNe Type Ic-BL sample, with the mean magnitude ($-18.51$ mag) shown with a solid line, and standard deviation (0.90 mag) range shown with dashed lines. }
    \label{maghist}
\end{figure}

We begin by determining $\rm{M_{r}^{max}}$ and $t_0$ for new events, using the GP-processed LCs.  We show a histogram of the peak $r$-band absolute magnitude distributions of our sample in Figure \ref{maghist}. The distribution ranges over 4 magnitudes from $-16.86$ mag to $-20.89$ mag. SN 2020wgz is a clear outlier with respect to the rest of the sample, with $\rm{M}_r^{\rm{max}} = -20.89 \pm 0.11$ mag, placing it in the luminosity regime of SLSNe. We discuss this event's properties at length in \S \ref{2020wgz}. The average peak magnitude is $\rm{\overline{M}_{r}^{\rm{max}}} = -18.51 \pm 0.15$ mag, with an associated 1$\sigma$ standard deviation of 0.90 mag (shown in the Figure as dashed lines). The average absolute magnitude derived is consistent with that found in the iPTF sample of \citet{taddia2019iptf} for events without an associated GRB ($\rm{\overline{M}}_r^{\rm{max}} = -18.6$ mag, with a 1$\sigma$ standard deviation of 0.5 mag). When excluding SN 2020wgz from the sample, we find a mean peak magnitude of $\overline{\rm{{M}}}_r^{\rm{max}} = -18.44 \pm 0.14$ mag, and a 1$\sigma$ standard deviation of 0.82 mag. The uncertainty that we report is derived from the uncertainty in photometric observations,  and these magnitudes are all corrected for MW extinction. 

We then derive the $g-r$ colors for each of the events in the sample 10 days after peak: $(g-r)_{10}$, to compare the color evolution of the sample. We find an average value of $\overline{(g-r)}_{10}$ = $1.00 \pm 0.06$ mag, with a 1$\sigma$ standard deviation of 0.43 mag. All of the events in the sample except for SN 2020ayz, SN 2018gep, SN 2020wgz, and SN 2020zg have $(g-r)_{10}$ values consistent within the average and 1$\sigma$ standard deviation, showing that the sample has broadly consistent color evolution.

We then investigate the effects of the Malmquist bias on the peak $r$-band absolute magnitudes of the sample. The Malmquist bias arises because more luminous objects can be detected out to greater distances than less luminous objects. Because our sample is magnitude-limited, it is likely that ZTF failed to detect fainter SNe that occurred at greater distances, and we quantify this effect below. 

\begin{figure}
    \centering
    \includegraphics[width=\linewidth]{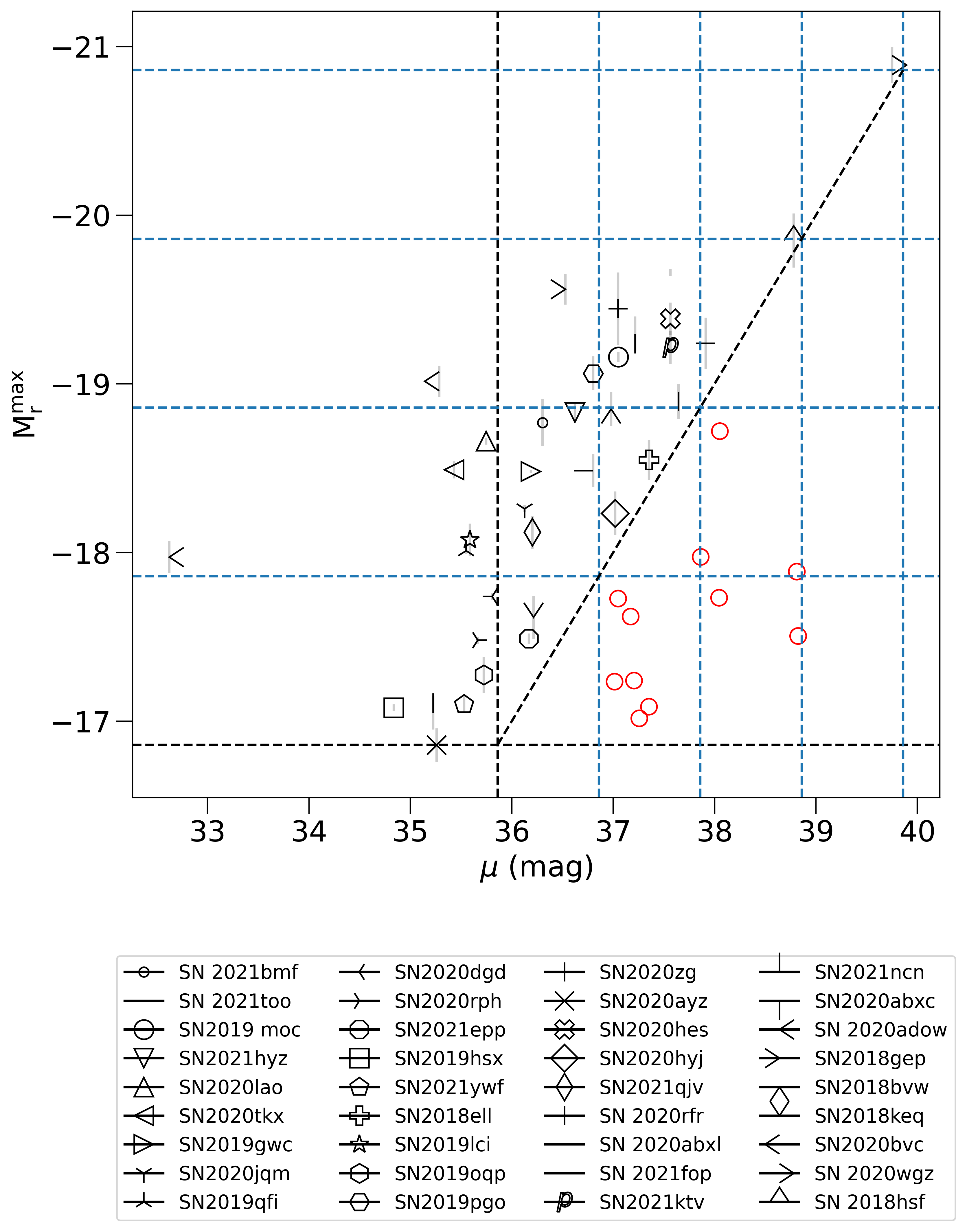}
    \caption{Investigation of the Malmquist bias on the sample. $\rm{M_{r}^{max}}$ and the distance modulus ($\mu$) for the events are plotted, with the minimum absolute magnitude (-16.86 mag) of the sample shown as a black dashed horizontal line. The distance modulus where the sample is complete ($\mu = 35.86$ mag) is shown as a black dashed vertical line. One magnitude by one magnitude bins are marked by dashed blue lines, where the cutoff distance for the sample is marked by a black dashed vertical line. The random simulated events created through the method described in \citet{Richardson2014} are shown as red circles. The overall effect of the Malmquist bias is a 0.2 mag decrease in the overall average $\rm{M_{r}^{max}}$ of the sample.  }
    \label{malmquist}
\end{figure}

We plot the peak magnitudes and distance modulus ($\mu$) for each event in the sample in Figure \ref{malmquist}. The faintest peak absolute $r$-band magnitude of the sample is $-16.86$. The BTS survey detects objects with an apparent magnitude as faint as $\sim$ 19 mag, so we use this value as the upper limit for SN detectability for the BTS survey. The minimum peak absolute $r$-band magnitude and upper limit for SN detectability sets $\mu = 35.86$ as the distance modulus for which we observe the complete peak magnitude distribution. We then utilize the method described in \citet{Richardson2014} to simulate the magnitude of missing SNe randomly in one magnitude bins, cutting off the sample at the furthest event. The results of the simulation are shown in Figure \ref{malmquist}. We calculate a new average $r$-band peak absolute magnitude of $-18.3$ mag, with a $1\sigma$ dispersion of 0.91 mag. Therefore, the Malmquist bias produces a difference of $\sim$ 0.2 mag in the average peak magnitude distribution for the sample. This is the same difference found in \citet{taddia2019iptf} for their sample. We note that this difference is just an estimate, and more complicated effects (e.g., the distribution of events with respect to their position in the Milky Way plane, the assumption that there are no events fainter than the faintest in our sample for the whole population) were not accounted for. 

\begin{table*}[]
\caption{Empirical parameters, explosion epochs, and peak $r$-band times for the SNe Ic-BL sample derived through the methods outlined in \S \ref{LCs}. We draw from results in previous works where applicable (described in \S \ref{empiricalparam}.)}
\begin{tabular}{l|l|l|l|l|l|l|l}
\hline
\hline

ZTF name     & SN name     & $(g-r)_{10}$ (mag)        & $\Delta m_{15}$ (mag)   & $\Delta m_{-10} $ (mag)    & $\rm{M_{r, peak}}$            & $t_{\rm{exp,}r}$ (days)                   & $t_0$ (JD)         \\
\hline
 ZTF18aaqjovh & SN 2018bvw & - & 0.63 (0.05) & 1.42 (0.04) & $-18.85$ (0.10) & $-8.09^{+2.04}_{-2.04}$ & 2458248.80 \\ 
        ZTF18abhhnnv & SN 2018ell & 1.13 (0.49) & 0.87 (0.09) & 0.65 (0.12) & $-18.55$ (0.12) & $-16.15^{+0.8}_{-0.84}$ & 2458330.71 \\ 
        ZTF18abukavn & SN 2018gep & 0.51 (0.03) & 1.15 (0.02) & - & $-19.56$ (0.09) & $-4.09^{+0.01}_{-0.01}$ & 2458374.74 \\ 
        ZTF18acbvpzj & SN 2018hsf & - & - & 0.59 (0.36) & $-19.85$ (0.16) & $-12.43^{+2.43}_{2.43}$ & 2458432.83 \\
        ZTF18acxgoki & SN 2018keq & 1.26 (0.22) & 0.33 (0.11) & 0.78 (0.17) & $-17.62$ (0.13) & $-14.07^{+4.1}_{-4.1}$ & 2458475.68 \\
        ZTF19aawqcgy & SN 2019hsx & 1.32 (0.47) & 0.88 (0.09) & 0.65 (0.11) & $-17.08$ (0.02) & $-15.63^{+0.38}_{-0.53}$ & 2458647.57 \\
        ZTF19aaxfcpq & SN 2019gwc & 0.90 (0.14) & 0.88 (0.06) & 1.14 (0.06) & $-18.48$ (0.01) & $-12.78^{+0.46}_{-0.46}$ & 	2458651.08 \\ 
        ZTF19abfsxpw & SN 2019lci & 0.87 (0.39) & - & 0.07 (0.03) & $-18.07$ (0.10) & $-18.1^{+0.47}_{-0.2}$ & 2458693.71 \\ 
        ZTF19ablesob & SN 2019moc & 1.00 (0.11) & 0.77 (0.13) & 0.21 (0.14) & $-19.16$ (0.03) & $-20.02^{+0.27}_{-3.23}$ & 2458716.26 \\ 
        ZTF19abqshry & SN 2019oqp & 0.89 (0.16) & 0.69 (0.19) & 0.21 (0.07) & $-17.27$ (0.11) & $-22.45^{+0.41}_{-0.41}$ & 2458737.71 \\ 
        ZTF19abupned & SN 2019pgo & 1.39 (0.1) & 0.85 (0.04) & 0.43 (0.04) & $-19.06$ (0.10) & $-15.85^{+0.38}_{-0.66}$ & 2458743.91 \\ 
        ZTF19abzwaen & SN 2019qfi & 1.03 (0.53) & 0.92 (0.16) & 0.76 (0.39) & $-18.01$ (0.02) & $-15.09^{+1.4}_{-1.4}$ & 2458754.06 \\ 
        ZTF20aafmdzj & SN 2020zg & 0.37 (0.08) & 0.37 (0.22) & - & $-19.45$ (0.21) & $-5.06^{+0.95}_{-0.95}$ & 2458867.74 \\ 
        ZTF20aaiqiti & SN 2020ayz & 1.75 (0.13) & 0.81 (0.03) & 0.52 (0.06) & $-16.86$ (0.10) & $-14.33^{+0.32}_{-0.51}$ & 2458887.97 \\ 
        ZTF20aalxlis & SN 2020bvc & 1.00 (0.02) & 0.84 (0.03) & 0.48 (0.02) & $-19.02$ (0.09) & $-18.11^{+0.89}_{-0.89}$ & 2458901.06 \\ 
        ZTF20aapcbmc & SN 2020dgd & 1.10 (0.8) & 0.75 (0.37) & 0.28 (0.14) & $-17.74$ (0.02) & $-18.03^{+2.5}_{-2.5}$ & 	2458914.55 \\ 
        ZTF20aaurexl & SN 2020hes & 0.77 (0.13) & 1.02 (0.09) & 0.41 (0.03) & $-19.39$ (0.10) & $-16.03^{+4.97}_{-4.97}$ & 2458964.99 \\
        ZTF20aavcvrm & SN 2020hyj & 0.92 (0.11) & 0.66 (0.02) & 0.66 (0.06) & $-18.23$ (0.13) & $-15.01^{+0.98}_{-0.98}$ & 2458969.94 \\ 
        ZTF20aazkjfv & SN 2020jqm & 1.13 (0.5) & 0.56 (0.08) & 0.55 (0.08) & $-18.26$ (0.02) & $-17^{+1}_{-1}$ & 2458996.71 \\ 
        ZTF20abbplei & SN 2020lao & 0.92 (0.07) & 0.83 (0.08) & 2.88 (0.1) & $-18.66$ (0.02) & $-10.6^{+0.99}_{-0.99}$ & 	2459004.42 \\ 
        ZTF20abrmmah & SN 2020rfr & 1.40 (0.26) & 0.05 (0.11) & 2.21 (0.14) & $-18.90$ (0.10) & $-6.99^{+1.96}_{-1.96}$ & 2459078.86 \\ 
        ZTF20abswdbg & SN 2020rph & 0.96 (0.27) & 0.48 (0.08) & 0.49 (0.06) & $-17.48$ (0.02) & $-19.88^{+0.02}_{-0.02}$ & 	2459092.84 \\ 
        ZTF20abzoeiw & SN 2020tkx & 1.01 (0.15) & 0.72 (0.13) & 0.79 (0.22) & $-18.49$ (0.05) & $-12.77^{+4.54}_{-4.54}$ & 2459117.00 \\ 
        ZTF20achvlbs & SN 2020wgz & 0.06 (0.05) & 0.64 (0.04) & 2.53 (0.09) & $-20.89$ (0.11) & $-10.04^{+0.15}_{-0.15}$ & 2459140.01 \\ 
        ZTF20acvcxkz & SN 2020abxl & 1.45 (0.31) & 0.8 (0.05) & 0.45 (0.03) & $-19.24$ (0.15) & $-13.25^{+0.01}_{-0.01}$ & 2459200.97 \\ 
        ZTF20acvmzfv & SN 2020abxc & 0.90 (0.14) & 0.56 (0.05) & 0.69 (0.05) & $-19.30$ (0.10) & $-12.95^{+0.47}_{-0.88}$ & 2459203.73 \\ 
        ZTF20adadrhw & SN 2020adow & 0.59 (0.01) & 0.99 (0.01) & 1.9 (0.01) & $-17.97$ (0.09) & $-10.59^{+2.4}_{-2.4}$ & 2459218.95 \\ 
        ZTF21aagtpro & SN 2021bmf & 0.72 (0.13) & 1.14 (0.17) & 0.13 (0.06) & $-18.77$ (0.14) & $-23.76^{+5.68}_{-5.52}$ & 2459265.62 \\ 
        ZTF21aaocrlm & SN 2021epp & 1.30 (0.23) & 0.32 (0.04) & 0.4 (0.06) & $-17.49$ (0.03) & $-15.12^{+1.48}_{-1.48}$ & 2459292.33 \\ 
        ZTF21aapecxb & SN 2021fop & 0.74 (0.15) & 0.68 (0.08) & 2.08 (0.36) & $-18.49$ (0.10) & $-7.05^{+5.01}_{-5.01}$ & 2459292.75 \\ 
        ZTF21aartgiv & SN 2021hyz & 0.84 (0.19) & 1.04 (0.19) & 1.52 (0.08) & $-18.83$ (0.05) & $-12.88^{+0.94}_{-0.94}$ & 2459319.81 \\ 
        ZTF21aaxxihx & SN 2021ktv & 1.15 (0.29) & 0.53 (0.09) & 0.92 (0.07) & $-19.22$ (0.10) & $-12.04^{+2.94}_{-2.94}$ & 2459344.79 \\ 
        ZTF21abchjer & SN 2021ncn & 1.01 (0.13) & 1.1 (0.10) & 1.19 (0.15) & $-17.05$ (0.10) & $-10.02^{+0.01}_{-0.01}$ & 2459364.97 \\
        ZTF20abcjdwu & SN 2021qjv & 1.31 (0.08) & 0.87 (0.10) & 2.83 (0.20) & $-18.12$ (0.10) & $-10.01^{+0.02}_{-0.02}$ & 2459389.83 \\ 
        ZTF21abmjgwf & SN 2021too & 1.20 (0.18) & 0.59 (0.15) & 0.22 (0.08) & $-19.66$ (0.02) & $-23.23^{+0.41}_{-0.41}$ & 2459434.59 \\ 
        ZTF21acbnfos & SN 2021ywf & 1.1 (0.6) & 0.68 (0.42) & 0.44 (0.25) & $-17.10$ (0.05) & $-17.47^{+0.49}_{-0.49}$ & 2459485.95 \\ 
\end{tabular}
\label{LCtable}
\end{table*}

\begin{figure}
    \centering
    \includegraphics[width=.99\linewidth]{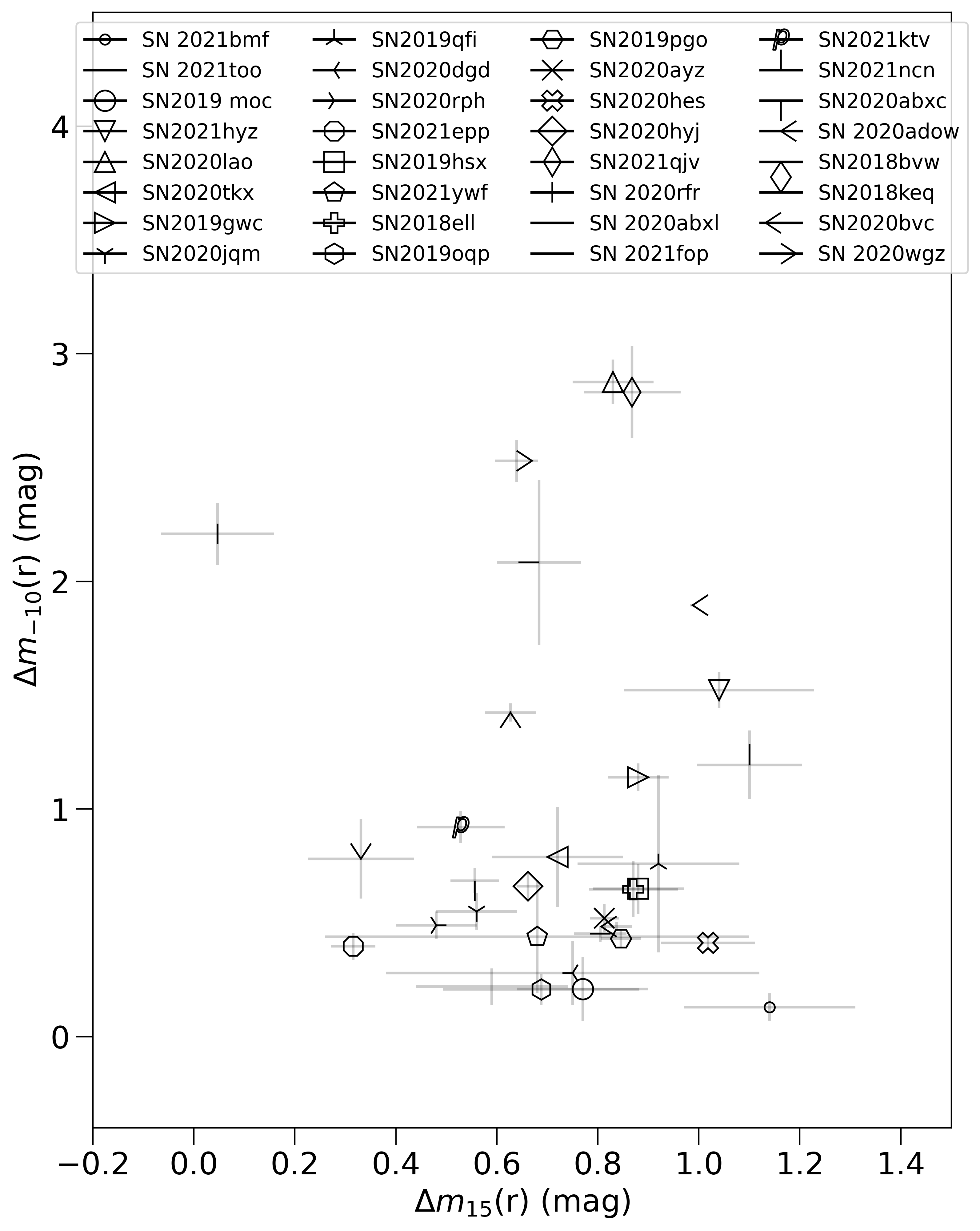}
    \caption{$\Delta m_{-10}(r)$ plotted against $\Delta m_{15}(r)$ for 32 events in the sample. We found no correlations between the two parameters, with a $p$-value of 0.98. }
    \label{declinevsrise}
\end{figure}

We also derive the decline parameter $\Delta m_{15}(r)$ and rise parameter $\Delta m_{-10}(r)$, which is the difference in magnitudes in the $r$-band from the peak to 15 days after the peak, and from 10 days before the peak to the peak, for every event in the sample. In Figure \ref{declinevsrise}, we show $\Delta m_{-10}(r)$ plotted against $\Delta m_{15}(r)$ for our sample, with the exception of SN 2018hsf, SN 2019lci, SN 2020zg, and SN 2018gep,  (the first three events lacked sufficient photometry at the times necessary to calculate these parameters while SN 2018gep had a rise time from the explosion epoch to peak quicker than 10 days). Through a Spearman rank coefficient test, we do not find any correlation between the two parameters, with a $p$-value of 0.98. This contrasts with the results found in \citet{taddia2019iptf}, who found that fast-rising objects also are fast-decliners, with a $p$-value of 0.06. However, they only tested the correlation for 12 of the best-sampled events in their sample, while we test the correlation using 32 events, utilizing the GP interpolations and Bazin fits. Therefore, we show that when removing observational biases with respect to the best-sampled LCs, that fast-rising SNe Ic-BL are not necessarily also fast decliners.

\begin{figure}
    \centering
    \includegraphics[width=.99\linewidth]{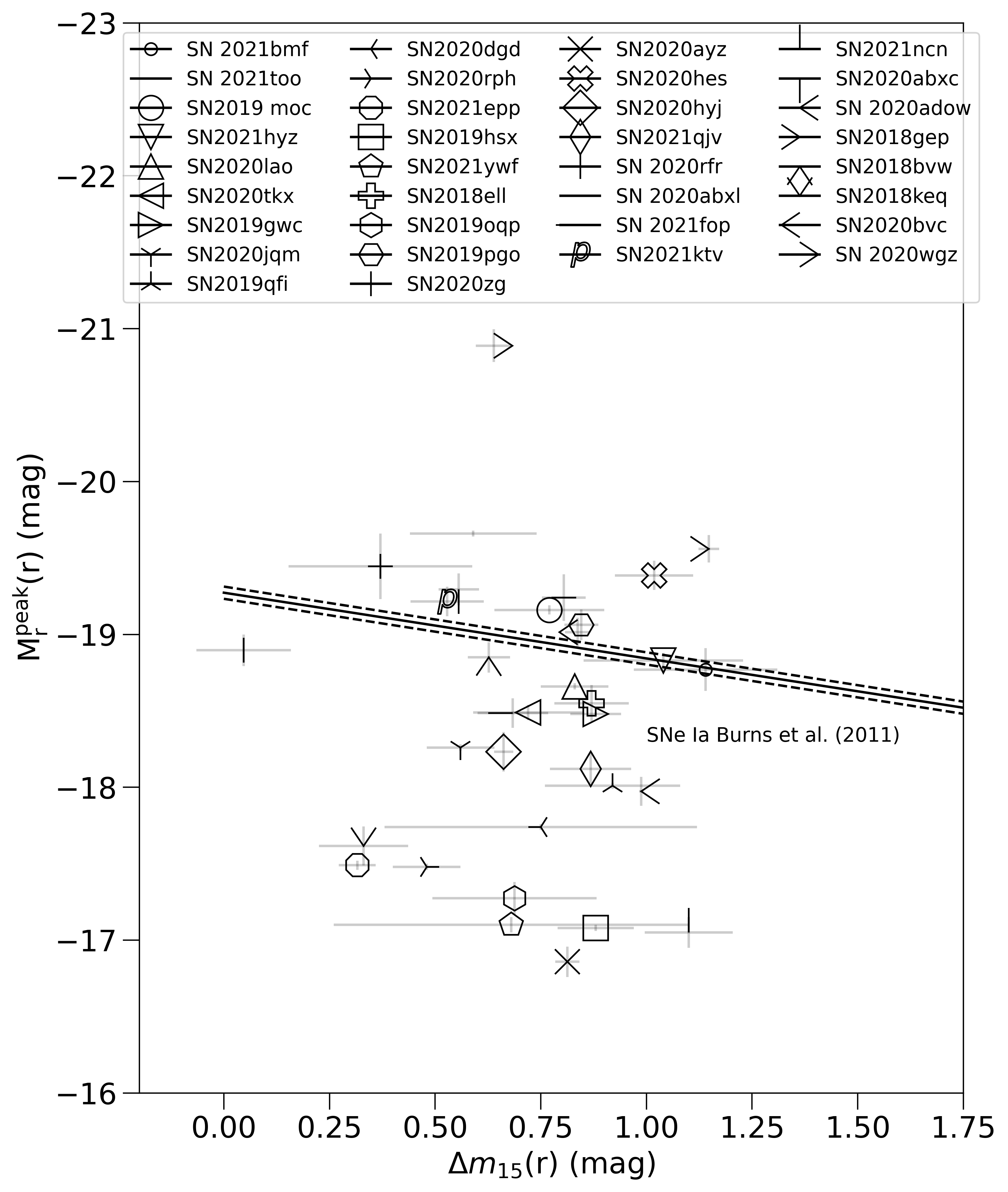}
    \caption{$\rm{M}_r^{\rm{max}}$  plotted against $\Delta m_{15}(r)$ for 34 events in the sample. There is no evidence for a Phillips relation, which is shown for SNe Ia as the black line with the dashed lines representing the errors from \citet{Burns2011}. }
    \label{Phillips}
\end{figure}

We then test for the presence of a Phillips Relation \citep{Phillips1993}, between $\rm{M}_r^{\rm{max}}$ and  $\Delta m_{15}(r)$.  A relation between the decline rate and luminosity was established for SNe Ia through this relation \citep{Burns2011} and was also found for GRB-SNe \citep{Cano2014, Li2014}. We show these two parameters plotted against each other for 34 events in our sample (with the exception of SN 2018hsf and SN 2019lci), as well as the Phillips Relation for SNe Ia from \citet{Burns2011} in Figure \ref{Phillips}. We do not find this relation present in our sample, consistent with the findings from \citet{taddia2019iptf}. Therefore, less luminous SNe do not necessarily decline faster than more luminous events.

\subsection{Explosion epochs}
\label{4.3}

When sufficient early-time photometry is available, we determine the explosion epoch ($t_{\rm{exp,}r}$) and rise time ($\tau_{\rm{rise}}$) of every newly presented SN through Monte Carlo approaches, after characterizing the early emission of the transients in $g$ and $r$ bands through a power law following the methodology of \cite{Miller_2020}\footnote{\href{https://github.com/adamamiller/ztf_early_Ia_2018}{https://github.com/adamamiller/ztf\_early\_Ia\_2018}.} The power law is fit from the estimated baseline up to 40 percent of the maximum flux determined by the Bazin fit. For those SNe which do not have sufficient early-time photometry, we take the average of the first detection and the last non-detection in ZTF forced photometry as the explosion epoch, and determine the error through the half-width. SN 2018hsf did not have a ZTF non-detection prior to the first detection, due to the instrument not operating for some time period before. For this case, we use the last ATLAS non-detection in $o$ band to calculate the explosion epoch. We calculate the rise time $\rm{\tau_{rise}}$ obtained for our sample, measured from the derived explosion epoch to the peak epoch in $r$-band, along with the average value and 1$\sigma$ standard deviation. We find $\rm{\overline{\tau}_{rise}} = 14.0 \pm 0.8$ days, with a 1$\sigma$ standard deviation of 5.81 days. This is consistent with the value obtained in \citet{taddia2019iptf}, who found $\rm{\overline{\tau}_{rise}} = 15$ days with a 1$\sigma$ standard deviation of 6 days.

%\begin{figure}
    %\centering
    %\includegraphics[width=.99\linewidth]{Risetimecdf.pdf}
    %\caption{Rise time CDF of the SNe Ic-BL sample in blue, with the mean (14.08 days) shown as a solid line, and 1$\sigma$ standard deviation (5.81 days) shown as dashed lines. The CDF of the iPTF sample from \citet{taddia2019iptf} is shown in green with its mean (15 days) and 1$\sigma$ standard deviation (6 days) shown as well.}
    %\label{risetimecdf}
%\end{figure}

\begin{figure*}
    \centering
    \includegraphics[width = 0.4\linewidth]{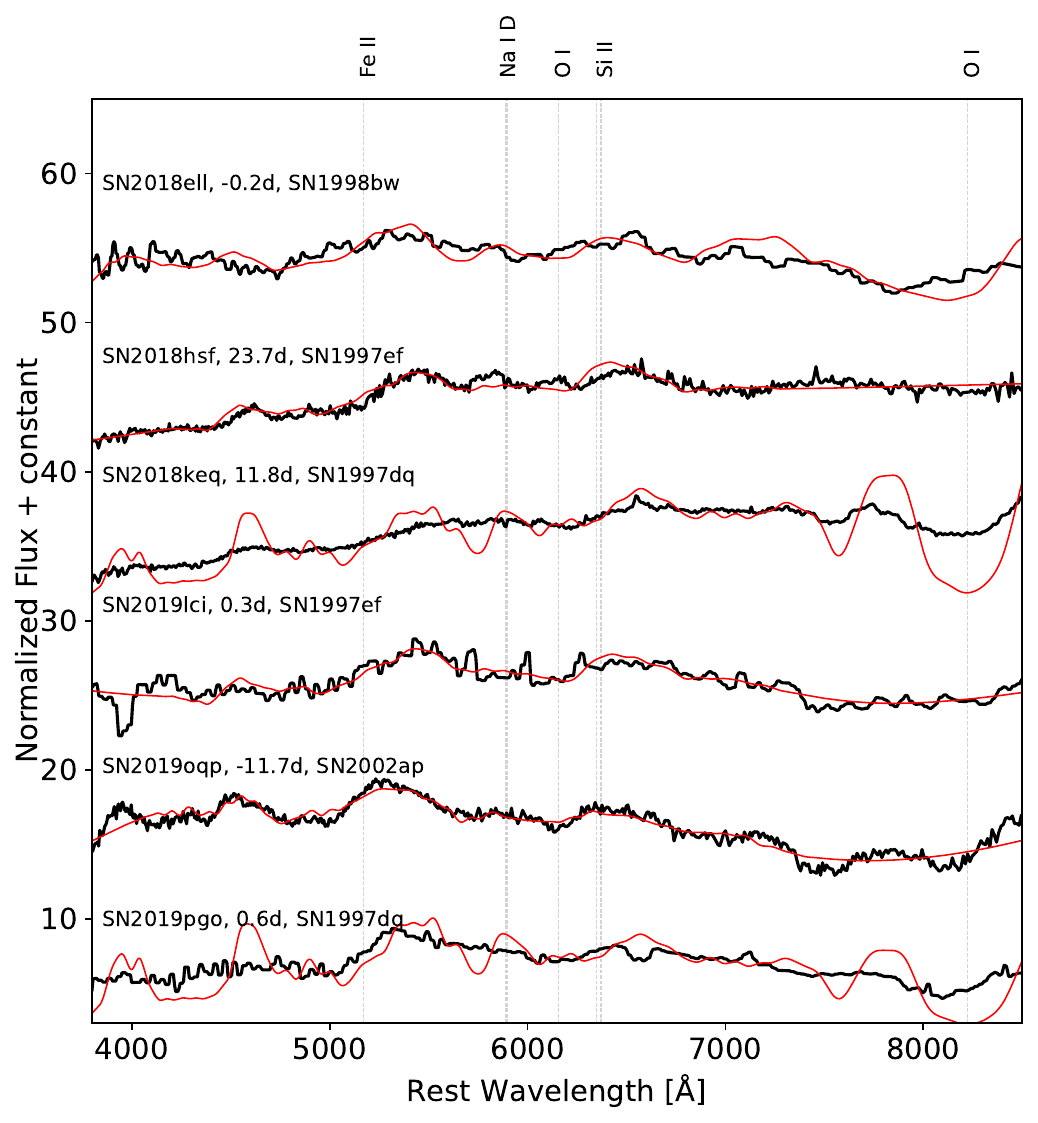}
    \includegraphics[width = 0.4\linewidth]{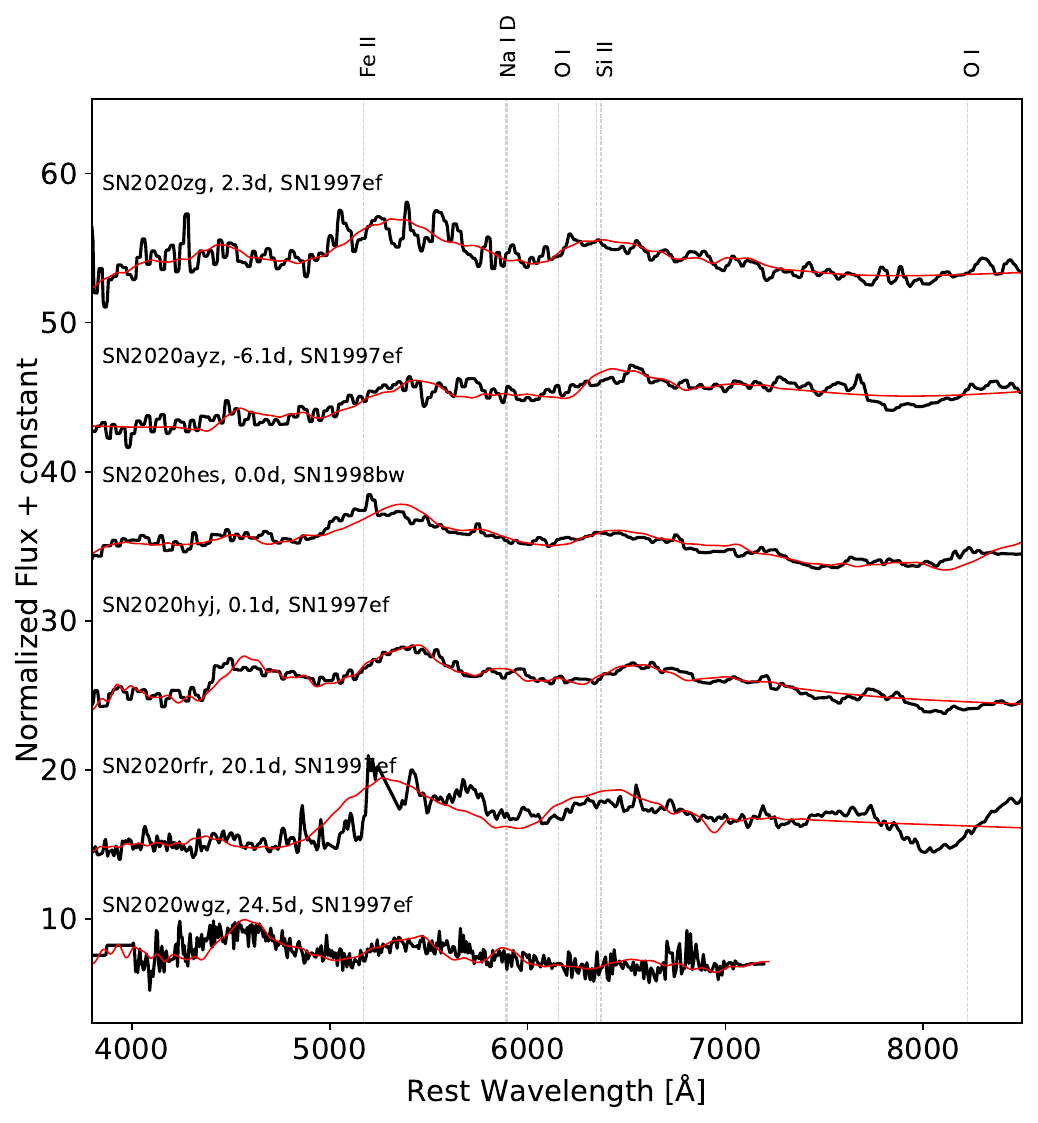}
     \includegraphics[width = 0.4\linewidth]{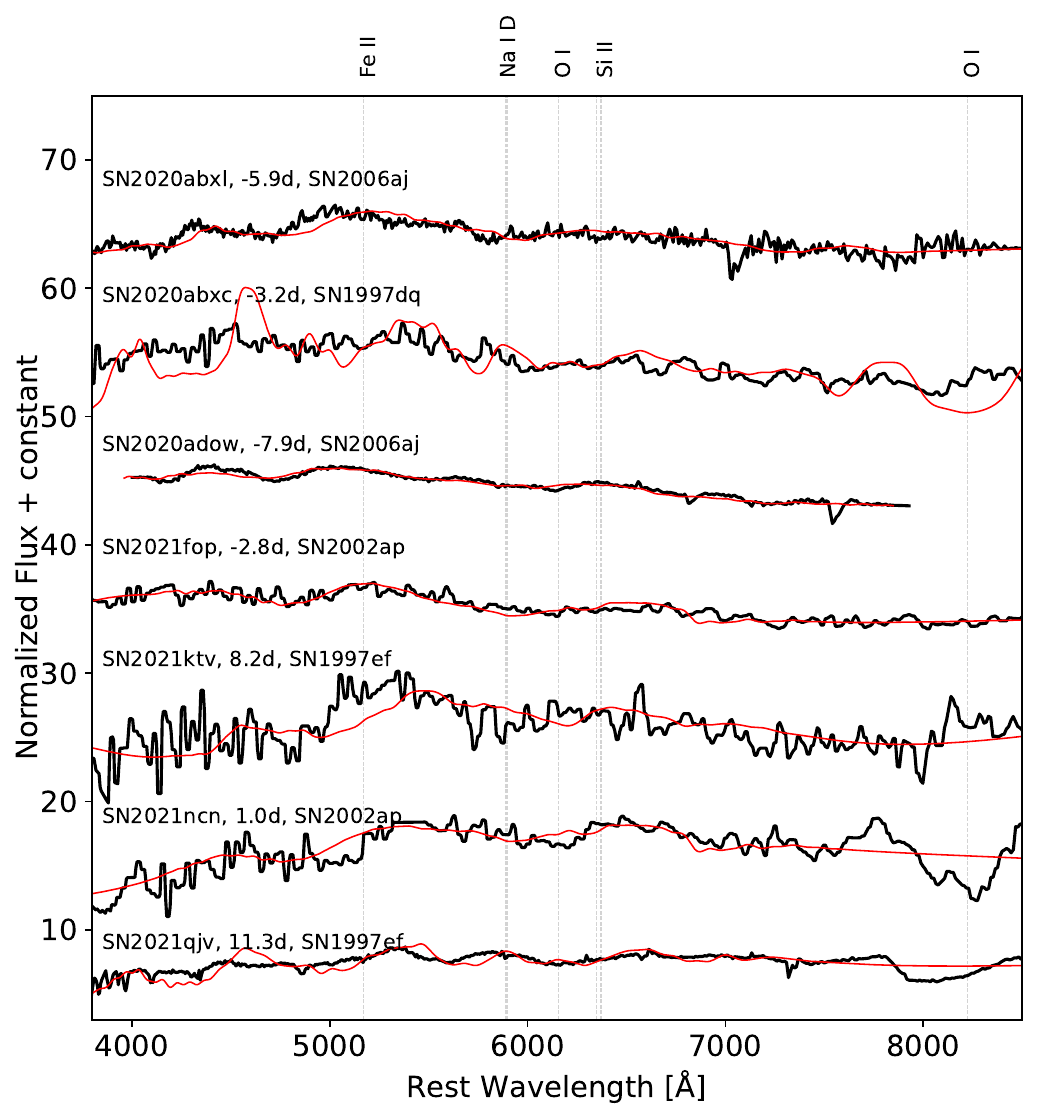}
    \caption{Photospheric phase spectra (black) and the SNID best match templates (red) for the Ic-BL sample that has not been presented in previous works. The text shows the name of the transient, the observer-frame spectroscopic phase since $r$-band maximum, and the name of the transient template used for the fitting.}
    \label{classification}
\end{figure*}

\section{Supernova Spectra}
\label{spectra}

We present the photospheric phase spectra for each of the new events in our sample in Figure \ref{classification}, not presented in previous works. In \S \ref{Observations}, we detail the instruments used to obtain spectra of the events in our sample. All of the spectra were reduced in a standard manner, with wavelength and flux calibrations. All spectra will be made publicly available on WISeREP \citep{WiseRep}.

 As shown in Figure \ref{classification}, none of the spectra display hydrogen and helium features, while they all share characteristic broad absorption features. Their broad features are the result of the blending of multiple absorption lines due to Doppler broadening effects, and the Fe II and Si II absorption lines are indicated in the Figure. 

 \subsection{Photospheric Velocities}
 In addition to the broadened features, the center of the absorption troughs are blueshifted relative to the rest frame Fe II and Si II absorption line wavelengths, due to the high velocity of the ejecta towards the observer. \citet{Modjaz2016} showed that the blueshifted absorption velocity of the Fe II line at 5169 $\AA$ is a good proxy for the photospheric expansion velocity ($\rm{v_{ph}}$), and we calculate these velocities for each of the events in our sample. We use a similar method to \citet{Anand2024}, beginning by using \texttt{WOMBAT} to remove host galaxy emission lines and telluric features, and use \sw{SESNSpectraPCA} \citep{SESNSpectraPCA} to smooth the spectra. Finally, we utilize \sw{SESNSpectraLib} \citep{Modjaz2016, Liu2016} to fit for the Fe II absorption velocity, by convolving a blueshifted Gaussian with a Type Ic SN template, and measuring the blueshift of the Fe II feature with respect to these templates.

In order to derive $\rm{v_{ph}}$ at the peak of the SNe LCs, we use the highest quality spectrum available closest to the peak of the SNe to measure the velocities. For some events, this is different than the spectrum used for classification purposes. The full set of spectra for each SN in our sample is shown in the Appendix. Through this method, we were able to obtain either an estimate or constraint of the peak $\rm{v_{ph}}$ for 26 out of the 36 events in our sample (10 events had spectra that lacked sufficent SNR to obtain an accurate velocity measurement). For six of the events, we could only obtain velocity measurements more than 15 days post peak light, and therefore their velocities are not representative of the peak velocity. All the velocities are reported in Table \ref{explosiontable}, where we report the velocities derived for new events, along with those already presented in literature. In Figure \ref{vejhist}, we show a histogram of the distribution of the peak $\rm{v_{ph}}$, without the events with spectra taken greater than 15 days from peak. We find an average value of 16,100 $\pm$ 1,100 km $\rm{s^{-1}}$, with a 1$\sigma$ standard deviation of 5,600 km $\rm{s^{-1}}$. In Figure \ref{vejevol}, we show the evolution of $\rm{v}_{\rm{ph}}$ over time for the sample, compared to other GRB-SNe, X-ray flash (XRF)-SNe, a ``normal" Type Ic SN, the iPTF sample of \citet{taddia2019iptf}, and the samples from \citet{Anand2024} and \citet{Corsi2024}. We see that our sample's evolution is broadly consistent with that of the iPTF sample and the other single object events shown, with the exception of the ``normal" Type Ic SN 1994I. 

\subsection{Na I Equivalent Width}
\label{NaI}
Out of the 36 events in the sample, 20 had spectra from either DBSP or LRIS, which are the highest resolution spectra in the sample (though still low-resolution in general). We measure the equivalent width (EW) of the Na I absorption feature for every event in the sample, as this feature has been shown to be a proxy for the amount of host-galaxy extinction present \citep{Stritzinger2018}. 14 events, including the majority of events in \citet{Corsi2024} and \citet{Anand2024} had Na I EWs consistent to 0 within error bars. There are five new events not previously analyzed in literature that have EWs greater than 0: SN 2018hsf, SN 2018keq, SN 2019lci, SN 2020ayz, and SN 2020bvc. One event from \citet{Anand2024}, SN 2019hsx, showed a significant Na I feature, which is at odds with their conclusion that no events in the sample demonstrated the feature. The EWs are shown in Table \ref{hostextinct}.

It is possible to convert the EWs to host-galaxy extinctions through the relation from \citet{Stritzinger2018}: $A_{\rm{V}}^{\rm{host}}\rm{[mag]} = 0.78(\pm0.15) \times EW_{\rm{Na \, I}}$. However, this relation only holds strongly when using high-resolution spectra for these measurements; when using low-resolution spectra, \citet{Poznanski2011} showed that even though a weak correlation exists, the large scatter makes any relation between the two quantities useless. Therefore, it is possible that these six events have additional host galaxy extinction; however due to the lack of high-resolution spectra, we cannot quantify the amount or say with certainty that a significant amount of extinction is present. Therefore, we compute a conservative upper limit on the amount of host-galaxy extinction for these events, and show the results in Table \ref{hostextinct}. However, given the lack of constraints, we do not correct for host-galaxy extinction for any of these events during our analysis.

\begin{table}[]
    \centering
    \begin{tabular}{c|c|c}
      Event & $EW_{\rm{Na \, I}}$ & $< A_{\rm{V}}^{\rm{host}}\rm{[mag]}$ \\ \hline
    SN 2018hsf & $1.23 \pm 0.14$ & 1.3\\
SN 2018keq & $1.01 \pm 0.33$ & 1.2\\
SN 2019hsx & $0.80 \pm 0.39$ & 1.1\\
SN 2019lci & $0.47 \pm 0.22$ & 0.6 \\
SN 2020ayz & $1.52 \pm 0.13$ & 1.5 \\
SN 2020bvc & $0.55 \pm 0.17$ & 0.7  
    \end{tabular}
    \caption{Na I EW and resulting host galaxy extinction upper limits derived for events that show evidence of Na I absorption.}
    \label{hostextinct}
\end{table}

%There are many caveats in using the Na I EW that likely produce systematic uncertainties in the host galaxy extinction greater than the statistical errors we report, due to the use of low-resolution spectra for these measurements \citep{Poznanski2011}. Therefore, the errors we report for these events do not include these systematic effects. 

%We convert the equivalent widths to host extinctions through the relation from \citet{Stritzinger2018}: $A_{\rm{V}}^{\rm{host}}\rm{[mag]} = 0.78(\pm0.15) \times EW_{\rm{Na \, I}}$. We then correct these six events for their host-galaxy extinction. We simply assume the mean value, as

\begin{figure}
    \centering
    \includegraphics[width=\linewidth]{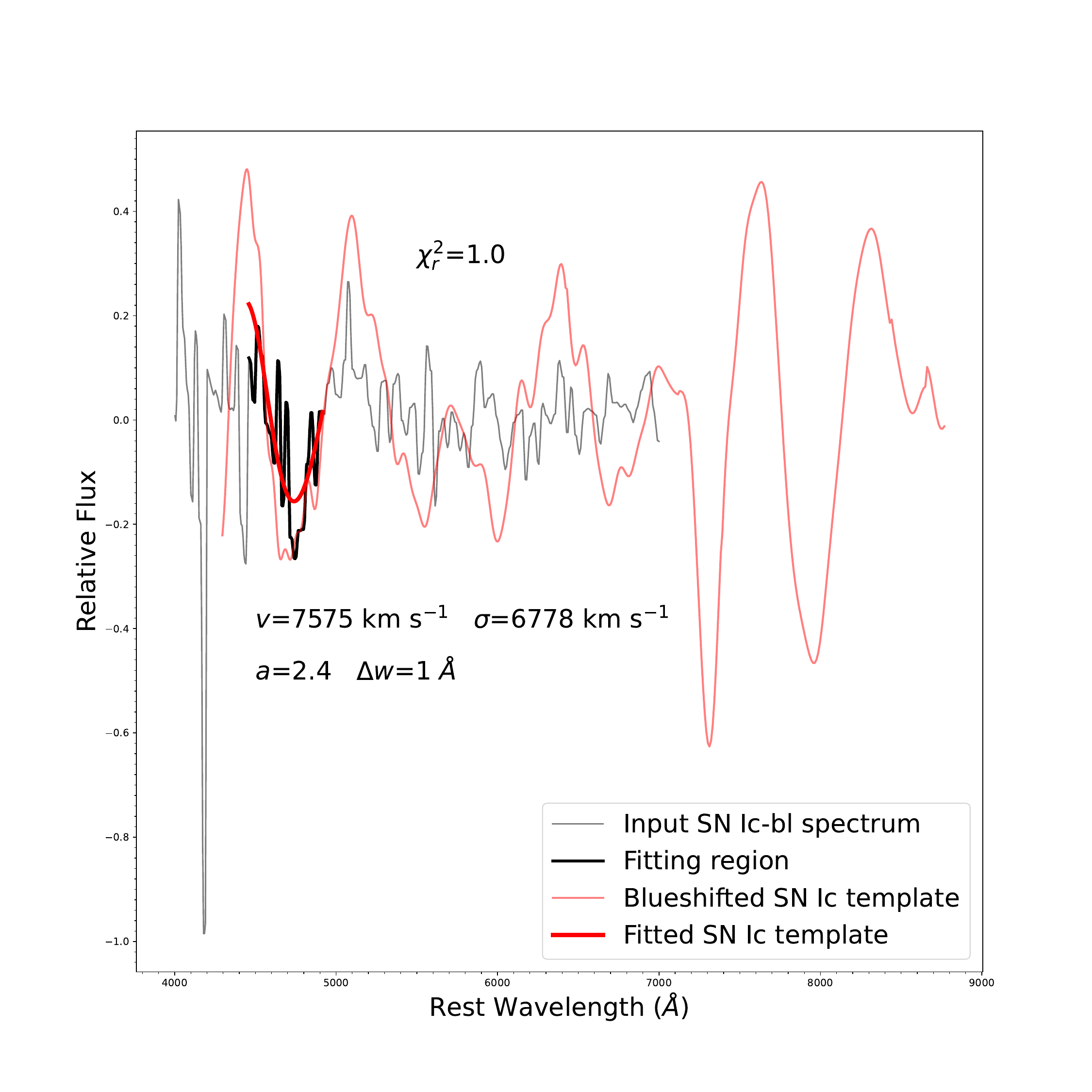}
    \caption{Example of the convolution fitting of the Fe II feature for the photospheric phase spectra of SN 2018hsf. The spectrum of the transient is shown in gray, with the fitting region bolded in black. The blueshifted Ic template is shown in red, with the fitting region bolded in red. The reduced chi-square value for the fit is shown ($\chi_r^2$), along with the blueshifted velocity with respect to the Ic template ($v$), the Doppler broadened line width velocity ($\sigma$), the amplitude ($a$), and the wavelength range ($\delta w$).}
    \label{fig:enter-label}
\end{figure}

\begin{figure}
    \centering
    \includegraphics[width = \linewidth]{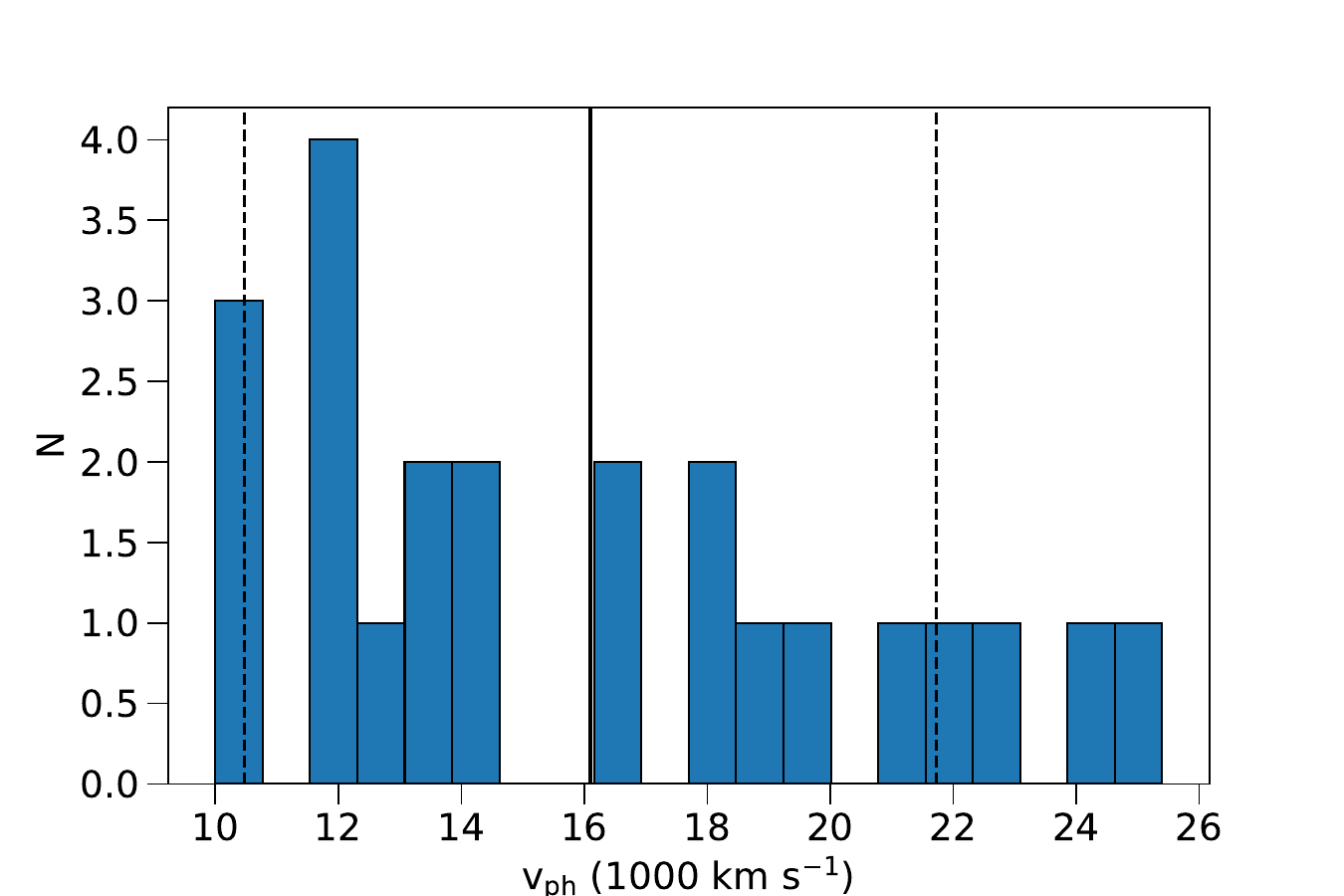}
    \caption{The peak $\rm{v_{ph}}$ for the sample, with the mean velocity (16,100 km s$^{-1}$) shown with a solid line, and 1$\sigma$ standard deviation (5,600 km s$^{-1}$) range shown with dashed lines.}
    \label{vejhist}
\end{figure}

\begin{figure}
    \centering
    \includegraphics[width = \linewidth]{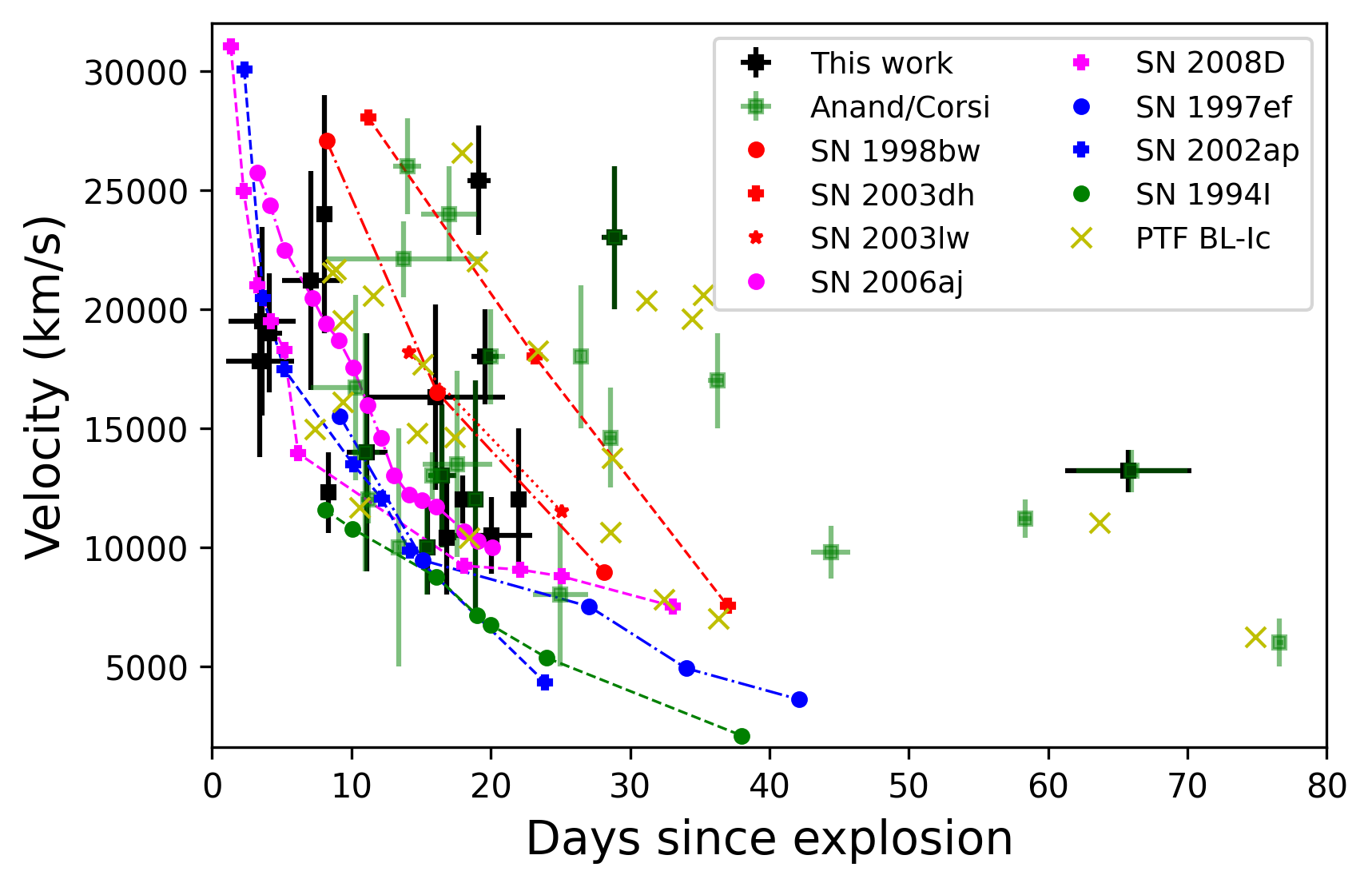}
    \caption{SN velocities measured from the Fe II 5169$\rm \AA$ line as a function of the spectroscopic phase for every event in the sample that a velocity was able to be obtained (26 events). The black points indicate the velocities at the spectral epoch, plotted with measured velocities of SNe Ic-BL from literature, the iPTF sample \citep{taddia2019iptf}, and the samples from \citet{Anand2024} and \citet{Corsi2024}. Red symbols represent GRB-SNe \citep{Iwamoto1998,Mazzali2003,Mazzali2006b}; magenta represents XRF/X-ray transient-SNe \citep{Mazzali2006a,Pian2006,Modjaz2009}; blue represents SNe Ic-BL \citep{Mazzali2000,Mazzali2002}; and green represents the ``normal'' Type Ic SN 1994I \citep{Sauer2006}. }
    \label{vejevol}
\end{figure}
 
%\subsection{SN Line velocity}

\section{Bolometric LCs and Properties}
\label{Bollabel}

%In order to calculate the bolometric luminosty, distance is an important ingredient.

\subsection{Bolometric light curves and peak luminosites}
\label{sec:lum}
\begin{figure}
    \centering
    \includegraphics[width=.99\linewidth]{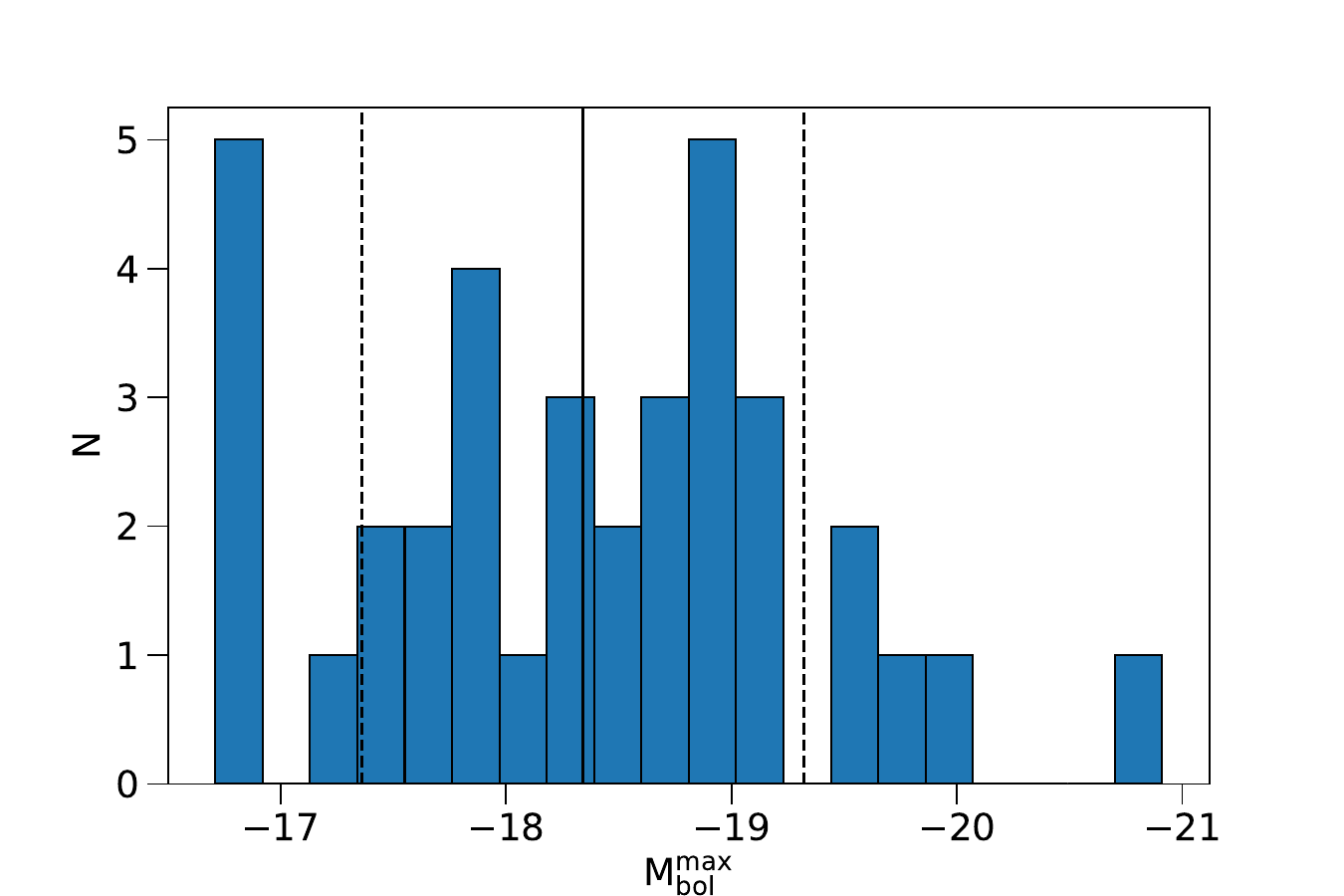}
    \caption{$\rm{M_{bol}^{max}}$ of all SNe in the sample, with the mean magnitude (-18.34 mag) shown with a solid line, and 1$\sigma$ standard deviation range (0.98 mag) shown with dashed lines. }
    \label{bolmaghist}
\end{figure}

There is a lack of complete multi-band coverage for the majority of events in our sample -- in particular the coverage in the $i$ band is sparse. Therefore, due to the sufficient coverage in $g$ and $r$ bands, we use the $g-r$ colors calculated in \S \ref{empiricalparam} along with the bolometric correction (BC) coefficients of \cite{Lyman14, lyman16} to compute bolometric LCs for our sample. According to \cite{Lyman14}, stripped-envelope supernovae possess a BC as follows:
\begin{equation}
  \text{BC}_{g} = 0.054 - 0.195 \times (g-r) - 0.719 \times (g-r)^{2}.
  \label{eq:bc_se_sl}
\end{equation}
 This BC is valid after the initial shock-breakout phase. We calculate the $\rm{BC}_g$ coefficient for every epoch in our sample using the GP-interpolated data. Then, using the definition of BC coefficients:

\begin{equation}
  \text{BC}_{x} = \rm{M_{\rm{bol}}} - \rm{M_{x}},
\label{eq:bc}
\end{equation}
where $x$ is the relevant filter, $\rm{M}_{\rm{bol}}$ is the absolute bolometric magnitude, and $\rm{M}_{x}$ is the absolute magnitude in the relevant filter, we calculate $\rm{M}_{\rm{bol}}$ at every epoch. In Figure \ref{bolmaghist}, we show the distribution of $\rm{M_{bol}^{max}}$ for the sample. The distribution ranges from $-16.28$ to $-20.91$ mag, and we find an average $\overline{\rm{M}}_{bol}^{max}$ of $-18.34 \pm 0.16$ mag, with a 1$\sigma$ standard deviation of 0.98 mag. This is consistent with the value found in \citet{taddia2019iptf}, who found $\rm{\overline{M}}_{bol}^{max} = -18.5$ with a 1$\sigma$ standard deviation of  0.5 mag. When excluding SLSN SN 2020wgz from the sample, we find an average $\overline{\rm{M}_{bol}^{max}}$ of $-18.27 \pm 0.15$ mag, with a 1$\sigma$ standard deviation of 0.89 mag.

\begin{figure*}
    \centering
    \includegraphics[width = 0.8\linewidth]{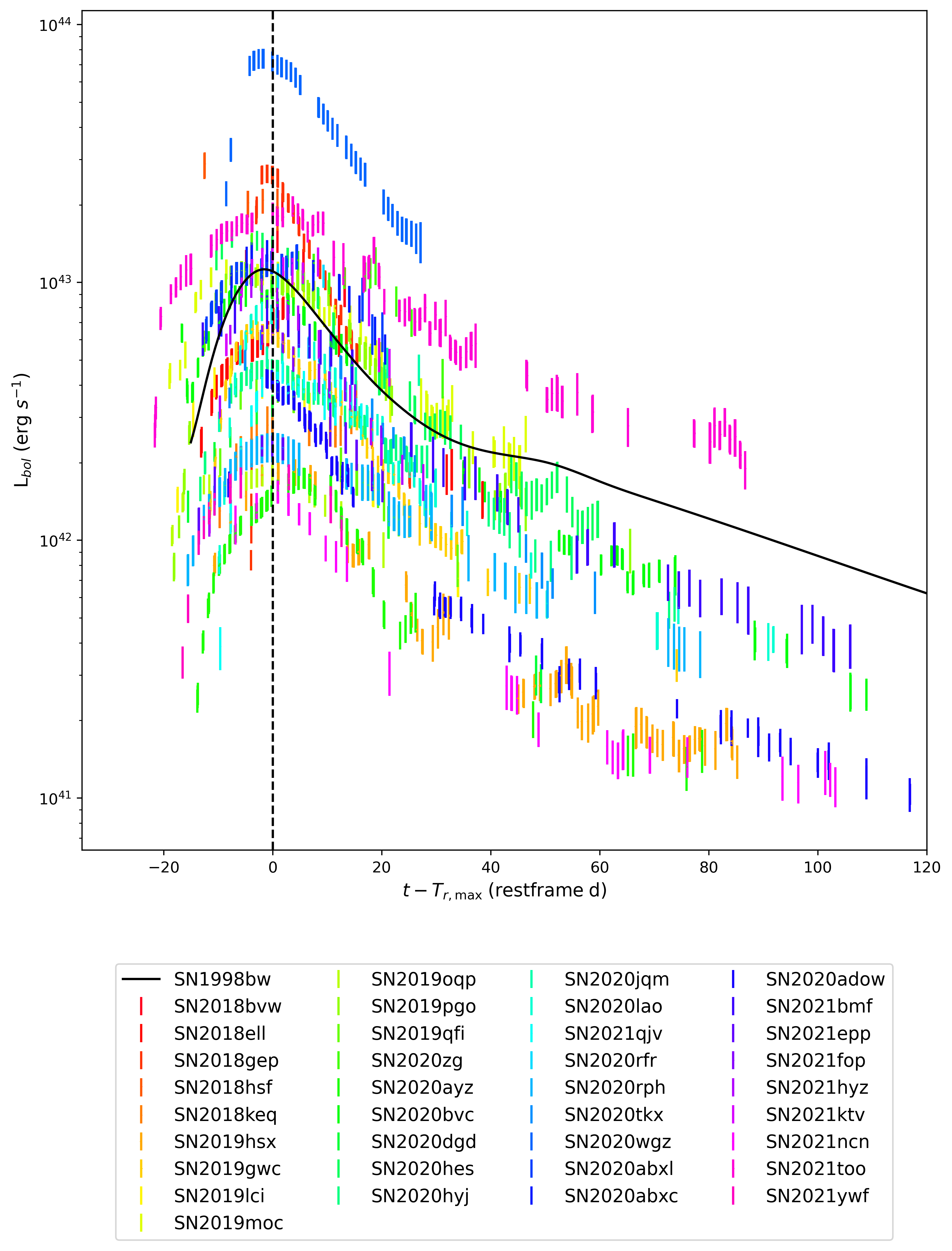}
    \caption{Bolometric luminosity LCs for the SNe Ic-BL sample, calculated using Lyman BC coefficients (described in \S \ref{sec:lum}). The bolometric luminosity LC of SN 1998bw \citep{98bwpaper} is also shown as comparison. }
    \label{bolLCs}
\end{figure*}

We then convert $\rm{M_{\rm{bol}}}$ to a bolometric luminosity. To calculate uncertainties we include the photometric uncertainty on the peak magnitude that we estimate from the GP analysis and a 15\% correction to take into account MW extinction, a peculiar velocity correction uncertainty of 150~km~s$^{-1}$, and the uncertainty of the Hubble constant, $\pm3$~km~s$^{-1}$~Mpc$^{-1}$. In Figure \ref{bolLCs}, we show the bolometric luminosity LCs of our sample.

\subsection{Blackbody Temperature and Radii Evolution}
\label{blackbody}
Because we have multi-band photometry, we are able to investigate the spectral energy distribution (SED) evolution over time for the events in our sample. In particular, we calculate how the blackbody temperature and radius evolve over time, through fitting a diluted blackbody function to the SEDs:

\begin{equation}
    F_{\lambda}=(R/d)^2 \cdot \epsilon^2 \cdot \pi \cdot B(\lambda, T) \times 10^{-0.4 \cdot A_{\lambda}}.
\end{equation}

In this function, $F_{\lambda}$ is the flux at wavelength $\lambda$, $B$ is the Planck function, $A_{\lambda}$ is the extinction, $T$ is the temperature, $R$ is the radius, $d$ is the distance, and $\epsilon$ is the dilution factor \citep{E96, Hamuy01, D05} representing a correction between the BB distribution we fit to the observed fluxes. 
%Photometry in the $i$ band is also acquired for some events directly, and for events that do not have sufficient $i$ band coverage, we use  interpolated $i$ band data (see \S \ref{interpolation}).

\begin{figure*}
    \centering
    \includegraphics[width = 0.8\linewidth]{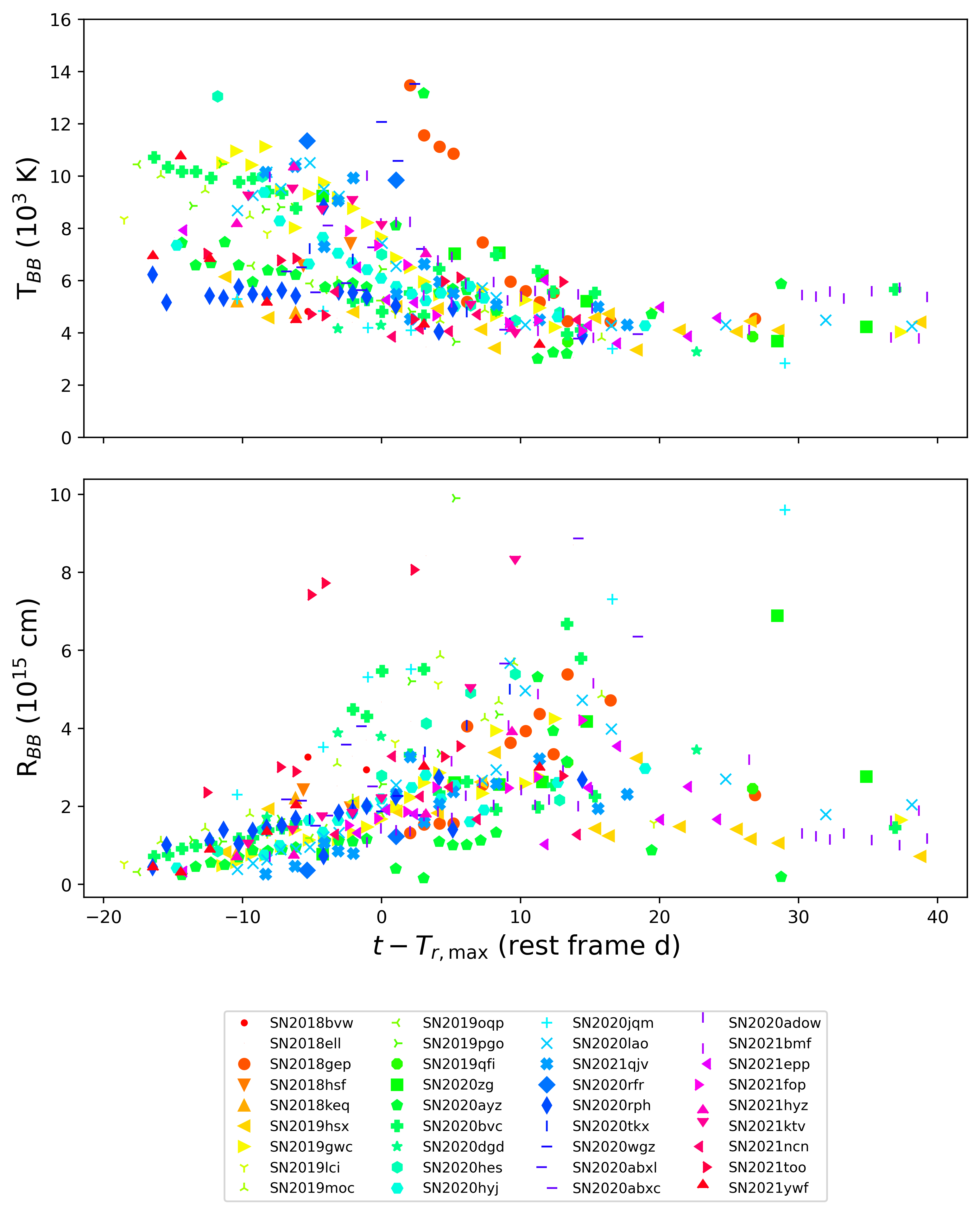}
    \caption{Blackbody temperatwure and radii evolution for the SNe Ic-BL sample, calculated through the methods reported in \S \ref{blackbody}.}
    \label{blackbodyfigure}
\end{figure*}

The SEDs consist of photometry binned within 0.5 days of each other, using $g$, $r$, and $i$-band data. For some cases, we also have \textit{Swift} host-corrected UVOT photometry (detailed in \S \ref{Swiftobs}), and include those points in the SEDs when available. We estimate the luminosity by integrating the blackbody distributions from 2000 $\AA$ to 20000 $\AA$. The blackbody temperature and radius evolution over time are shown in Figure \ref{blackbodyfigure}. We find that the temperature shows a progressive decline until $\sim 5$ days after peak, when the distribution begins to flatten out between 4000 and 7000 K, which is similar to what is found in \citet{taddia2019iptf}. We find that the radius increases to around 10 days after peak and then decreases for the majority of the sample, again consistent to what is found in \citet{taddia2019iptf}. 

\section{Explosion Properties}
\label{explosionlabel}

\begin{table*}[]
\caption{Explosion properties and photospheric velocities for the SNe Ic-BL sample. Values reported include those newly derived and those presented in previous works.}
\centering
\begin{tabular}{l|l|l|l|l|l|l}
\hline
\hline
ZTF name     & SN name     & $\rm{M_{Ni}} \, \rm{(M_\odot})$                    & $\rm{E_{K}} \, (10^{51} \rm{erg})$                & $\rm{M_{ej}} \, \rm{(M_\odot})$                & $\rm{v_{ph}}$ ($10^3$ km s$^{-1}$  &  Spectrum phase  (d)     \\
\hline
ZTF18aaqjovh & SN 2018bvw & $0.52^{+0.05}_{-0.05}$ & $4.45\pm2.7$ & $1.66\pm0.5$ & $21.2\pm4.6$ & -1 \\ 
        ZTF18abhhnnv & SN 2018ell & $0.31^{+0.01}_{-0.01}$ &  $11.44\pm3.84$ & $3\pm0.35$ & $25.4\pm2.3$ & 3.2\\ 
        ZTF18abukavn & SN 2018gep & $0.61^{+0.01}_{-0.01}$ & $0.7\pm0.6$ & $0.2\pm0.04$ & $24\pm5$ &4 \\ 
        ZTF18acbvpzj & SN 2018hsf & $0.7^{+0.13}_{-0.1}$ & $>0.07$ & $>0.31$ & $8.2\pm0.94$ & 23.7 \\ 
        ZTF18acxgoki & SN 2018keq & $0.16^{+0.05}_{-0.05}$ & - & - & - & - \\ 
        ZTF19aawqcgy & SN 2019hsx & $0.07^{+0.01}_{-0.01}$ & $0.99\pm0.5$ & $1.64\pm0.43$ & $10\pm2$ & -0.2 \\ 
        ZTF19aaxfcpq & SN 2019gwc & $0.22^{+0.01}_{-0.01}$ & $>0.44$ & $>0.6$ & $11.1\pm0.9$ & 46.0 \\ 
        ZTF19abfsxpw & SN 2019lci & $0.24^{+0.01}_{-0.01}$ & - & - & -  & -\\ 
        ZTF19ablesob & SN 2019moc & $0.52^{+0.01}_{-0.02}$ & $3.48\pm1.85$ & $2.09\pm0.5$ & $16.8\pm3.9$ & -9.7\\ 
        ZTF19abqshry & SN 2019oqp & $0.08^{+0}_{-0}$ & - & - & - & -\\ 
        ZTF19abupned & SN 2019pgo & $0.55^{+0.01}_{-0.01}$ & $0.89\pm0.6$ & $1.37\pm0.4$ & $10.4\pm2.4$ & 1\\ 
        ZTF19abzwaen & SN 2019qfi & $0.13^{+0.01}_{-0.01}$ & $>0.7$ & $>1.22$ & $9.59\pm1.2$ & 29.3 \\ 
        ZTF20aafmdzj & SN 2020zg & $0.5^{+0.01}_{-0.01}$ & - & - & - & - \\ 
        ZTF20aaiqiti & SN 2020ayz & $0.07^{+0}.01_{-0.01}$ & $1.85\pm0.92$ & $2\pm0.3$ & $12.3\pm1.7$ & $-6.1$ \\ 
        ZTF20aalxlis & SN 2020bvc & $0.41^{+0.01}_{-0.01}$ & $5.15\pm2.45$ & $2.39\pm0.45$ & $19\pm2.5$ & -14\\ 
        ZTF20aapcbmc & SN 2020dgd & $0.13^{+0.03}_{-0.03}$ & $3.07\pm2.42$ & $2.81\pm1.5$ & $13.5\pm3.9$ &-0.43 \\ 
        ZTF20aaurexl & SN 2020hes & $0.58^{+0.01}_{-0.01}$ & $2\pm1.9$ & $1.26\pm0.36$ & $16.3\pm3.9$ & 0\\ 
        ZTF20aavcvrm & SN 2020hyj & $0.22^{+0.01}_{-0.01}$ & - & - & - & -\\ 
        ZTF20aazkjfv & SN 2020jqm & $0.29^{+0.05}_{-0.04}$ & $5\pm3$ & $5\pm1$ & $13\pm3$ &-0.5\\ 
        ZTF20abbplei & SN 2020lao & $0.23^{+0.01}_{-0.01}$ & $2.48\pm0.71$ & $1.22\pm0.16$ & $18\pm2$ & 9 \\ 
        ZTF20abrmmah & SN 2020rfr & $0.47^{+0.03}_{-0.03}$ & - & - & - & -\\  
        ZTF20abswdbg & SN 2020rph & $0.07^{+0.01}_{-0.01}$ & $3.08\pm2.81$ & $3.83\pm1.59$ & $12\pm5$ & -1\\ 
        ZTF20abzoeiw & SN 2020tkx & $0.22^{+0.01}_{-0.01}$ & $>1.5$ & $>1.5$ & $13.2\pm0.9$ & 53\\ 
        ZTF20achvlbs & SN 2020wgz & $2.46^{+0.02}_{-0.02}$ & $>$ 0.12 & $>0.48$ & $11.1 \pm 0.3$ & 25\\ 
        ZTF20acvcxkz & SN 2020abxl & $0.4^{+0.01}_{-0.01}$ & - & - & - &- \\ 
        ZTF20acvmzfv & SN 2020abxc & $0.61^{+0.02}_{-0.02}$ & - & - & - &- \\ 
        ZTF20adadrhw & SN 2020adow & $0.14^{+0.02}_{-0.02}$ & $2.2\pm1.1$ & $1\pm0.2$ & $19.5\pm3.96$ & -7.9\\ 
        ZTF21aagtpro & SN 2021bmf & $0.98^{+0.16}_{-0.17}$ & $23.63\pm16.14$ & $8.05\pm5.37$ & $21.9\pm1.5$ & $-10.0$ \\ 
        ZTF21aaocrlm & SN 2021epp & $0.12^{+0.02}_{-0.02}$ & $6\pm5$ & $5\pm2$ & $14\pm5$ & -4 \\ 
        ZTF21aapecxb & SN 2021fop & $0.19^{+0}_{-0}$ & - & - & - & - \\ 
        ZTF21aartgiv & SN 2021hyz & $0.29^{+0.01}_{-0.02}$ & $>4$ & $>1.3$ & $23\pm3$ & 16 \\ 
        ZTF21aaxxihx & SN 2021ktv & $0.48^{+0.01}_{-0.01}$ & $0.63\pm0.4$ & $0.95\pm0.2$ & $10.5\pm1.6$ & 8.2 \\ 
        ZTF21abchjer & SN 2021ncn & $0.05^{+0.01}_{-0.01}$ & - & - & - & - \\ 
        ZTF20abcjdwu & SN 2021qjv & $0.17^{+0.01}_{-0.01}$ & $1\pm1$ & $1.16\pm0.35$ & $12\pm3$ & 11.3\\ 
        ZTF21abmjgwf & SN 2021too & $0.92^{+0.03}_{-0.03}$ & $6.42\pm2.09$ & $5.06\pm0.78$ & $14.4\pm2.1$ & 5.4 \\ 
        ZTF21acbnfos & SN 2021ywf & $0.06^{+0.01}_{-0.01}$ & $0.9\pm0.3$ & $1.1\pm0.2$ & $12\pm1$ & 0.5\\    
\end{tabular}
\label{explosiontable}
\end{table*}

After computing bolometric LCs, we fit the LCs up to the peak luminosity (usually 20 to 60 rest-frame days from first detection) to analytic models from \citet{arnett1982}. In these models, the instantaneous heating rate from the decay of $^{56}$Ni and $^{56}$Co is equivalent to the peak bolometric luminosity of the SN. The model also assumes spherical symmetry in the explosion, and full gamma-ray trapping of the ejecta. In these models, the nickel mass ($\rm{M_{Ni}}$) and characteristic photon diffusion time scale ($\tau_m$) are free parameters. These parameters are important probes of the explosion mechanisms, as the amount of $\rm{M_{Ni}}$ powers the bolometric LC, while $\tau_m$ is a proxy for the rise timescale of the SN, along with relating to the kinetic energy ($\rm{E_{K}}$) and ejecta mass ($\rm{M_{ej}}$) of the SN. We note that \citet{Corsi2024} and \citet{Anand2024} used the same methods to derive the explosion parameters for events in their sample, so for overlapping events we report the values derived in their works.

\begin{figure}
    \centering
    \includegraphics[width = \linewidth]{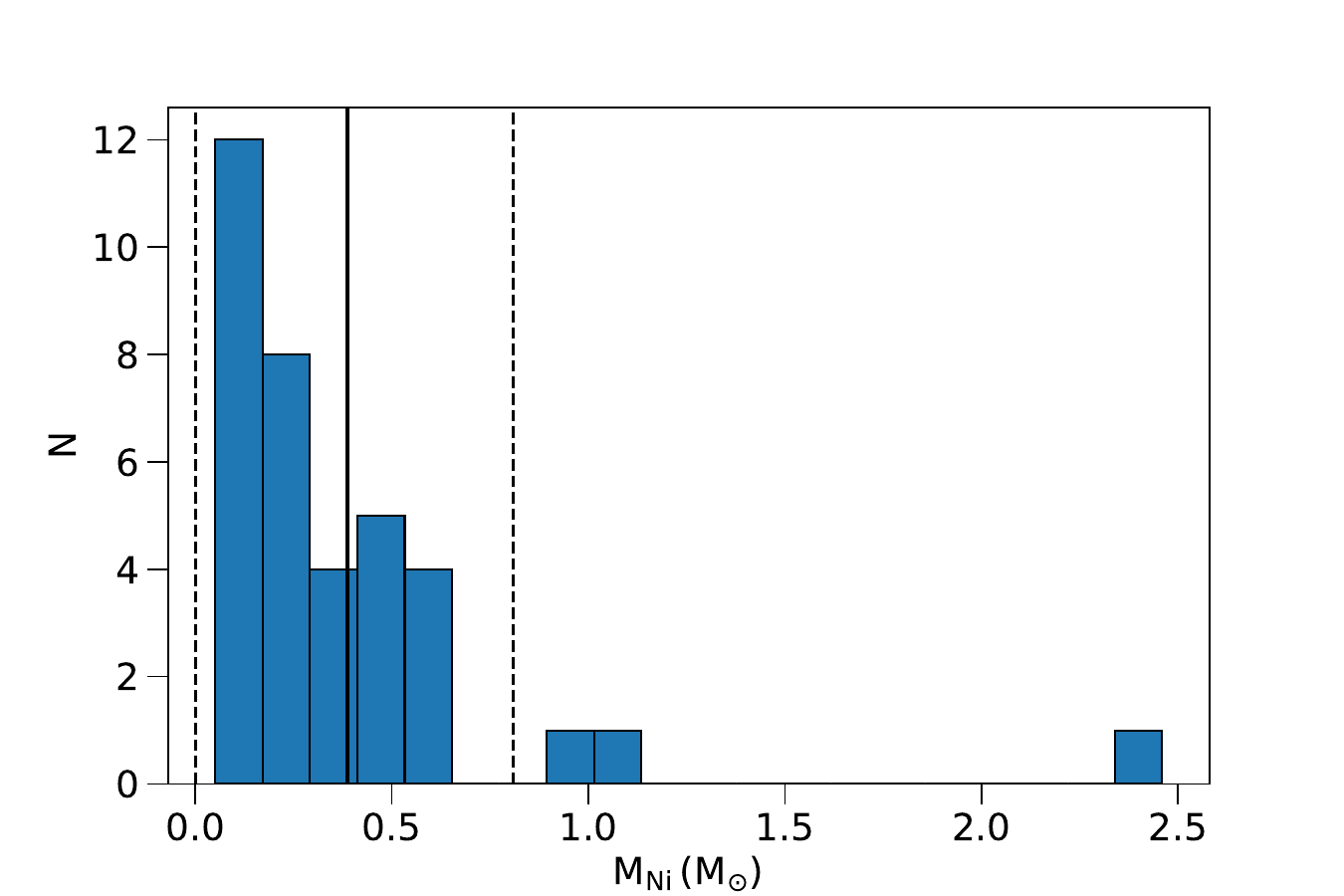}
     \includegraphics[width = \linewidth]{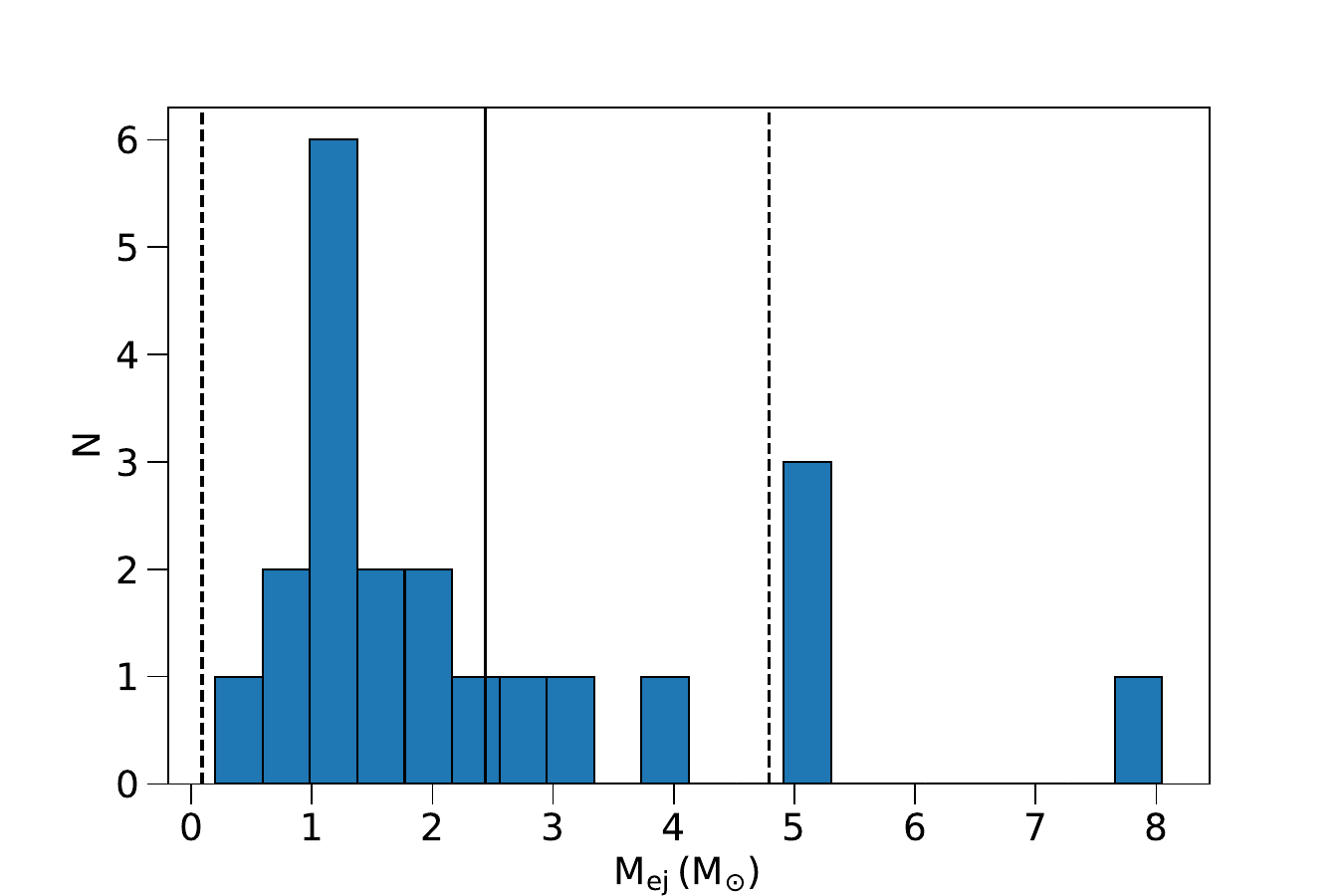}
    \includegraphics[width = \linewidth]{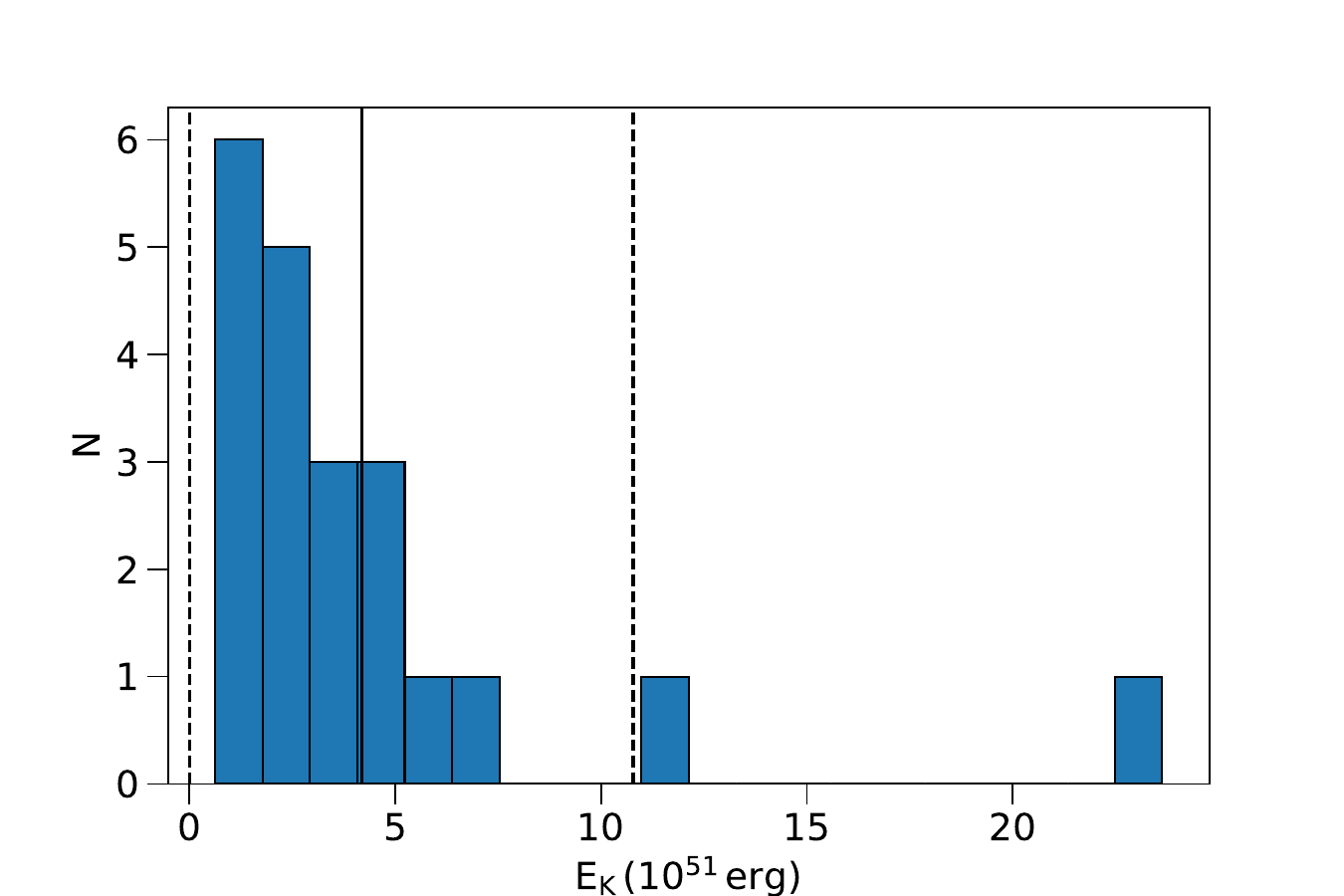}
    \caption{Histograms of the explosion properties derived for the Ic-BL sample. The mean values and 1$\sigma$ dispersions for the properties are shown with solid and dashed lines respectively. \textit{Top panel}: The $\rm{M_{Ni}}$ distribution for the SNe Ic-BL sample. \textit{Middle panel}: The $\rm{M_{ej}}$ distribution for the SNe Ic-BL sample. \textit{Lower panel}: The $\rm{E_{K}}$ distribution for the SNe Ic-BL sample.}
    \label{Progenhisto}
\end{figure}

We use HAFFET to generate semi-analytic bolometric luminosity LCs corresponding to different $\rm{M_{Ni}}$ and $\tau_m$ values, and use MCMC techniques to find the best-fit and 16, 50, and 84\% confidence interval values corresponding to each event not presented in \citet{Corsi2024} and \citet{Anand2024}. The distribution of $\rm{M_{Ni}}$  masses derived is shown in Figure \ref{Progenhisto}, and the values along with their statistical uncertainties are reported in Table \ref{explosiontable}. The values range from 0.05 to 2.46 $\rm{M_{\odot}}$, with a mean of $0.37^{+0.08}_{-0.06}  \, \rm{M_{\odot}}$ and a 1$\sigma$ standard deviation of 0.42. This is consistent with the normal SNe Ic-BL SN sample from \citet{taddia2019iptf}, who found a mean value of 0.31 $\rm{M_{\odot}}$ with a 1$\sigma$ standard deviation of 0.17 $\rm{M_{\odot}}$. 12 events in our sample were also included in \citet{Rodriguez2023}, who performed a systematic study of the iron yield in different classes of stripped-envelope SNe. They find an average Nickel mass of SNe Ic-BL of $0.14 \pm 0.02$, which is around a factor of 2 less than the value we derived. This is due to a different fitting procedure, as they calculated the Nickel masses based on the radioactive decay tail, rather than the peak of the LC using the \citet{arnett1982} model. Due to the Malmquist bias (see \S \ref{empiricalparam}, where the peak absolute magnitude distribution is 0.2 mag lower after accounting for this effect), the corrected Nickel mass distribution is 17\% lower than what we measure \citep{taddia2019iptf}, which is a small effect. When we exclude the SLSN SN 2020wgz from the sample, we find a new mean of $0.32 \pm  0.03 \, \rm{M_{\odot}}$, and 1$\sigma$ standard deviation of 0.21 $\rm{M_{\odot}}$. 

We note that for SN 2020bvc, the $\rm{M_{Ni}}$ we derive ($0.41 \pm 0.01$ $\rm{M_{\odot}}$) is inconsistent with what is derived in \citet{20aalxlis}  of $0.13 \pm 0.01$ $\rm{M_{\odot}}$. The reason for this discrepancy is that \citet{20aalxlis} used the radioactive decay tail to derive $\rm{M_{Ni}}$, rather than the peak of the LC. However, the $\rm{M_{Ni}}$ we derive is consistent with the results from \citet{Rho2021}, who found $\rm{M_{Ni}} \sim 0.4 \, \rm{M_{\odot}}$. This value is likely more accurate as they used hydrodynamical LC modeling to derive this value. The value we derive for SN 2018gep ($\rm{M_{Ni}}$ = $0.13 \pm 0.01$ $\rm{M_{\odot}}$) is consistent with what is found in \citet{18abukavn}, where they find $\rm{M_{Ni}} < 0.3 \, \rm{M_{\odot}}$. The value we derive for SN 2018bvw ($\rm{M_{Ni}}$ = $0.52 \pm 0.05$ $\rm{M_{\odot}}$) is slightly larger than the value derived in \citet{18aaqjovh} of $\sim 0.3 $ $\rm{M_{\odot}}$, though they did not report any errors in their work. 

\begin{figure*}
    \centering
    \includegraphics[width = 0.7\linewidth]{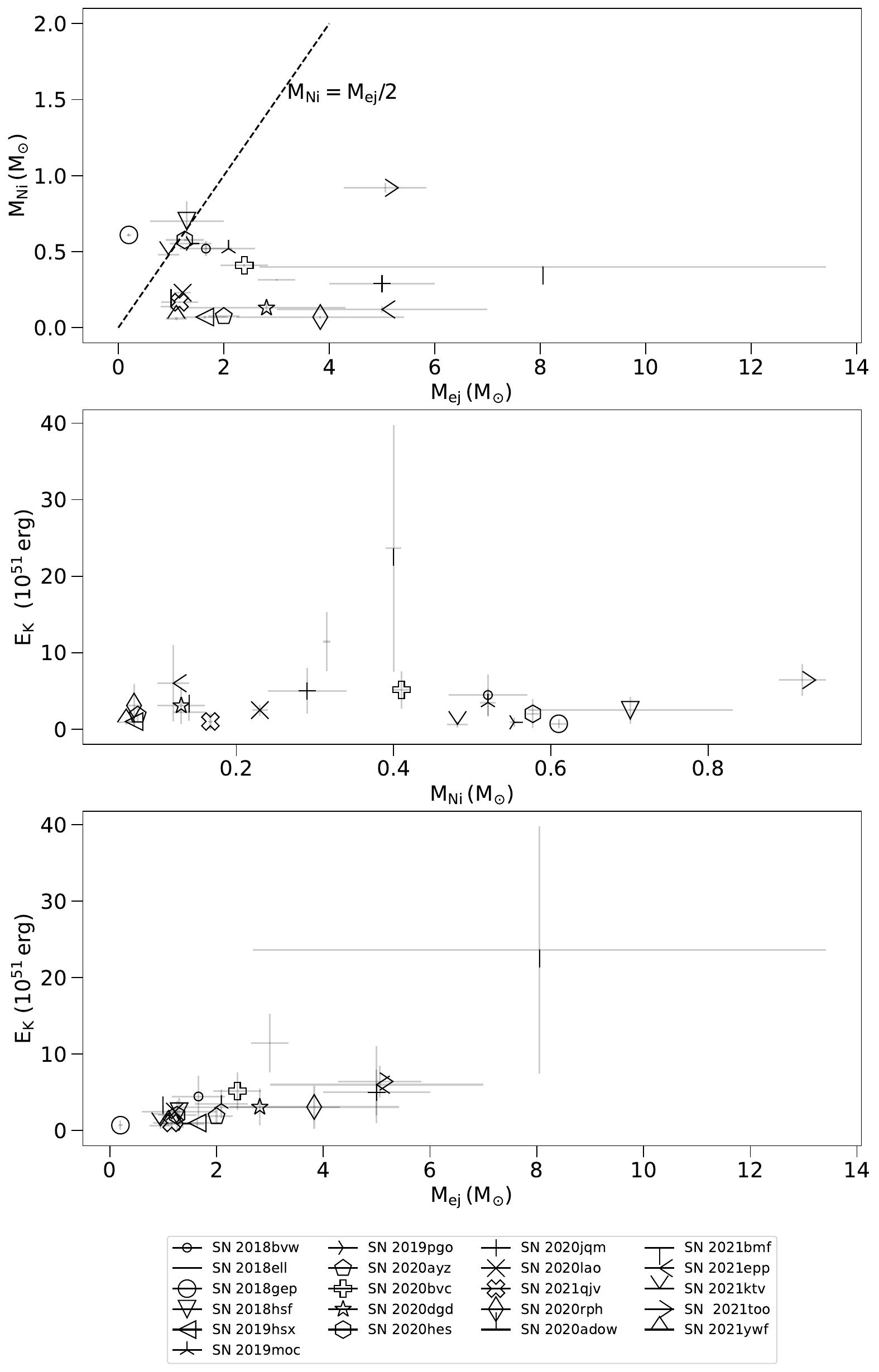}
    \caption{Explosion properties ($\rm{M_{Ni}}$,  $\rm{M_{ej}}$, and $\rm{E_{K}}$) plotted against each other. \textit{Top panel}: $\rm{M_{Ni}}$ plotted against $\rm{M_{ej}}$. We do not find any correlations between the two, with a $p$-value of 0.86. The dashed line indicates the boundary of $\rm{M_{Ni}} = \rm{M_{ej}}/2$. Events above that boundary are likely not soley powered by radioactive decay \citep{kasen2010}. \textit{Middle panel}: $\rm{E_{K}}$ plotted against $\rm{M_{Ni}}$. We do not find any correlation between the two, with a $p$-value of 0.75. \textit{Bottom panel}: $\rm{E_{K}}$ plotted against $\rm{M_{ej}}$. We find a significant correlation between the two, with a $p$-value of 0.007.}
    \label{explosioncomparison}
\end{figure*}

%\begin{figure}
    %\centering
    %\includegraphics[width = \linewidth]{Nimasscdf.pdf}
    % \includegraphics[width = \linewidth]{Mejcdf.pdf}
    %\includegraphics[width = \linewidth]{KEcdf.pdf}
    %\caption{CDFs of the explosion properties derived for the Ic-BL sample in navy, compared with those of normal Typie Ic-BL SNe \citet{Taddia2016} in green. The mean values and 1$\sigma$ dispersions for the properties are shown with solid and dashed lines respectively. \textit{Top panel}: The $\rm{M_{Ni}}$ CDF for the Ic-BL sample. \textit{Middle panel}: The $\rm{M_{ej}}$ CDF. \textit{Lower panel}: The $\rm{E_{K}}$ CDF for the Ic-BL sample (because a negative $\rm{E_{K}}$ is unphysical, we show the lower-limit at 0).}
    %\label{cdfs}
%\end{figure}

Armed with $\tau_m$ and $\rm{v_{ph}}$, we then derive the $\rm{M_{ej}}$ and $\rm{E_{\rm{KE}}}$ for each of the events in our sample using the equations from \citet{Lyman2016}. $\rm{M}_{\rm{ej}}$ is described as

\begin{equation}
 \rm{M_\mathrm{ej}} = \frac{\tau_\mathrm{m}^2\beta c v_\mathrm{sc}}{ 2\kappa_\mathrm{opt}}\textrm{\! ,}
 \label{eq4}
\end{equation}
and $\rm{E_{\rm{KE}}}$ is described as 
\begin{equation}
  \rm{E_\mathrm{KE}}  = \frac{3 v_\mathrm{sc}^2 \rm{M_\mathrm{ej}}}{10}\textrm{\! ,}
 \label{eq5}
\end{equation}
where $\beta = 13.8$ is a constant, $c$ is the speed of light, $\kappa_{\mathrm{opt}}$ is a constant, average optical opacity, and $v_{\rm{sc}}$ is the photospheric velocity $\rm{v_{ph}}$ at maximum light. $\kappa_{\mathrm{opt}}$ varies in the literature for stripped-envelope SNe, and we adopt the value used by \citet{Chugai2000}, \citet{Tartaglia2021}, and \citet{Barbarino2021}, $\kappa_{\mathrm{opt}} = 0.07$\,cm$^{2}$\,g$^{-1}$ (which was shown to accurately recreate hydrodynamical LCs of observed stripped-envelope SNe; \citealt{Taddia2018a}).
 
Because a measurement of $\rm{v_{ph}}$ is needed to derive $\rm{M_\mathrm{ej}}$ and $\rm{E_{\rm{KE}}}$, we were able to derive values for the 20 events that we estimated $\rm{v_{ph}}$ at peak for (described in \S \ref{spectra}). For the six events that did not have spectra within 15 days of the peak epoch, we derive lower limits on their $\rm{M}_\mathrm{ej}$ and $\rm{E}_{\rm{KE}}$. We present the values in Table \ref{explosiontable}, and show their distributions in Figure \ref{Progenhisto}. The $\rm{M_{ej}}$ values range from 0.2 to 8.05 $\rm{M_{\odot}}$, with a mean of $2.45^{+0.47}_{-0.41}  \, \rm{M_{\odot}}$ and 1$\sigma$ standard deviation of 2.35 $\rm{M_{\odot}}$. This is consistent with the normal Ic-BL sample from \citet{taddia2019iptf}, who found a mean value of 4.0 $\rm{M_{\odot}}$ with a 1$\sigma$ standard deviation of 2.9 $\rm{M_{\odot}}$. 
$\rm{E_{\rm{KE}}}$ ranges from $6.6 \times 10^{50} $ erg to $2.4 \times 10^{52} $ erg, with an average of $4.02^{+1.37}_{-1.00} \times 10^{51} $ erg and 1$\sigma$ standard deviation of $6.60 \times 10^{51} $ erg. This is consistent with the distribution found in \citet{taddia2019iptf}, who found an average of $7 \times 10^{51} $ erg with a  1$\sigma$ standard deviation of $5.8 \times 10^{51} $ erg. 

%In addition to the histogram distributions, we also show the CDFs for the explosion properties derived, along with those from \citet{taddia2019iptf} for the normal Ic-BL population, in Figure \ref{cdfs}.

In Figure \ref{explosioncomparison}, we plot the three explosion parameters against each other to test for the presence of any correlations using the Spearman rank test. We find no significant correlations between $\rm{M_{Ni}}$ and $\rm{M_{ej}}$ ($p$-value of 0.86), no correlation between $\rm{M_{Ni}}$  and $\rm{E_{K}}$ ($p$-value of 0.75), and a significant correlation between $\rm{M_{\rm{ej}}}$ and  $\rm{E_{K}}$ ($p$-value of 0.007). The correlation between $\rm{M_{\rm{ej}}}$ and  $\rm{E_{K}}$ is expected due to the relationships shown in Eq. \ref{eq4} and \ref{eq5}. 

We also show the boundary of $\rm{M_{Ni}} = \rm{M_{ej}}/2$, which is the boundary used by \citet{taddia2019iptf} that indicates a shift from an event powered by radioactive decay up to its peak luminosity, to one that must be powered by additional mechanisms, including the spindown of a magnetar \citep{kasen2010}. We note that there is one event (SN 2018gep) that is clearly in this regime, and \citet{18abukavn} showed that this event is best modeled as a shock breakout event in a dense CSM. There are two events whose error bars extend into this regime (SN 2019pgo and SN 2020hes), and one event that has $\rm{M_{Ni}} > \rm{M_{ej}}/2$, SN 2021ktv. However, SN 2021ktv's error bars for $\rm{M_{ej}}$ make it uncertain that it solely exists in this regime, so we cannot firmly establish the presence of any additional powering mechanisms for these three events through this method. SN 2018hsf possesses a relatively high  $\rm{M_{Ni}} = 0.7 \, \rm{M_{\odot}}$, and the lower limits derived for  $\rm{M_{\rm{ej}}}$ find $\rm{M_{\rm{ej}}} > 0.31$. Therefore, it is possible that SN 2018hsf possesses additional powering mechanisms. SN 2020wgz also likely exists in this regime, as it possesses an extremely high $\rm{M_{Ni}} = 2.46 \, \rm{M_{\odot}}$, and also has a luminosity that places it in the regime of SLSN.  However, only lower limits on its $\rm{M_{\rm{ej}}}$ and  $\rm{E_{K}}$ were derived due to a lack of a velocity measurement close to peak light. We provide more details on this event in \S \ref{2020wgz}. 

\subsection{SN 2020wgz -- a SLSN-I?}
\label{2020wgz}
SLSNe are a sub-class of SNe that can reach peak magnitudes of $\sim -21$ mag, a factor of 100 more than normal CCSNe \citep{Quimby, GalyamLuminous}. A subset of SLSNe do not show hydrogen features in their spectra, and are classified as SLSN-I. These events have slow rise times and peak much later than normal CCSNe (typically 20 - 100 days after explosion). SLSN-I spectra possess a blue continuum, often along with Fe II, Mg II, and O II absorption lines. Due to their high luminosity and spectral composition, the spin-down of magnetar is the most favored physical model for describing these events \citep{Dessartmnras2012}. 

SN 2020wgz possesses a peak absolute bolometric magnitude of $\rm{M_{bol}^{max}} = -20.91 \pm 0.09$ mag, placing it in the luminosity class of SLSNe. The two spectra obtained near peak light shown in the Appendix show no hydrogen features, along with a blue featureless continuum. The spectra are too low-quality/noisy to make out the presence of any absorption features. Therefore, SN 2020wgz's LC and early-time spectra have similarities to SLSN-I. However, SN 2020wgz has a rapid-rise time of $\sim$ 10 days, differing from normal SLSN-I. Furthermore, a spectrum taken 21 days after peak shows a transition to a redder continuum, along with the development of broad spectral features, which enabled a classification of SN Ic-BL. 

SN 2011kl is a SLSN-I that was associated with an ultra-long GRB 111209A \citep{Greiner2015}. To date, it is the only SN associated with a GRB that is not a SNe Ic-BL. SN 2020wgz possesses a peak absolute magnitude around one order of magnitude higher than SN 2011kl ($\rm{M_{bol}^{max}} \sim -20.0$ mag), and has a similar rise time to SN 2011kl ($\sim$ 14 days). However, SN 2011kl's spectra do not show the same transition to a redder continuum and the development of broad absorption features that SN 2020wgz does. LC modeling of SN 2011kl showed that it likely was solely powered by a magnetar. 

This spectral transition was seen in a few other objects (e.g., \citealt{Pastorello2010}), and most recently in SLSN-I SN 2017dwh \citep{Blanchard2019}, which was a Nickel-rich SLSN-I possessing a similar but slightly fainter peak absolute magnitude of $\sim$ $-20.5$ mag, and a slightly longer rise time of $\sim$ 19 days. \citet {Blanchard2019} found that the high amount of Nickel in the explosion may be due to enhanced production, or due to more efficient mixing of Fe group elements in the outer ejecta. They concluded that SN 2017dwh was likely powered by a combination of the spindown of a magnetar and radioactive decay. This event provides evidence that the progenitors for SLSN-I and SNe Ic-BL might be similar for certain cases.

In \S \ref{explosionlabel}, we determined that SN 2020wgz had too high of a Nickel mass to be powered purely by the \citet{arnett1982} model. Utilizing the spectral similarities to SN 2017dwh, we use the open-source electromagnetic transient Bayesian fitting software package \texttt{redback} to fit a \texttt{general magnetar driven supernova} model \citep{general_magnetar} to the observed $r$ and $g$-band LCs. This model utilizes the combination of both the spindown of a magnetar along with radioactive decay to power the LC. We show the best-fit model in Figure \ref{2020wgzmodel}, along with the corner plot in the Appendix. We see that the model fits the LC well in both the $g$ and $r$ bands, showing that it is likely SN 2020wgz was powered by a magnetar, at least to some extent.

\begin{figure*}
\centering
    \includegraphics[width=0.9\linewidth]{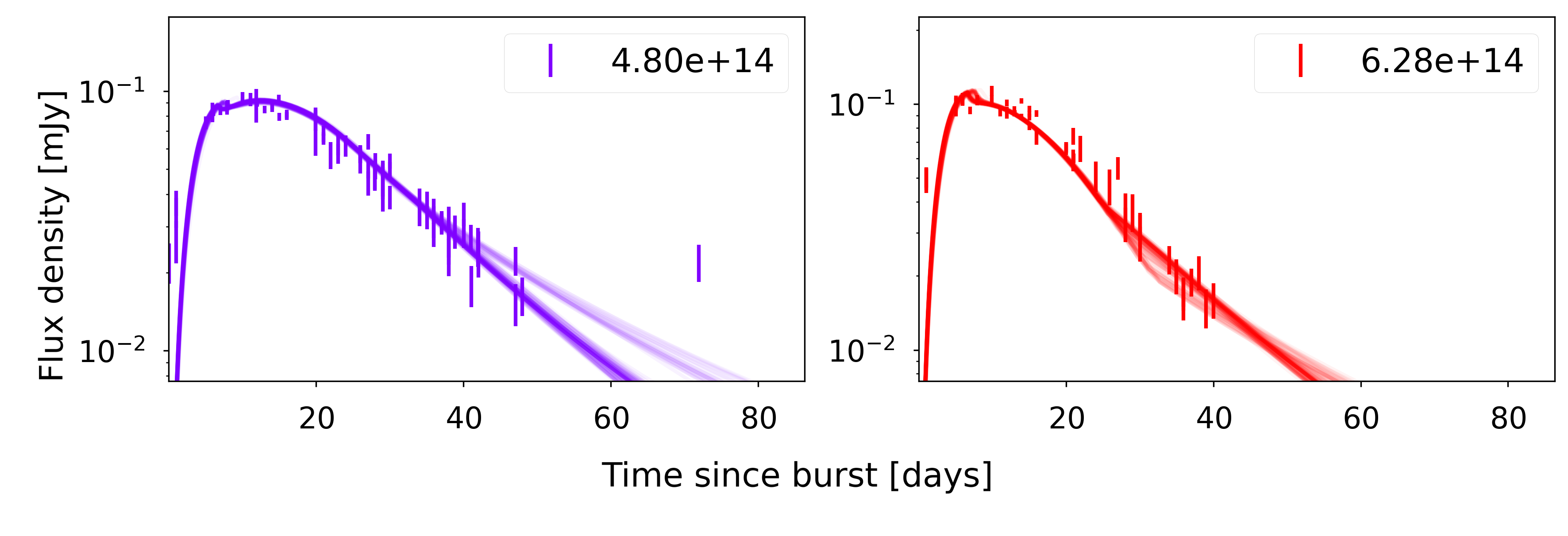}
    \caption{The best-fit magnetar  powered model fit to SN 2020wgz, in the $r$ band on the left, and $g$ band on the right.}
    \label{2020wgzmodel}
    
\end{figure*}

\section{Discussion in Context of Multi-Wavelength Observations}
\label{Multiwavelengthlabel}

There are eight events in the sample that have transient radio counterparts (SN 2018bvw, SN 2018gep, SN 2020tkx, SN 2020jqm, SN 2020bvc, SN 2020adow, SN 2021bmf, and SN 2021ywf). SN 2018bvw, SN 2018gep, and SN 2020tkx are subjects of single-object papers, and the physical mechanisms of their explosions are discussed at length in those works. SN 2018bvw likely is a transition event between engine-driven GRB-SNe \citep{18aaqjovh} and normal SNe Ic-BL, and the event shows possible signs of harboring a mildly relativistic jet. SN 2018gep originates from a shock breakout in a massive shell of dense
circumstellar material without an accompanying relativistic jet \citep{18abukavn, Leung2018, Pritchard2021}, and SN 2020bvc shows a double peak LC due to shock cooling emission followed by radioactive decay, with the event showing similar multiwavelength features to low-luminosity GRB 060218/SN 2006aj. This event also shows signs of mildly relativistic ejecta  \citep{20aalxlis, Izzo2020}. Furthermore, \citet{Corsi2024} showed that SN 2020jqm's erratic radio LC and derived shock properties place it in the same regime as radio-loud CSM interacting SNe like PTF11qcj \citep{PTF11qcj}. The rest of the events only have one to two radio observations, making multi-wavelength modeling efforts for these objects rather non-constraining with respect to both model comparison and parameter estimation. However, we include multi-wavelength modeling efforts for SN 2020tkx, SN 2020adow, and SN 2021ywf in the Appendix. Modeling efforts for SN 2021bmf will be explored in a future work.

\subsection{Search for Gamma-rays}
Using the derived explosion epochs for the sample, we search for potential GRB coincidences in several online archives. \citet{Corsi2024} did this search for their sample of events, which includes 8 of the events in our sample. They found no \textit{Swift} or \textit{Fermi} GRBs spatially and temporally coincident to any of their events. The only coincident GRBs were from the WIND satellite on the Konus instrument, which does not provide localizations. They found that the number of events found within the explosions epochs were consistent with random fluctuations, and also found that no events with an explosion epoch constrained to less than 1 day had a coincident GRB within that time frame. Furthermore, \citet{18abukavn} and \citet{20aalxlis}  determined that SN 2018gep and SN 2020bvc had no GRB counterparts. \citet{18aaqjovh} found that SN 2018bvw had one possible GRB counterpart from Konus-WIND, corresponding to a low-luminosity GRB on 2018 May 3 03:41:01; however, they determine that this association is most likely due to chance.

We repeat this search for the 25 events in our sample not covered above. We find no spatially and temporally coincident \textit{Swift} or \textit{Fermi} GRBs for any of the events. There were 11 SNe that had KONUS-\textit{Wind} GRBs that were detected within their explosion epochs, corresponding to 31 coincident GRB events over a time period of 78.29 days searched. This rate is also consistent with random fluctuations. Therefore, we determine that no events in the sample have certain, coincident gamma-ray emission. 

\subsection{Comparison of Explosion Properties with Radio Observations}
\label{radiocomparison}

\begin{figure}
    \centering
    \includegraphics[width=.9\linewidth]{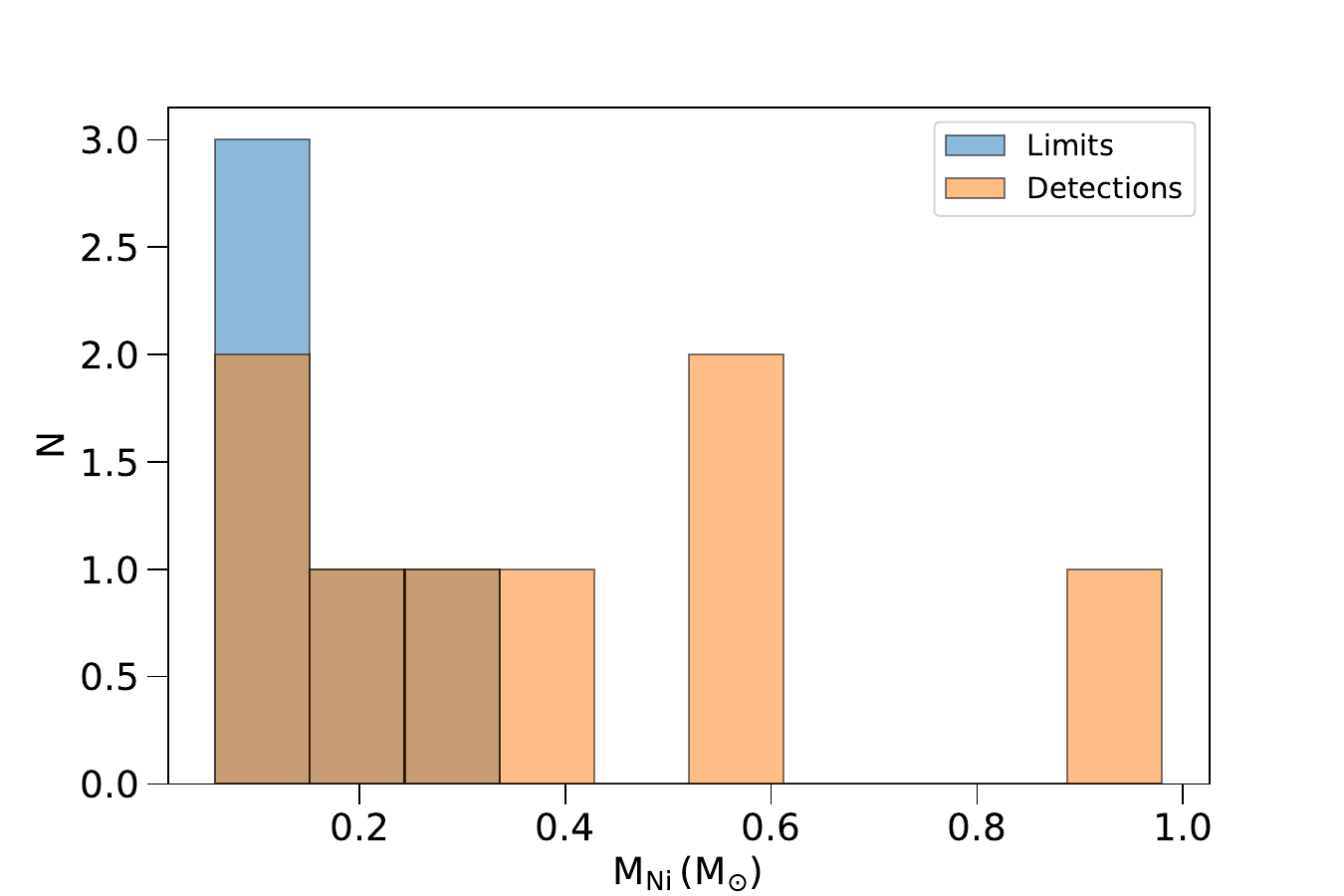}
    \includegraphics[width=.9\linewidth]{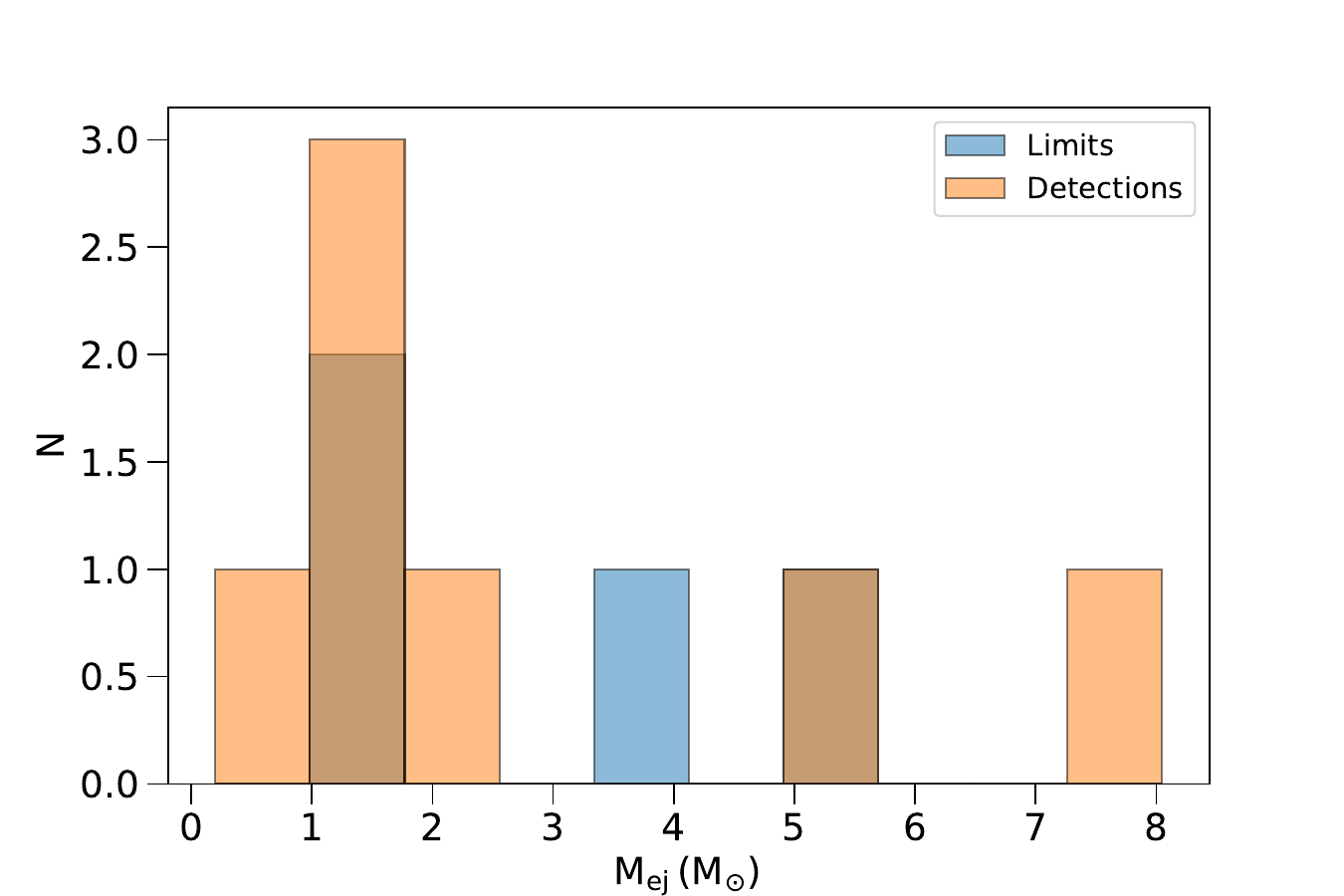}
    \includegraphics[width=.9\linewidth]{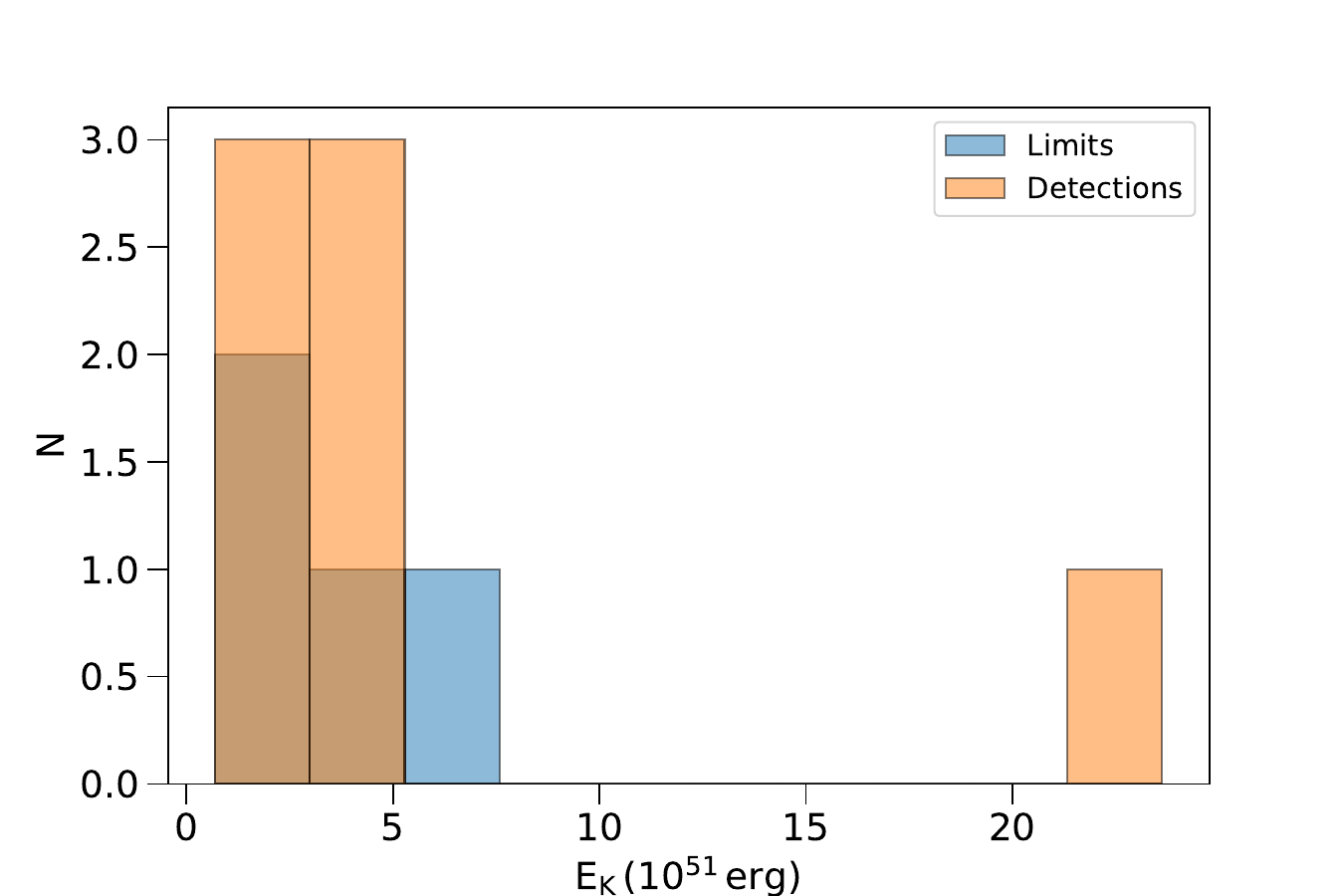}

    \includegraphics[width=.9\linewidth]{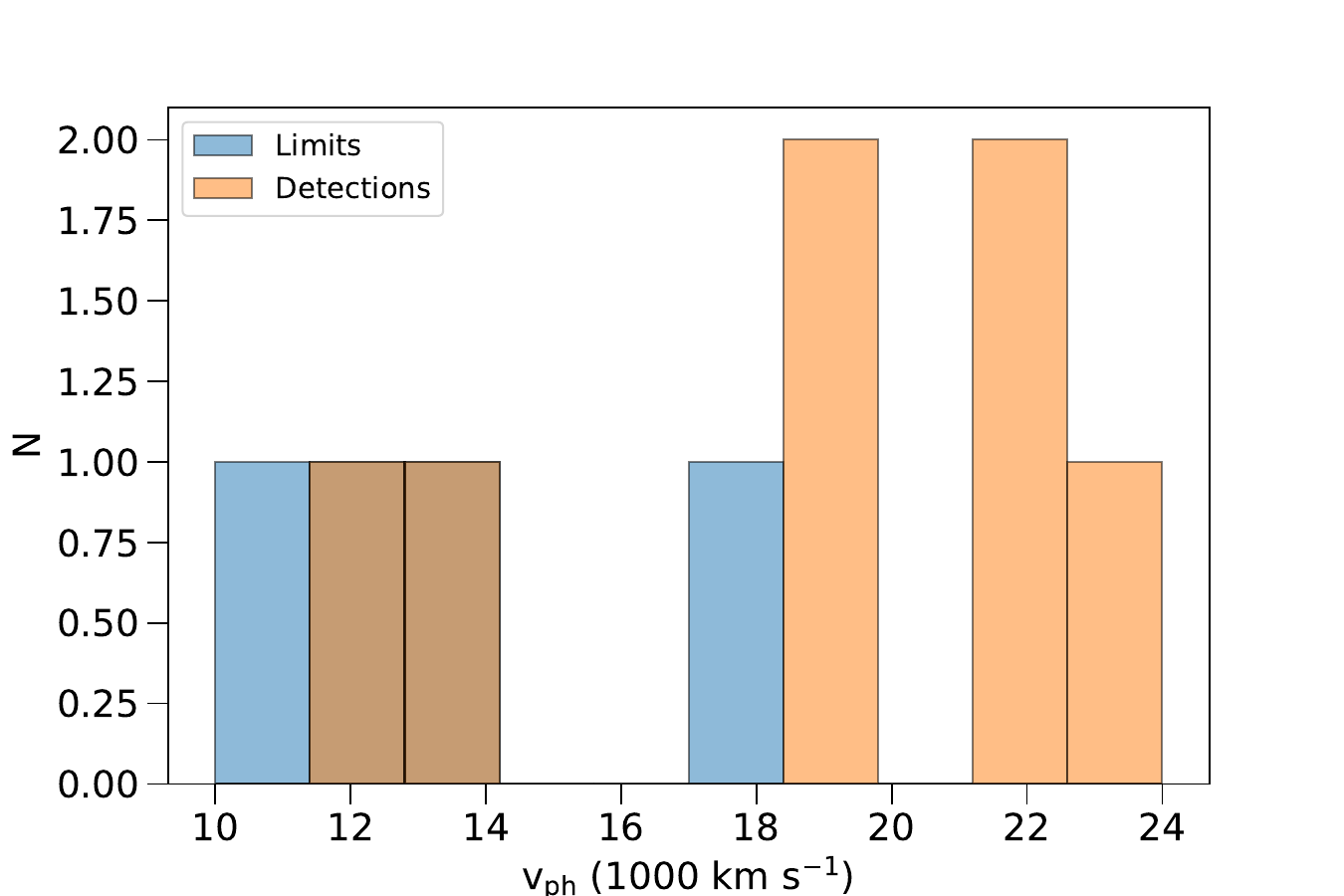}
    \caption{Histograms showing the distribution of explosion properties and $\rm{v_{ph}}$ for the population of events with radio observations. We distinguish events that have radio detections in orange, and events that have radio non-detections in blue. We find no statistical differences between the populations for any of the properties.} 
    \label{Proghisto}
\end{figure}

We then test whether statistical differences exist between the explosion properties for the subsample of 13 events that have radio observations, between the eight events with detections and five events with non-detections \citep{Corsi2024}.  In Figure \ref{Proghisto}, we show the distributions of $\rm{M_{Ni}}$, $\rm{M_{ej}}$, $\rm{E_{K}}$, and $\rm{v_{ph}}$ for the sub-sample with radio observations. Through running a two-sample Kolmogorov-Smirnov (K-S) test, we do not find statistical differences between the populations for any of the properties, though we note that due to the small sample sizes, moderately different populations cannot be ruled out. We note that for the $\rm{M_{ej}}$ and $\rm{E_{K}}$ analysis, we do not include events that only have lower limits derived. Specifically, we compute a K-S test statistic of 0.50 with a $p-$value of 0.31 for the $\rm{M_{Ni}}$ distribution, a K-S test statistic of 0.42 with a $p$-value of 0.66 for the $\rm{M_{Ej}}$ distribution, a K-S test statistic of 0.32 with a $p$-value of 0.89 for the $\rm{E_{K}}$ distribution, and a K-S test statistic of 0.71 with a $p$-value of 0.12 for the $\rm{v_{ph}}$ distribution.

\begin{figure}
    \centering
    \includegraphics[width=.8\linewidth]{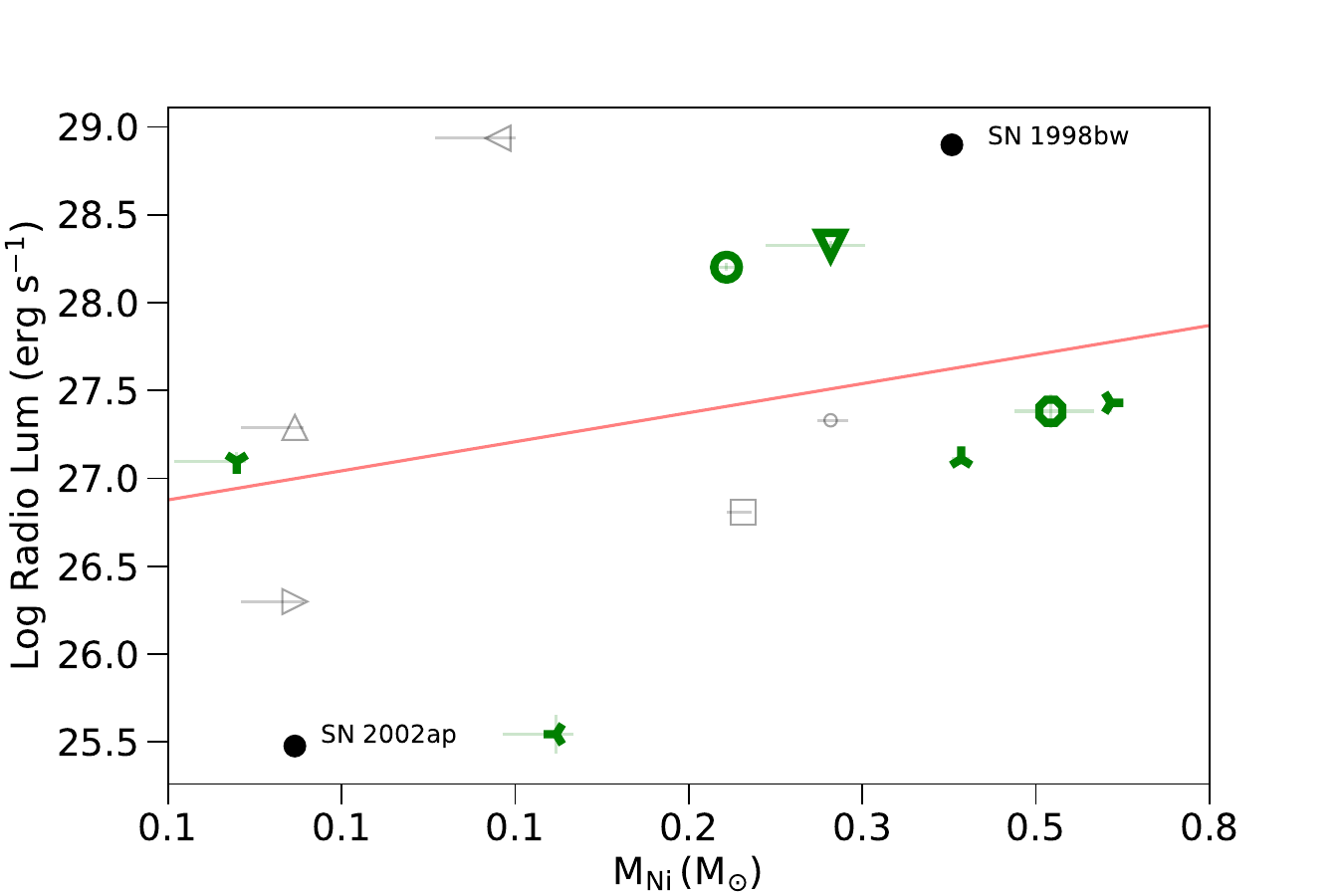}
    \includegraphics[width=.8\linewidth]{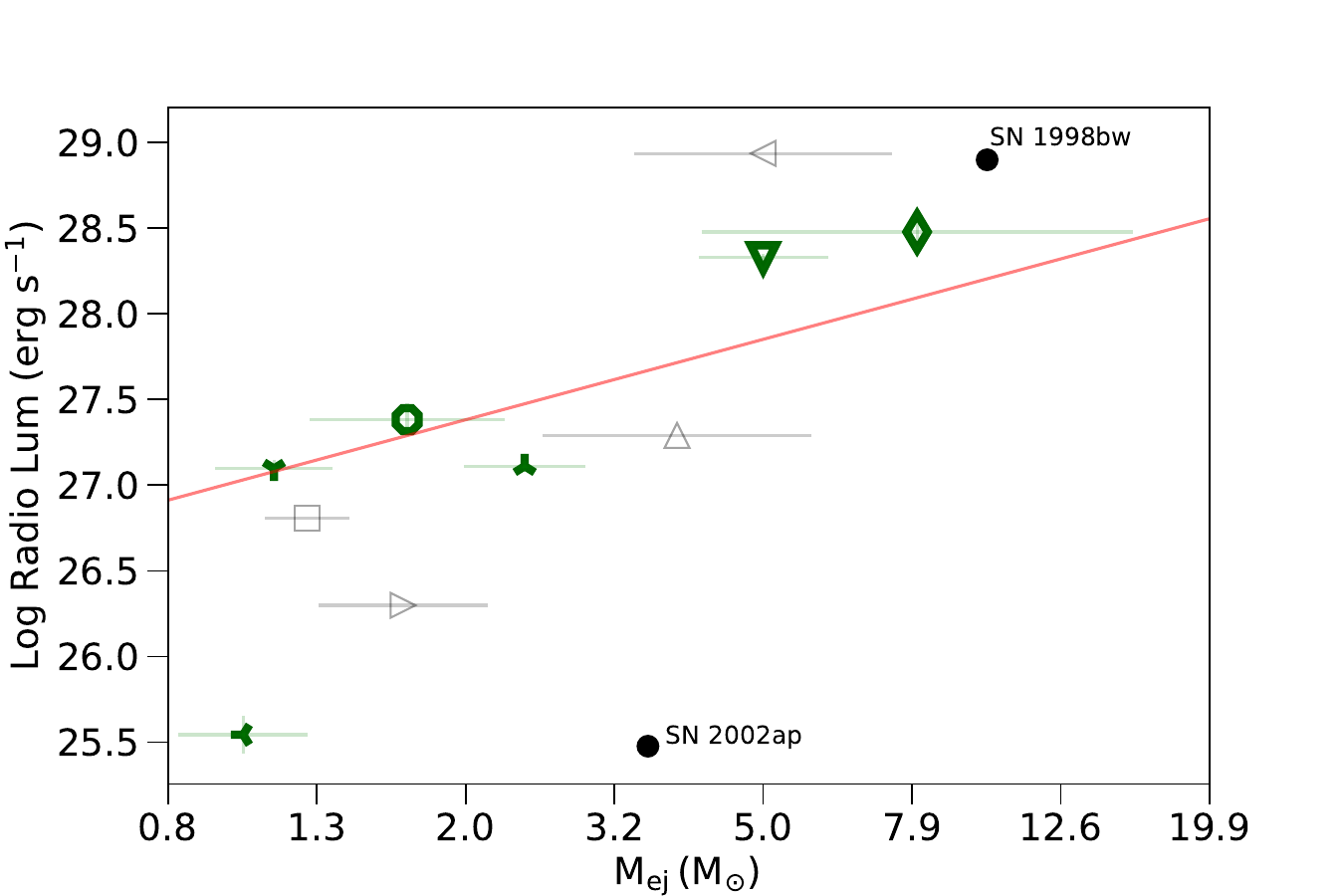}
    \includegraphics[width=.8\linewidth]{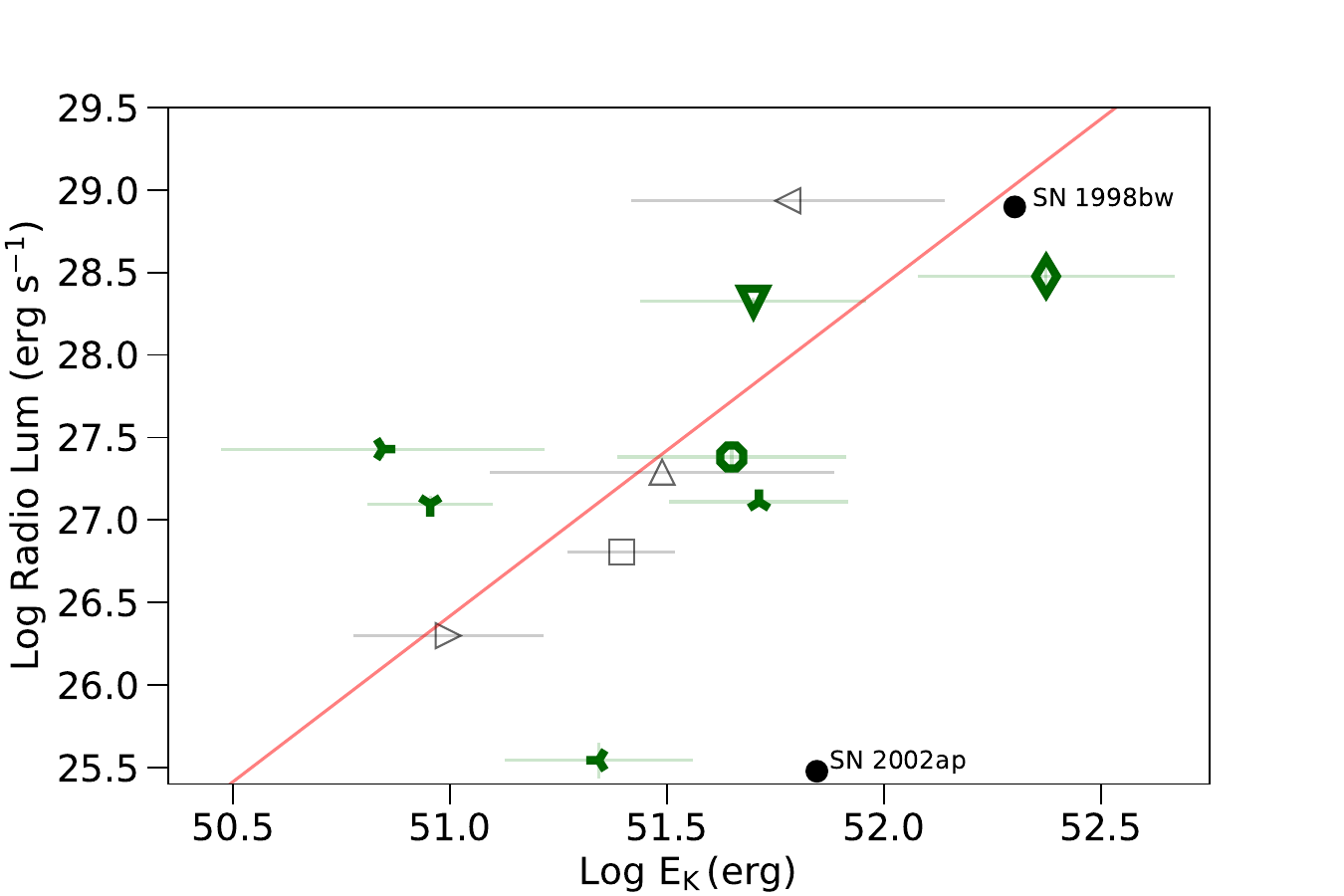}
    \includegraphics[width=.8\linewidth]{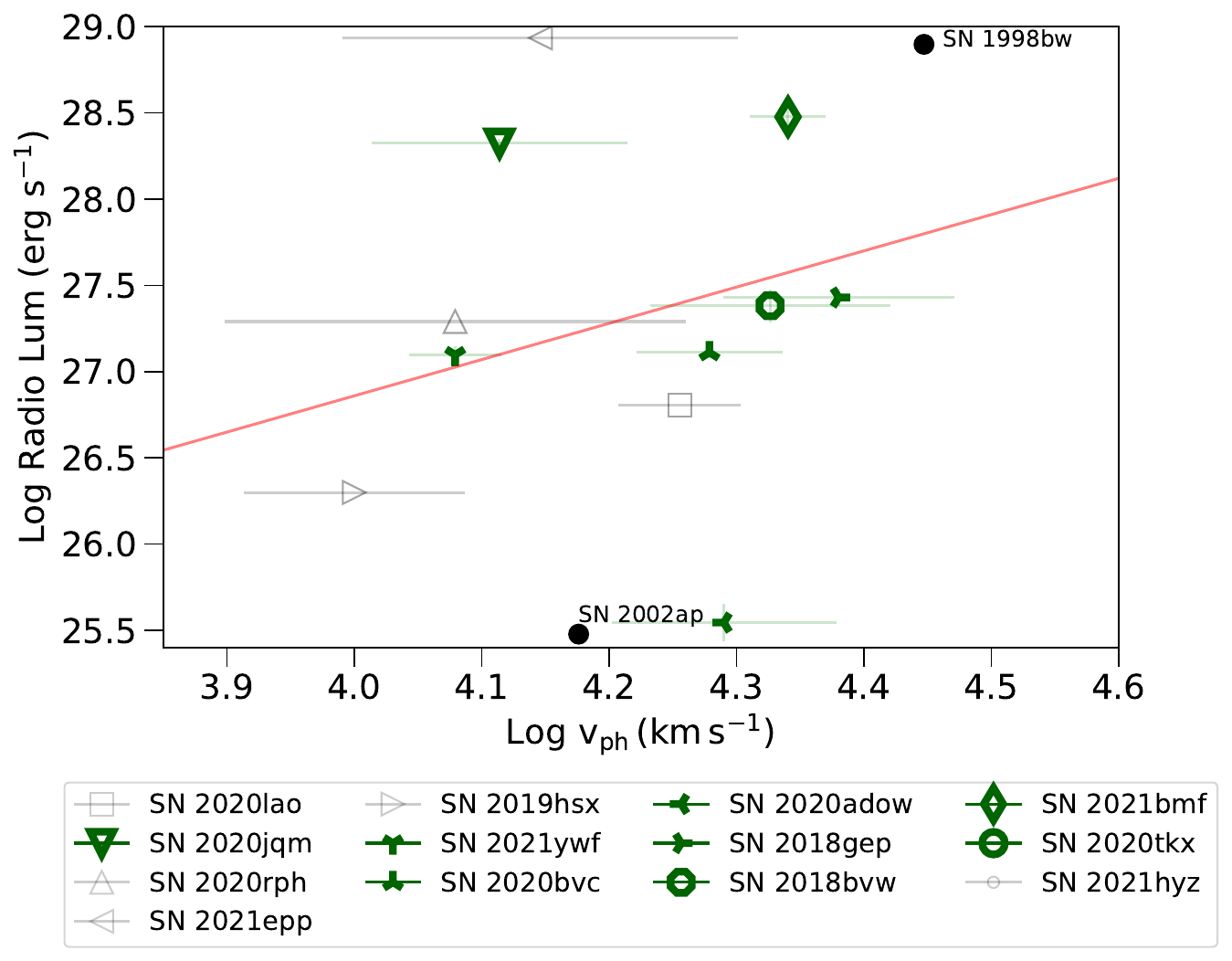}
    \caption{Best-fit Linmix linear regression model fitting between the peak radio luminosities and the explosion properties and photospheric velocities shown in red. Black points indicate limits while green points indicate detections. We find no statistically significant correlations  between the peak radio luminosities and any of the parameters. The parameters for SN 1998bw \citep{Sollerman2000, Weiler2001} and SN 2002ap \citep{Mazzali2002, berger2002} are shown as reference.}
    \label{Linmix}
\end{figure} 

We also investigate whether correlations exist between the peak radio luminosity and explosion properties. If correlations do exist, they may point to a connection between the optically described explosion properties and the existence of radio emission, possibly from accompanying off-axis relativistic jets. In order to check this, we extrapolate the peak radio fluxes for every event to 6 GHz through assuming a spectrum $F_\nu \propto \nu^{-1} $, and convert this flux to a luminosity using the distance of the SN. Once armed with the radio luminosities, we use a linear regression code, \texttt{Linmix}, that takes into account both detections and non-detections when fitting linear models to the data \citep{Kelly2007}. \texttt{Linmix} assumes symmetric error bars, so for $\rm{M_{Ni}}$ values that have asymmetric error bars, we assume the larger error bar value for both the positive and negative error. 

We show the linear regression results in Figure \ref{Linmix}.
%For $M_{Ej}$, we calculate a linear correlation coefficeint of $0.38 \pm 0.31$. Again in the top panel of Figure \ref{Ejmasscorrelation}, we show the best-fit model, with the dark black points represent the limits, while the green points represent the detections. We find a Pearson coefficient of $0.78 \pm 0.07$, with a p value of 0.22, and show the four events with radio detections in the 2 parameter space in the bottom panel of Figure \ref{Ejmasscorrelation}. Though the Linmix regression model does not indicate a robust linear correlation, we do find a significant linear correlation between $M_{Ej}$ and the radio luminosity, when just looking at the four detections. It is important to note that SN 2018hom and SN 2020tkx both only have lower limits for their ejecta masses due to the lack of spectra available at peak. Therefore, we assume the lower limit values as the ejecta masses, but this correlation may be improved even more if the actual values were able to be calculated. 
For $\rm{M_{Ni}}$, we calculate a linear correlation coefficient of $0.32 \pm 0.32$, with a $p$-value of 0.33 showing there are no significant correlations present between the radio luminosity and the $\rm{M_{Ni}}$. For $\rm{M_{ej}}$, we calculate a linear correlation coefficient of $0.46\pm 0.32$, with a $p$-value of 0.18, while for $\rm{E_{K}}$ we calculate a linear correlation coefficient of $0.58 \pm 0.33$ with a $p$-value of 0.11, again showing a lack of significant correlations. Finally, for $\rm{v_{ph}}$, we calculate a linear correlation coefficient of $0.17\pm 0.42$, with a $p$-value of 0.70. We note that we do not include events that have lower limits for this analysis. 

Therefore, we find that there are no statistically significant correlations between the radio luminosities and the explosion properties derived from optical data of SNe Ic-BL. We note that there do seem to be slightly positive correlations between the radio luminosities and $\rm{M_{ej}}$ and $\rm{E_{K}}$; however, these correlations are not statistically significant. Future radio observations of SNe Ic-BL will increase the sample of radio detected-events, and future analyses can use this work as a stepping stone to probe deeper into the relationship between events' optical and radio properties.

\section{Summary and Conclusions}
We analyzed 36 SNe Ic-BL from ZTF between March 2018 and August 2021 as part of the BTS survey, building upon iPTF's sample reported by \citet{taddia2019iptf}. We present the ZTF LCs, along with spectra obtained for every event. Below we present a summary of some of our findings:

\begin{itemize}
    \item{We find the average peak absolute magnitude in $r$ band of the sample $\overline{\rm{M}}_r^{\rm{max}} = -18.51 \pm 0.15$ mag, with a 1$\sigma$ standard deviation of 0.90 mag. This is consistent with the distribution from \citet{taddia2019iptf}}.
    \item{We calculate expansion velocities for each event with sufficient SNR spectra, to estimate the photospheric velocity at peak, and find an average of 16,100 $\pm$ 1,100 km $\rm{s^{-1}}$, with a 1$\sigma$ standard deviation of 5,600 km $\rm{s^{-1}}$.}
    \item{ We also utilize the \citet{arnett1982} model to determine the explosion properties for each of the events. We determine an average $\rm{M_{Ni}}$ of $0.39^{+0.08}_{-0.06}$ $\rm{M_{\odot}}$ and 1$\sigma$ standard deviation of 0.42 $\rm{M_{\odot}}$, an average $\rm{M_{ej}}$ of $2.45^{+0.47}_{-0.41}$ $\rm{M_{\odot}}$ and 1$\sigma$ standard deviation of 2.35 $\rm{M_{\odot}}$, and an average $\rm{E_{\rm{KE}}}$ of  $4.02^{+1.37}_{-1.00} \times 10^{51} $ erg and 1$\sigma$ standard deviation of $6.5 \times 10^{51} $ erg. All these values are consistent with the averages found in the previous sample of \citet{taddia2019iptf}.}
    \item{SN 2020wgz has the luminosity and spectral features of a SLSN-I that transitions to a SN Ic-BL. Its LC fits well to a magnetar and nickle-powered combined model.}
    \item{We do not find associated gamma-ray emission for any of the events, through searches of archival data from gamma-ray satellites.}
    \item{We find that SN 2018gep's radio detections reported in \citet{18abukavn} are from the transient and not its host galaxy. We also search the VLASS survey for any transient radio emission for the SNe, and find SN 2021bmf displays transient radio emission over the three epochs of the survey. }
    \item{When comparing the optical properties of the radio-loud and radio-quiet population, we find no correlations between the radio luminosity and optical properties of the sample, showing evidence that the optically inferred explosion parameters alone are not sufficient to make inferences about relativistic jet formation mechanisms in SNe Ic-BL.}
    
\end{itemize} 

Disentangling the picture of SNe Ic-BL and their different physical mechanisms continues to be a challenging task, and this work shows that purely studying the global optical properties of SNe Ic-BL will limit our understanding of these explosions. Future radio studies are integral to understanding this landscape. These studies should follow up a large fraction of optically discovered SNe Ic-BL at early times across multiple epochs, enabling the creation of robust radio LCs or deep, constraining upper limits for a large amount of events. Furthermore, the next generation of X-ray space based missions like Einstein Probe \citep{EP} and SVOM \citep{SVOMpaper} can be utilized to discover low-luminosity GRBs and X-ray Flash events that peak at lower peak energies than normal GRBs. This will allow for more opportunities to study these understudied events' associated SNe Ic-BL and characterize a new subset of events within the landscape of relativistic stellar explosions.
\label{Conclusion}

\section*{Acknowledgements} 
 G.P.S. thanks Alexander Dittman, Cole Miller, Isiah Holt, Muhammad Mousa, and Maryam Modjaz for useful discussions regarding this work. G.P.S. thanks Cristina Barbarino, Tassilo Schweyer, Joonas Viuho, Amanda Djupvik for obtaining spectra from NOT. G.P.S. also thanks Simi Bhullar for her moral support throughout the paper-writing process. SY acknowledges the funding from the National Natural Science Foundation of China under Grant No. 12303046 and the Henan Province High-Level Talent International Training Program. S. A. acknowledges generous support from the David and Lucile Packard Foundation and the National Science Foundation GROWTH PIRE grant No. 1545949.  A.C. acknowledges support from the National Science Foundation via AST-2431072, and from NASA Swift GI program via the award 80NSSC23K0314. S. Schulze is partially supported by LBNL Subcontract NO.\ 7707915. N.S. acknowledges support from the Knut and Alice
Wallenberg foundation through the “Gravity Meets Light” project (PIs: Rosswog \& Jerkstrand). 

This work was supported by the GROWTH project \citep{2019PASP..131c8003K} funded by the National Science Foundation under Grant No 
1545949. MMK acknowledges generous support from the David and Lucille Packard Foundation. Some of the data presented herein were obtained at the W.M. Keck Observatory, which is operated as a scientific partnership among the California Institute of Technology, the University of California and the National Aeronau- tics and Space Administration. The Observatory was made possible by the generous financial support of the W.M. Keck Foundation. The authors wish to recognize and acknowledge the very significant cultural role and reverence that the summit of Mauna Kea has always had within the indigenous Hawaiian community. We are most fortunate to have the opportunity to conduct observations from this mountain. SED Machine is based upon work supported by the National Science Foundation under Grant No. 1106171. Based on observations obtained with the Samuel Oschin Telescope 48-inch and the 60-inch Telescope at the Palomar Observatory as part of the Zwicky Transient Facility project. 

ZTF is supported by the National Science Foundation under Grant No. AST-2034437 and a collaboration including Caltech, IPAC, the Weizmann Institute of Science, the Oskar Klein Center at Stockholm University, the University of Maryland, Deutsches Elektronen-Synchrotron and Humboldt University, the TANGO Consortium of Taiwan, the University of Wisconsin at Milwaukee, Trinity College Dublin, Lawrence Livermore National Laboratories, IN2P3, University of Warwick, Ruhr University Bochum and Northwestern University. Operations are conducted by COO, IPAC, and UW. The ZTF forced-photometry service was funded under the Heising-Simons Foundation grant \#12540303 (PI: Graham). The Gordon and Betty Moore Foundation, through both the Data-Driven Investigator Program and a dedicated grant,
provided critical funding for SkyPortal.The Liverpool Telescope is operated on the island of La Palma by Liverpool John Moores University in the Spanish Observatorio del Roque de los Muchachos of the Instituto de Astrofisica de Canarias with financial support from the UK Science and Technology Facilities Council.

\appendix

\section{Discovery Paragraphs}
Below we present the discovery paragraphs for every event. We note here that the first ZTF photometry reported is not from forced photometry, but from ZTF's nightly alerts that are reported to TNS. 

\subsection{SN\,2018bvw}
We refer the reader to \citet{18aaqjovh} for details about this event.

\subsection{SN\,2018ell}
The first ZTF photometry of SN\,2018ell (ZTF18abhhnnv) was obtained on 2018 July 17 ($\mathrm{JD}=2458316.72$). This first detection was in the $r$ band with a host-subtracted magnitude of $20.09\pm0.19$, at $\alpha=16^{h}49^{m}57.02^{s}$, $\delta=+27\degr38\arcmin26.94\arcsec$ (J2000.0, throughout). The discovery was reported to TNS \citep{2018ell} two days later on 2018 July 19, with a note saying the latest non-detection from ZTF was five days prior to discovery (July 12, $g > 20.81$). The event was first classified as a Type Ia SN based on a P60 SEDM spectrum obtained on 2018 July 31 \citep{2018ell}, but was reclassified two years later as a Type Ic-BL SN \citep{Dahiwale2020}. The SN exploded in the outskirts of the spiral galaxy  WISEA J164957.78+273828.3, with a well-established redshift of $z = 0.0638$. 

\subsection{SN\,2018gep}
We refer the reader to \citet{18abukavn,Pritchard2021, Leung2018}  for details about this event. 

\subsection{SN\,2018hsf}
The first ZTF photometry of SN\,2018hsf (ZTF18acbvpzj) was obtained on 2018 October 31 ($\mathrm{JD}=2458422.83$). This first detection was in the $r$ band, with a host-subtracted magnitude of  $19.58\pm 0.16$  at $\alpha=02^{h}40^{m}12.78^{s}$, $\delta=-19\degr58\arcmin44.94\arcsec$.% (J2000.0). 
The discovery was reported to TNS \citep{2018hsf}, with no prior non-detection from ZTF. The transient was classified as a Type Ic-BL event based on a LRIS spectrum obtained on 2018 December 4 \citep{2018hsfclassification}. The SN exploded in the plane of spiral galaxy PSO J040.0536-19.9798, and a redshift of $z = 0.119$ was determined through narrow host lines from the LRIS spectrum.

\subsection{SN\, 2018keq}
The first ZTF photometry of SN\,2018keq (ZTF18acxgoki) was obtained on 2018 December 17 ($\mathrm{JD}=2458469.59$). This first detection was in the $r$ band with a host-subtracted magnitude of $19.03\pm0.10$, at $\alpha=23^{h}22^{m}41.97^{s}$, $\delta=+21\degr00\arcmin43.17\arcsec$. The discovery was reported to TNS \citep{2018keq} two days later on 2018 December 19, with a note saying the latest non-detection from ZTF was 4 days prior to discovery ($r > 20.2$ mag). The event was classifed as a Type Ic-BL SN \citep{2018keqclassification} based on a spectrum obtained by LRIS on 2019 January 04. The SN exploded in the spiral galaxy SDSS J232241.80+210042.6, with a well-established redshift of $z = 0.038$.

%\subsection{SN\,2018etk}
%We refer the reader to \citet{Corsi2024} for details about this event.

%\subsection{SN\,2018hxo}
%We refer the reader to \citet{Corsi2024} for details about this event.

%\subsection{SN\,2018hom}
%We refer the reader to \citet{Corsi2024} for details about this event.

%\subsection{SN\,2018kva}
%We refer the reader to \citet{Anand2024} for details about this event.

%\subsection{AT\,2018jex}
%We refer the reader to \citet{Corsi2022} for details about this event.

\subsection{SN\,2019hsx}
We refer the reader to \citet{Anand2024} for details about this event.

\subsection{SN\,2019gwc}
We refer the reader to \citet{Anand2024} for details about this event.

\subsection{SN\,2019lci}
The first ZTF photometry of SN\,2019lci (ZTF19abfsxpw) was obtained on 2019 July 11 ($\mathrm{JD}=2458675.79$) with the P48. This first detection was in the $g$ band, with a host-subtracted magnitude of  $19.15\pm0.12$, at $\alpha=16^{h}31^{m}01.62^{s}$, $\delta=8\degr28\arcmin23.74\arcsec$. The discovery was reported to TNS \citep{2019lci} three days later on 2019 July 14, with a note saying that the latest non-detection was two hours prior to discovery ($g > 20.0$ mag). The transient was classified as a Type Ic-BL event based on a P60 SEDM spectra obtained on 2019 July 16 \citep{2019lciclassification}. The SN exploded in the outskirts of spiral galaxy WISEA J163101.63+082829.7, with a well-established redshift of $z = 0.0292$.

\subsection{SN\,2019moc}
We refer the reader to \citet{Anand2024} for details about this event.

\subsection{SN\,2019oqp}
The first ZTF photometry of SN\,2019oqp (ZTF19abqshry) was obtained on 2019 August 21 ($\mathrm{JD}=2458716.66$). This first detection was in the $r$ band, with a host-subtracted magnitude of  $20.20\pm0.15$, at $\alpha=16^{h}49^{m}57.02^{s}$, $\delta=27\degr38\arcmin26.94\arcsec$ . The discovery was first reported to TNS by ATLAS \citep{2019oqp} on 2021 August 25 ($c = 19.32$). The transient was classified as a Type Ic-BL event based on a DBSP spectrum obtained on 2019 August 27 \citep{2019oqpclassification}. The SN exploded in the outskirts of the spiral galaxy MCG +08-30-042 with a redshift of $z = 0.0308$.

%\subsection{SN\,2019odp}
%Our first ZTF photometry of SN\,2019odp (ZTF19abqwtfu) was obtained on 2019 August 21 ($\mathrm{JD}=2458716.88$). This first detection was in the $g$ band, with a host-subtracted magnitude of  $18.80\pm0.09$, at $\alpha=23^{h}07^{m}19.09^{s}$, $\delta=13\degr51\arcmin21.48\arcsec$ . The discovery was reported to TNS (\textbf{Discrepancy between TNS date and magnitude and what's on fritz)} \citep{2019odp}, with a note saying the last non-detection from ZTF was seven days prior to discovery (August 14, $g > 19.60$) . The transient was classified as a Type Ic-BL event based on a spectraum obtained on 2019 August 27 \citep{2019odpclassification} from the EFOSC2-NTT instrument on the ESO New Technology Telescope. The SN exploded at the edge of galaxy UGC 12373,  with a redshift of $z = 0.014$. Due to a lack of high cadence early-time photometry, we estimate the explosion date to be the average of the last non-detection made by ZTF with the first detection, on $\mathrm{JD_{explosion}^{SN 2019odp}} = 2458715.70
%\pm1.46$.

\subsection{SN\,2019pgo}
The first ZTF photometry of SN\,2019pgo (ZTF19abupned) on 2019 August 30 ($\mathrm{JD}=2458725.89$). The first detection was in the $g$ band, with a host-subtracted magnitude of $20.69 \pm 0.22$ at $\alpha=23^{h}53^{m}00.05^{s}$, $\delta=25\degr07\arcmin16.46\arcsec$ . The discovery was reported by the Tsinghua-NAOC Transient Survey to TNS two days later \citep{2019pgo} on 2019 September 1. The transient was classified as a Type Ic-BL event based on a spectrum obtained on 2019 September 18 \citep{2019pgoclassification} with SPRAT on the Liverpool Telescope. The SN exploded in the galaxy WISEA J235300.28+250717.2. The galaxy does not have a well-established redshift, but a redshift of $z = 0.0500$ was determined through narrow host lines from a Liverpool Telescope spectrum. 

\subsection{SN\,2019qfi}
We refer the reader to \citet{Anand2024} for details about this event.

%\subsection{SN\,2019xcc}
%We refer the reader to \citet{Anand2024} for details about this event.

\subsection{SN\,2020zg}
The first ZTF photometry of SN\,2020zg (ZTF20aafmdzj) was obtained on 2020 January 15 ($\mathrm{JD}=2458863.63$). The first detection was in the $g$ band, with a host-subtracted magnitude of $g = 17.91 \pm 0.02$ at $\alpha=4^{h}02^{m}36.40^{s}$, $\delta=-16\degr11\arcmin54.33\arcsec$ . The discovery was reported to TNS \citep{2020zg} with a note saying the latest not-detection by ZTF was two days prior to discovery on 2020 January 13 ($g > 19.32$). The transient was classified as a Type Ic-BL event \citep{2020zgclassification} based on a DBSP spectrun obtained on on 2020 Feburary 26. The SN exploded in the edge of galaxy  WISEA J040236.46-161152.0. The galaxy does not have a well-established redsift, but a redshift of $z = 0.0557$ was determined through narrow host lines from the DBSP spectrum. 

\subsection{SN\,2020ayz}
The first ZTF photometry of SN\,2020ayz (ZTF20aaiqiti) was obtained on 2020 January 26 ($\mathrm{JD}=2458874.83
$). The first detection was in the $r$ band, with a host-subtracted magnitude of $r = 19.77 \pm 0.13$ at $\alpha=12^{h}12^{m}04.90^{s}$, $\delta=32\degr44\arcmin01.73\arcsec$ . The discovery was reported to TNS \citep{2020ayz} on the same day ($g = 20.0$) with a note saying the latest non-detection by ZTF was three days prior to discovery on 2020 January 23 ($g > 20.23$). The transient was classified as a Type Ic-BL event \citep{2020zgclassification} based on a SEDM spectrum obtained on 2020 Feburary 03. The SN exploded in the center of the spiral galaxy MCG+06-27-025, with a well-established redshift of $z =0.025437$.

\subsection{SN\,2020bvc}
We refer the reader to \citet{20aalxlis, Rho2021, Izzo2020, Long2023} for details about this event.

\subsection{SN\,2020dgd}
We refer the reader to \citet{Anand2024} for details about this event.

\subsection{SN\,2020hes}
The first ZTF photometry of SN\,2020hes (ZTF20aaurexl) was obtained on 2020 April 14 ($\rm{JD}=2458953.93$). The first detection was in the $g$ band, with a host-subtracted magnitude of $g = 18.62 \pm 0.07$ at $\alpha=17^{h}47^{m}05.71^{s}$, $\delta=42\degr46\arcmin39.72\arcsec$. The discovery was reported to TNS \citep{2020hes} with a note saying the latest non-detection by ZTF was ten days prior to discovery on 2020 April 04 ($g > 20.60$). The transient was classified as a Type Ic-BL event \citep{2020hesclassification} based on a SEDM spectrum obtained on 2020 April 25. The SN exploded in the galaxy WISEA J174706.10+424640.5. The galaxy does not have a well-established redsift, but a redshift of $z = 0.0700$ was determined through narrow host lines from a DBSP spectrum. 

\subsection{SN\,2020hyj}
The first ZTF photometry of SN\,2020hyj (ZTF20aavcvrm) was obtained on 2020 April 16 ($\mathrm{JD}=2458955.92$). This first detection was in the $g$ band, with a host-subtracted magnitude of $20.33 \pm 0.21$ at $\alpha=16^{h}23^{m}47.22^{s}$, $\delta=29\degr58\arcmin58.38\arcsec$ . The discovery was reported to TNS \citep{2020hyj} on 2020 April 16 with a note saying the latest non-detection by ZTF was two days prior to discovery on 2020 April 14 ($g > 20.32$). The transient was classified as a Type Ic-BL event \citep{2020hyjclassification} based on a SEDM spectrum obtained on on 2020 April 29. No host galaxy was found upon inspection of archival images through NED. The redshift of $z = 0.055$ was determined through narrow host lines from the SEDM spectrum.

\subsection{SN\,2020jqm}
We refer the reader to \citet{Corsi2024} for details about this event.

\subsection{SN\,2020lao}
We refer the reader to \citet{Anand2024} for details about this event.

\subsection{SN\,2020rfr}
The first ZTF photometry of SN\,2020rfr (ZTF20abrmmah) was obtained on 2020 August 12 ($\mathrm{JD}=2459073.84$). This first detection was in the $g$ band, with a host-subtracted magnitude of $19.67 \pm 0.15$ at $\alpha=22^{h}39^{m}49.30^{s}$, $\delta=-06\degr26\arcmin15.92\arcsec$ . The discovery was reported to TNS \citep{2020rfr} with a note saying the latest non-detection by ZTF was four days prior to discovery on 2020 August 08 ($r > 19.96$). The transient was classified as a Type Ic-BL event \citep{2020rfrclassification} based on a LRIS spectrum obtained on on 2020 September 20. The SN exploded in spiral galaxy WISEA J223949.24-062616.9. The galaxy does not have a well-established redshift, but a redshift of $z=0.073$ was determined through narrow host lines from the LRIS spectrum.

\subsection{SN\,2020rph}
We refer the reader to \citet{Anand2024} for details about this event.

\subsection{SN\,2020tkx}
We refer the reader to \citet{Anand2024} for details about this event.

\subsection{SN\,2020wgz}
Our first ZTF photometry of SN\,2020wgz (ZTF20achvlbs) was obtained on 2020 October 08 ($\mathrm{JD}=2459130.99$). This first detection was in the $r$ band, with a host-subtracted magnitude of $20.08 \pm 0.24$ at $\alpha=08^{h}57^{m}33.32^{s}$, $\delta=62\degr34\arcmin00.07\arcsec$ , The discovery was reported to TNS \citep{2020wgz} with a note saying the latest non-detection by ZTF was one hour prior to the first detection. The SN has no visible galaxy counterpart, and a redshift of $z = 0.1785$ was determined through narrow host lines from a DBSP spectrum obtained on 2020 October 21.

\subsection{SN\,2020abxl}
The first ZTF photometry of SN\,2020abxl (ZTF20acvcxkz) was obtained on 2020 December 05 ($\mathrm{JD}=2459188.79$). This first detection was in the $r$ band, with a host-subtracted magnitude of $19.48 \pm 0.16$ at $\alpha=05^{h}04^{m}22.76^{s}$, $\delta=-14\degr02\arcmin46.42\arcsec$ . The discovery was reported to TNS \citep{2020abxl} by ATLAS on 2020 December 07 ($o = 19.65)$ with a note saying the latest non-detection by ATLAS was one day prior to discovery on 2020 December 06 ($o > 19.33$). The transient was classified as a Type Ic-BL event \citep{2020abxlclassification}  based on a spectrum obtained by the EFOSC2-NTT on the ESO New Technology Telescope on 2020 December 11. The SN exploded in spiral galaxy SDSS J223949.25-062616.3. The galaxy does not have a well established redshift, but a redshift of $z = 0.0815$ was determined through narrow host lines from the NTT spectrum.

\subsection{SN\,2020abxc}
Our first ZTF photometry of SN\,2020abxc (ZTF20acvmzfv) was obtained on 2020 December 07 ($\mathrm{JD}=2459190.65$). This first detection was in the $r$ band, with a host-subtracted magnitude of $19.22 \pm 0.09$ at $\alpha=01^{h}00^{m}34.04^{s}$, $\delta=-08\degr07\arcmin00.67\arcsec$ . The discovery was reported to TNS \citep{2020abxc} with a note saying the latest non-detection by ZTF was two days prior to discovery on 2020 December 05 ($g > 20.03$). The transient was classified as a Type Ic-BL event \citep{2020abxcclassification} based on a spectrum obtained by the IMACS instrument on the Walter Baade Magellan 6.5-m telescope on 2020 December 08. The SN exploded in the spiral galaxy WISEA J010033.88-080656.9. The galaxy does not have a well-established redshift, but a redshift of $z = 0.0600$ was determined through narrow host lines from the IMACS spectrum.

\subsection{SN\,2020adow}
Our first ZTF photometry of SN\,2020adow (ZTF20adadrhw) was obtained on 2020 December 27 ($\mathrm{JD}=2459210.7560$).
This first detection was in the $g$ band, with a host-subtracted magnitude of $15.84\pm0.03$, at $\alpha=08^{h}33^{m}42.26^{s}$, $\delta=+27\degr42\arcmin43.8\arcsec$ .
The transient was discovered by ASASSN already the day before, on 2020 December 26
\citep{2020TNSTR3922....1S}. The transient was classified as a Type Ic-BL on December 27 \citep{2020TNSCR3946....1Z} from a spectrum obtained on the Liverpool Telescope. The SN exploded in the spiral galaxy KUG 0830+278, with a well-established redshift of $z = 0.0075$.

%We estimate the explosion date as 
%$\mathrm{JD_{explosion}^{SN 2020adow}} = 2459206.98\pm3.86$ %based on the first ZTF detection and the last non-detection. 

\subsection{SN\,2021bmf}
We refer the reader to \citet{Anand2024} for details about this event.

%\subsection{SN\,2021xv}
%We refer the reader to \citet{Anand2024} for details about this event.

%\subsection{SN\,2021aug}
%We refer the reader to \citet{Corsi2024} for details about this event.

\subsection{SN\,2021epp}
We refer the reader to \citet{Corsi2024} for details about this event.

\subsection{SN\,2021fop}
The first ZTF photometry of SN\,2021fop (ZTF21aapecxb) was obtained on 2021 March 17 ($\mathrm{JD}= 2459290.71$). This first detection was in the $g$ band, with a host-subtracted magnitude of $18.61 \pm 0.05$ at $\alpha=07^{h}46^{m}42.91^{s}$, $\delta=07\degr12\arcmin38.70\arcsec$ . The discovery was reported to TNS \citep{2021fop} on 2021 March 15 by ATLAS ($o = 18.63$)  with a note saying the latest non-detection by ATLAS was eight days prior to discovery on 2021 March 07 ($o > 19.76$). The transient was classified as a Type Ic-BL event \citep{2021fopclassification} based on a spectrum obtained by the SNIFS instrument on the UH 88 inch telescope on 2021 March 20. The SN exploded in the spiral galaxy WISEA J074642.81+071238.0. The galaxy does not have a well established redshift, but a redshift of $z = 0.077$ was determined through narrow host lines from a spectrum on 2021 March 18 obtained by the SPRAT instrument aboard the Liverpool Telescope.

%\subsection{SN\,2021htb}
%We refer the reader to \citet{Corsi2024} for details about this event.

\subsection{SN\,2021hyz}
We refer the reader to \citet{Corsi2024} for details about this event.

\subsection{SN\,2021ktv}
The first ZTF photometry of SN\,2021ktv (ZTF21aaxxihx) was obtained on 2021 May 01 ($\mathrm{JD}= 2459335.69$). This first detection was in the $r$ band, with a host-subtracted magnitude of $19.38 \pm 0.13$ at $\alpha=11^{h}03^{m}03.89^{s}$, $\delta=08\degr51\arcmin39.75\arcsec$ . The discovery was reported to TNS \citep{2021ktv} on the same day by ATLAS ($o = 19.50$)  with a note saying the latest non-detection by ATLAS was two days prior to discovery on 2021 April 30 ($o > 19.4$). The transient was classified as a Type Ic-BL event \citep{2021ktvclassification} based on a spectrum obtained by the SNIFS instrument on the UH 88 inch telescope on 2021 May 19. The SN exploded in galaxy WISEA J110303.81+085140.9. The galaxy does not have a well-established redshift, but a redshift of $z = 0.07$ was determined through narrow host lines from a spectrum taken by DBSP on 2021 June 04.

\subsection{SN\,2021ncn}
The first ZTF photometry of SN\,2021ncn (ZTF21abchjer) was obtained on 2021 May 23 ($\mathrm{JD}= 2459357.97$). This first detection was in the $r$ band, with a host-subtracted magnitude of $18.58 \pm 0.06$ at $\alpha=22^{h}36^{m}32.93^{s}$, $\delta=25\degr45\arcmin40.58\arcsec$ . The discovery was reported to TNS \citep{2021ncn} with a note saying the latest non-detection by ZTF was three days prior to discovery on 2021 May 20 ($g > 19.64$). The transient was classified as a Type Ic-BL event \citep{2021ncnclassification} based on a spectrum obtained by the SEDM on 2021 May 30. The SN exploded in the outskirts of spiral galaxy IC 5233, with a well-established redshift of $z = 0.0246$ .

\subsection{SN\,2021qjv}
The first ZTF photometry of SN\,2021qjv (ZTF20abcjdwu) was obtained on 2021 June 18 ($\mathrm{JD}= 2459383.72
$). This first detection was in the $r$ band, with a host-subtracted magnitude of $19.38 \pm 0.18$ at $\alpha=15^{h}10^{m}47.04^{s}$, $\delta=49\degr12\arcmin18.14\arcsec$ . The discovery was reported to TNS \citep{2021qjv} two days earlier on 2021 June 16 ($g = 19.67$) with a note saying the latest non-detection by ZTF was two days prior to discovery on 2021 May 20 ($r > 20.57$). The transient was classified as a Type Ic-BL event \citep{2021qjvclassification} based on a spectrum obtained by LRIS on 2021 July 09. Two galaxies, WISEA J151047.04+491218.1 and SDSS J151046.89+491215.4 are spatially consistent with the location of the SN, both with a redshift of $z = 0.0383$  

\subsection{SN\,2021too}
We refer the reader to \citet{Anand2024} for details about this event.

\subsection{SN\,2021ywf}
We refer the reader to \citet{Anand2024} for details about this event.

%ZTF18abhhnnv_spectra.pdf ,ZTF18acbvpzj_spectra.pdf	,ZTF18acxgoki_spectra.pdf	,ZTF19aawqcgy_spectra.pdf	,ZTF19aaxfcpq_spectra.pdf	,ZTF19abfsxpw_spectra.pdf	,ZTF19ablesob_spectra.pdf	,ZTF19abqshry_spectra.pdf	,ZTF19abupned_spectra.pdf	,ZTF19abzwaen_spectra.pdf	,ZTF20aafmdzj_spectra.pdf	,ZTF20aaiqiti_spectra.pdf	,ZTF20aapcbmc_spectra.pdf	,ZTF20aaurexl_spectra.pdf	,ZTF20aavcvrm_spectra.pdf	,ZTF20aazkjfv_spectra.pdf,ZTF20abbplei_spectra.pdf,ZTF20abcjdwu_spectra.pdf,

%ZTF20abrmmah_spectra.pdf,ZTF20abswdbg_spectra.pdf,ZTF20abzoeiw_spectra.pdf,

%ZTF20achvlbs_spectra.pdf,ZTF20acvcxkz_spectra.pdf,

%ZTF20acvmzfv_spectra.pdf,

%ZTF20adadrhw_spectra.pdf,ZTF21aapecxb_spectra.pdf,ZTF21aartgiv_spectra.pdf,

%ZTF21aaxxihx_spectra.pdf,ZTF21abchjer_spectra.pdf,

%ZTF21abmjgwf_spectra.pdf,

%ZTF21acbnfos_spectra.pdf
\section{Spectral Sequences}
Spectra are all shown at their rest wavelengths, and fluxes have been normalized  to the median of the spectrum. The rest frame phase of the spectrum with respect to the $r$-band peak is also shown to the right of each individual spectrum. We show a log of the spectroscopic observations in Table \ref{spectratable}.

\begin{longtable}{l|l|l|l}
\caption{Spectral log of observations. Spectrum phase is in rest-frame with respect to $r$-band maximum}\\
\label{spectratable}\\
\hline
ZTF name     & SN name     & Spectrum Phase & Instrument  \\
\hline
        ZTF18abhhnnv & SN 2018ell & 0.7 & SEDM\\ 
        ZTF18acbvpzj & SN 2018hsf & -7.5 & SEDM\\ 
        & & 22.1 & LRIS \\ 
        ZTF18acxgoki & SN 2018keq & 0.8 & SEDM\\
        & & 12.3 & LRIS\\
        & & 186.6 & LRIS\\ 
        ZTF19aawqcgy & SN 2019hsx & -3.2 & SEDM\\
        & & 0.7 & SEDM\\
        & & 18.4 & DBSP\\
        & & 103.6 & LRIS\\ 
        ZTF19aaxfcpq & SN 2019gwc & -8.9 & SEDM\\
        & & -7.0 & SEDM\\
        & & -5.0 & SEDM\\
        & & -0.2 & SEDM\\
        & & 4.6 & SEDM\\
        & & 8.5 & SEDM\\
        & & 15.2 & DBSP\\
        & & 20.0 & SEDM\\
        & & 45.1 & DBSP\\ 
        ZTF19abfsxpw & SN 2019lci & -13.8 & SEDM\\
        & & -12.8 & SEDM\\
        & & -9.0 & SEDM\\
        & & 1.7 & SEDM\\
        & & 6.6 & DBSP\\
        & & 12.4 & SEDM\\
        & & 21.2 & NOT\\ 
        ZTF19ablesob & SN 2019moc & -13.6 & SEDM\\
        & & -9.8 & SEDM\\
        & & -8.9 & DBSP\\
        & & -6.0 & SEDM\\
        & & 7.2 & DBSP\\
        & & 65.0 & LRIS\\ 
        ZTF19abqshry & SN 2019oqp & -17.7 & SEDM\\
        & & -13.8 & DBSP\\
        & & -10.9 & NOT\\
        & & -8.0 & DBSP\\ 
        ZTF19abupned & SN 2019pgo & -9.0 & SEDM\\
        & & -6.1 & SEDM\\
        & & 1.5 & SPRAT\\
        & & 1.5 & SEDM\\
        & & 38.7 & LRIS\\ 
        ZTF19abzwaen & SN 2019qfi & -5.2 & SEDM\\
        & & 4.6 & SEDM\\
        & & 13.3 & SEDM\\
        & & 29.9 & LRIS\\ 
        ZTF20aafmdzj & SN 2020zg & 10.2 & SEDM\\
        & & 13.0 & SEDM\\
        & & 23.5 & SEDM\\ 
        ZTF20aaiqiti & SN 2020ayz & -8.3 & SEDM\\
        & & -6.3 & SEDM\\
        & & 7.4 & SEDM\\
        & & 18.1 & DBSP\\
         & & 133.2 & LRIS\\  
        ZTF20aapcbmc & SN 2020dgd & -7.3 & SEDM\\
        & & -0.5 & SEDM\\
        & & 0.4 & SEDM\\
        & & 16.0 & LRIS\\
        & & 28.6 & SEDM\\
        & & 106.1 & LRIS\\ 
        ZTF20aaurexl & SN 2020hes & -7.9 & SEDM\\
        & & 0.5 & SEDM\\
        & & 7.0 & SPRAT\\ 
        ZTF20aavcvrm & SN 2020hyj & -0.4 & SEDM\\
        & & 2.4 & SEDM\\
        & & 4.3 & SEDM\\ 
        ZTF20aazkjfv & SN 2020jqm & -0.4 & SEDM\\
        & & 1.6 & SEDM\\
        & & 4.4 & SEDM\\
        & & 10.2 & NOT\\
        & & 21.8 & SEDM\\ 
        ZTF20abbplei & SN 2020lao & -8.2 & SEDM\\
        & & -7.2 & DBSP\\
        & & -4.3 & SEDM\\
        & & 0.5 & SEDM\\
        & & 9.3 & SEDM\\
        & & 14.1 & SEDM\\
        & & 16.1 & SPRAT\\
        ZTF20abcjdwu & SN 2021qjv & -1.3 & SEDM\\
        & & 4.5 & SPRAT \\
        & & 15.1 & LRIS \\
        ZTF20abrmmah & SN 2020rfr & 2.5 & SEDM\\
        & & 3.4 & SEDM\\
        & & 7.1 & SEDM\\
        & & 28.6 & DBSP\\
        & & 32.3 & LRIS\\  
        ZTF20abswdbg & SN 2020rph & -6.2 & SEDM\\
        & & -5.2 & SEDM\\
        & & -0.4 & NOT\\
        & & -0.4 & SEDM\\
        & & 5.3 & SEDM\\
        & & 47.5 & LRIS\\
        & & 78.2 & LRIS\\ 
        ZTF20abzoeiw & SN 2020tkx & -5.2 & SPRAT\\
        & & -2.3 & SEDM\\
        & & -0.4 & SEDM\\
        & & 6.4 & SEDM\\
        & & 14.2 & SEDM\\
        & & 20.1 & SEDM\\
        & & 24.9 & SEDM\\
        & & 50.3 & SEDM\\
        & & 54.2 & SEDM\\ 
        ZTF20achvlbs & SN 2020wgz & -0.4 & SEDM\\
        & & 3.8 & DBSP\\
        & & 21.6 & DBSP\\ 
        ZTF20acvcxkz & SN 2020abxl & -5.1 & ePESSTO+  \\ 
        ZTF20acvmzfv & SN 2020abxc & -2.1 & SEDM\\ 
        ZTF20adadrhw & SN 2020adow & -6.4 & SPRAT \\ 
        ZTF21aagtpro & SN 2021bmf & -11.4 & P200\\
        & & 55.5 & SEDM\\
        & & 96.8 & P200\\
        & & 137.1 & LRIS\\
        & & 172.5 & LRIS\\ 
        ZTF21aaocrlm & SN 2021epp & -5.1 & ePESSTO+\\ 
        ZTF21aapecxb & SN 2021fop & -0.2 & SEDM \\ 
        ZTF21aartgiv & SN 2021hyz & -7.0 & SEDM\\
        & & 0.7 & SEDM\\
        & & 15.0 & SEDM\\ 
        ZTF21aaxxihx & SN 2021ktv & 6.3 & SEDM\\
        & & 7.3 & SPRAT\\
        & & 23.3 & DBSP\\ 
        ZTF21abchjer & SN 2021ncn & -5.3 & SEDM\\
         & & 0.5 & SEDM\\
         & & 2.5 & SEDM\\ 
        
        ZTF21abmjgwf & SN 2021too & -0.4 & DBSP \\
        & & 5.4 & LRIS\\ 
        ZTF21acbnfos & SN 2021ywf & -5.3 & SEDM\\
        & & -0.4 & DBSP\\
        & & 2.5 & DBSP\\
        & & 15.1 & DBSP\\
        & & 125.0 & LRIS\\    
\end{longtable}

\begin{figure}[h!]
 \centering
 $\begin{array}{c}
  \includegraphics[width=12.5cm]{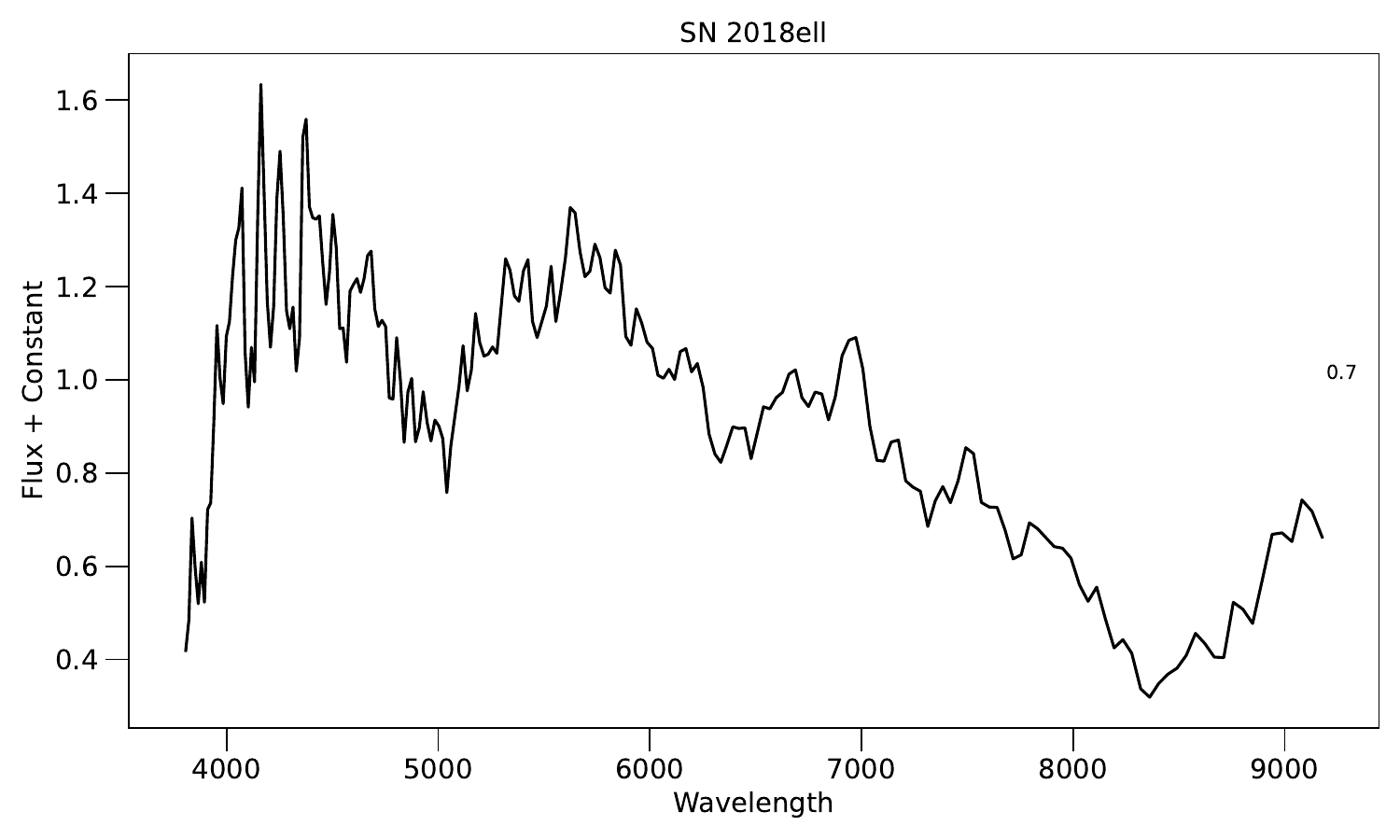 } \\
  \includegraphics[width=12.5cm]{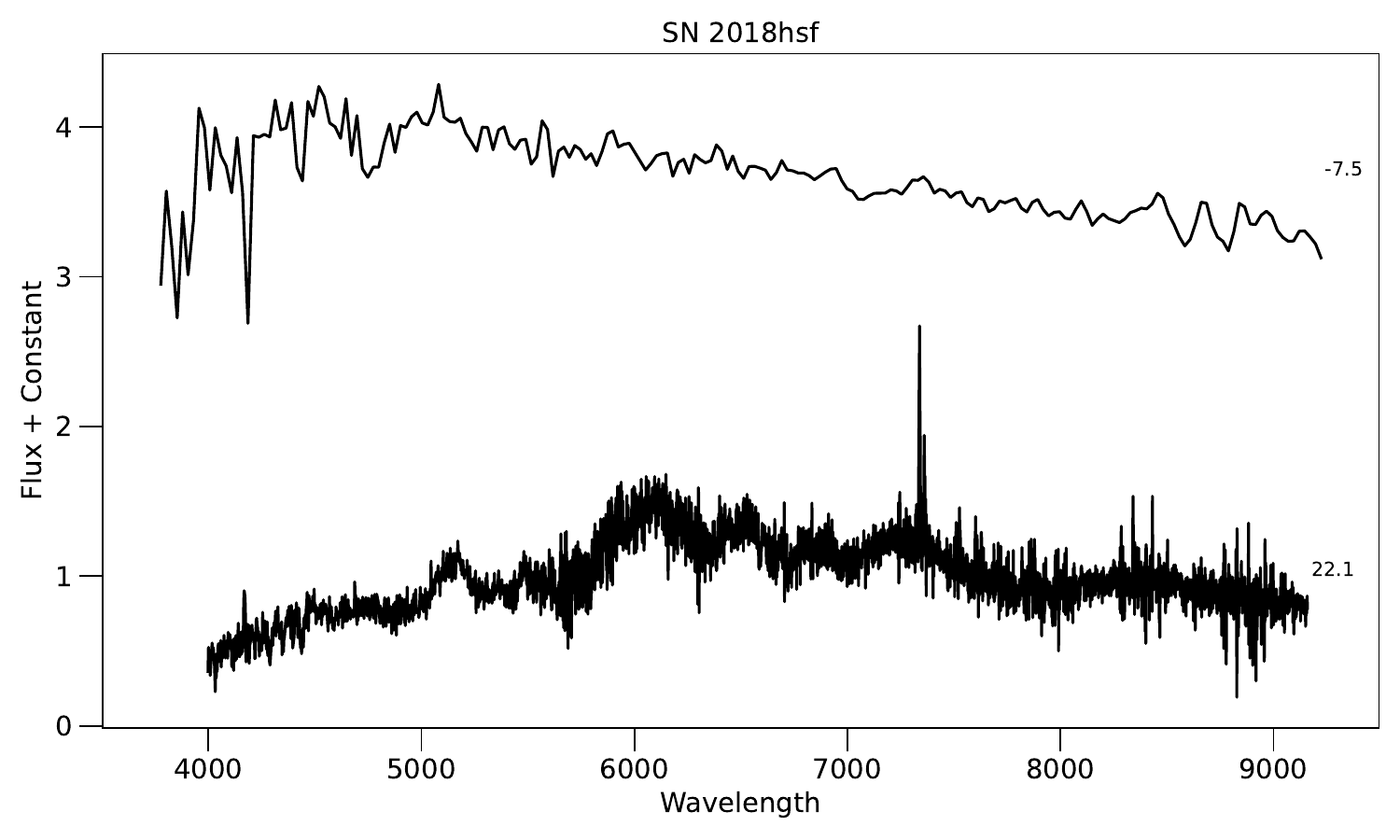} \\
  \includegraphics[width=12.5cm]{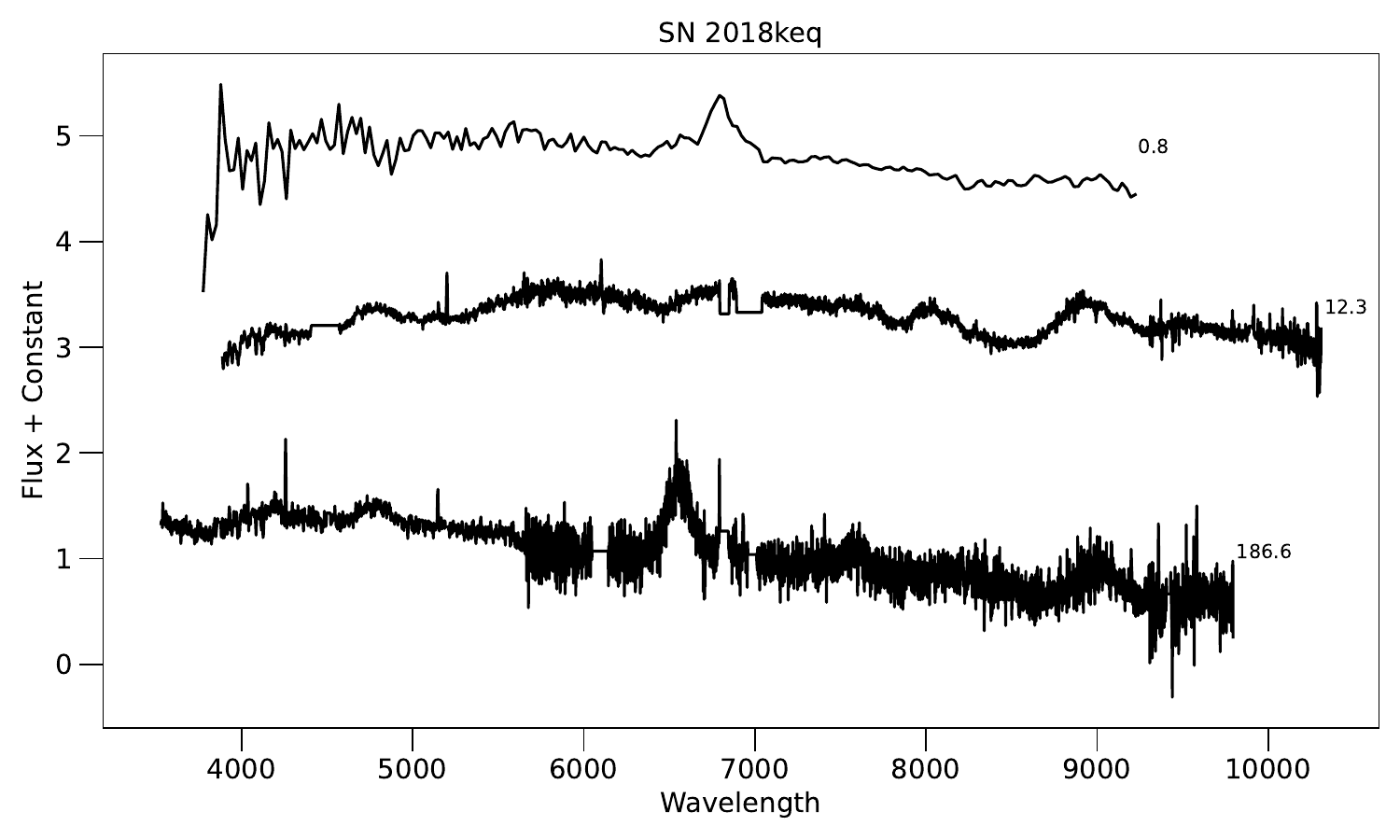} \\
    \end{array}$
  \caption{\label{spec_seq1}Spectral sequences of SN 2018ell, SN 2018hsf, and SN 2018keq. The phases next to each spectrum are the rest-frame days since the maximum $r$-band flux from the GP processing. The spectra for the rest of the events except for those presented in single-object works (SN 2018bvw, SN 2018gep, and SN 2020bvc) are presented in a similar format below.}
 \end{figure}

\begin{figure}[h!]
 \centering
 $\begin{array}{c}
      \includegraphics[width=12.5cm]{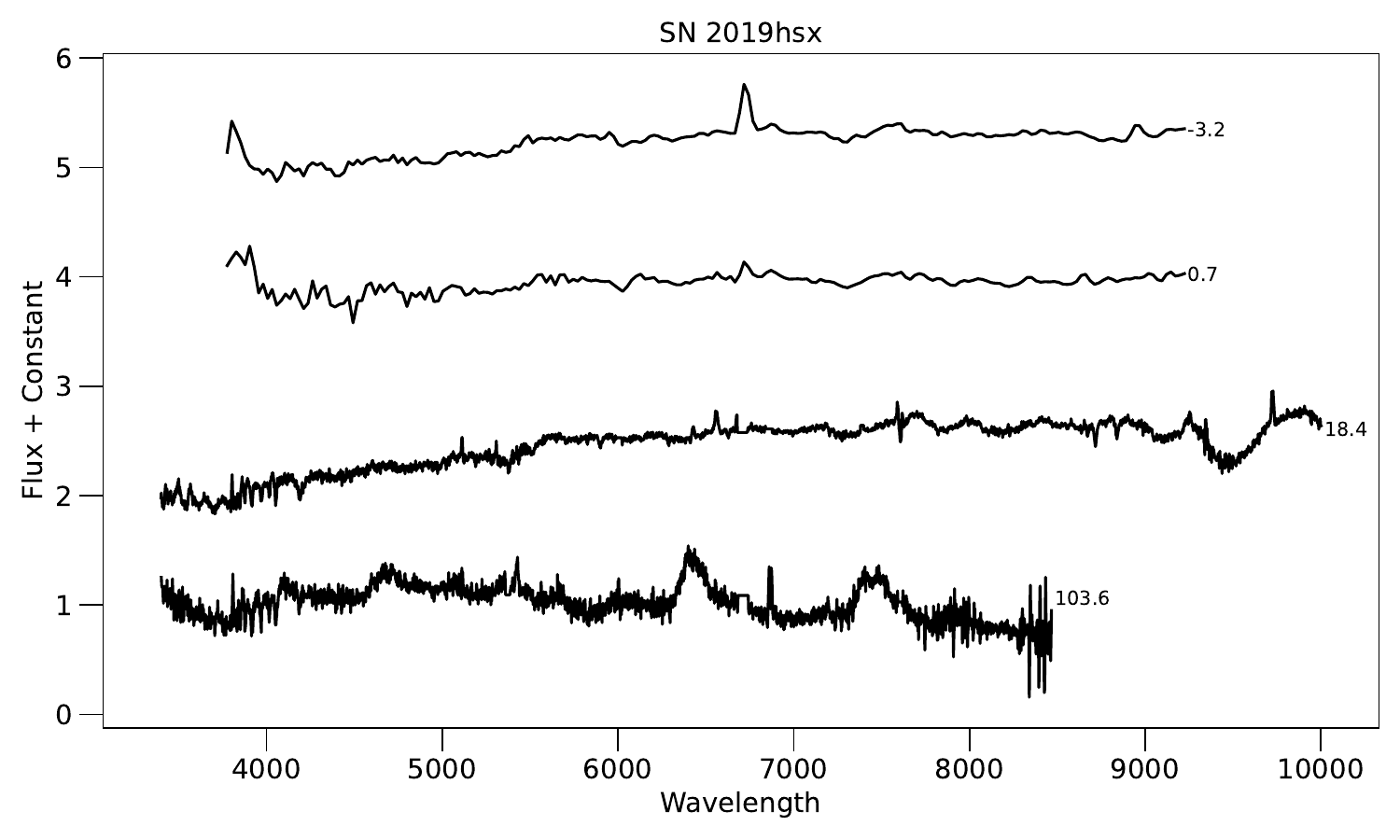} \\
  \includegraphics[width=12.5cm]{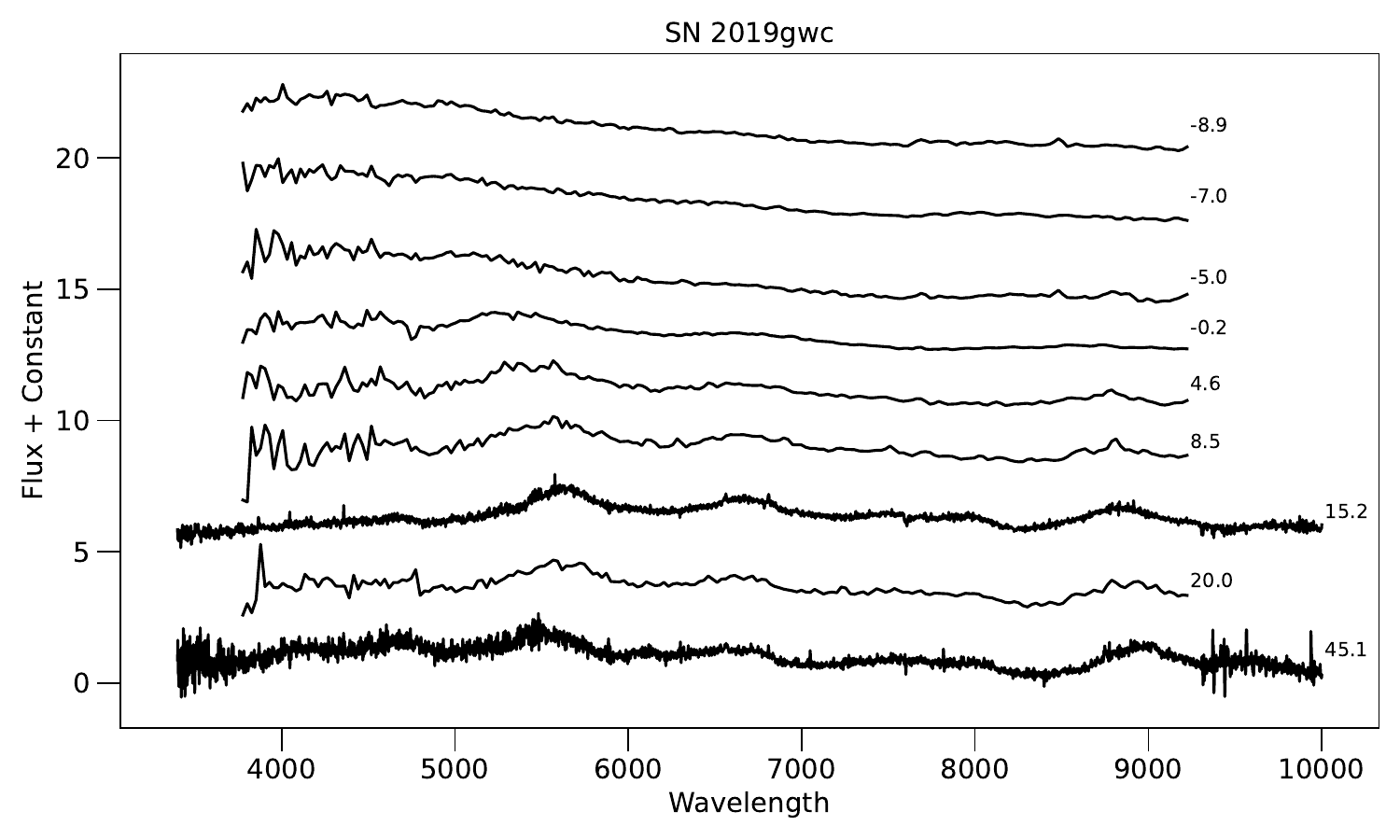} \\
  \includegraphics[width=12.5cm]{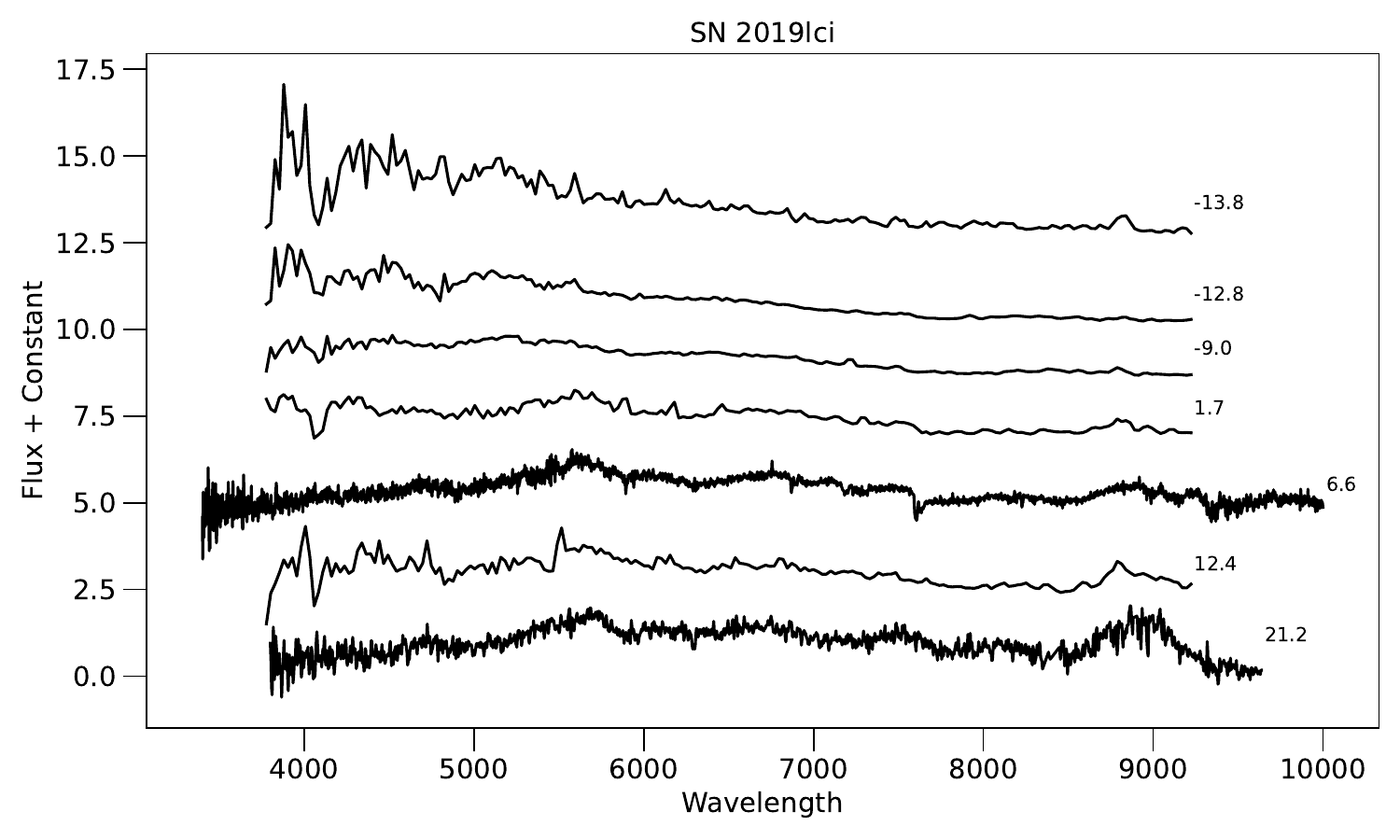} 
    \end{array}$
  \caption{\label{spec_seq2}Spectral sequences of SN 2019hsx, SN 2019gwc, SN 2019lci.}
 \end{figure}

\begin{figure}[h!]
 \centering
 $\begin{array}{c}
      \includegraphics[width=12.5cm]{ 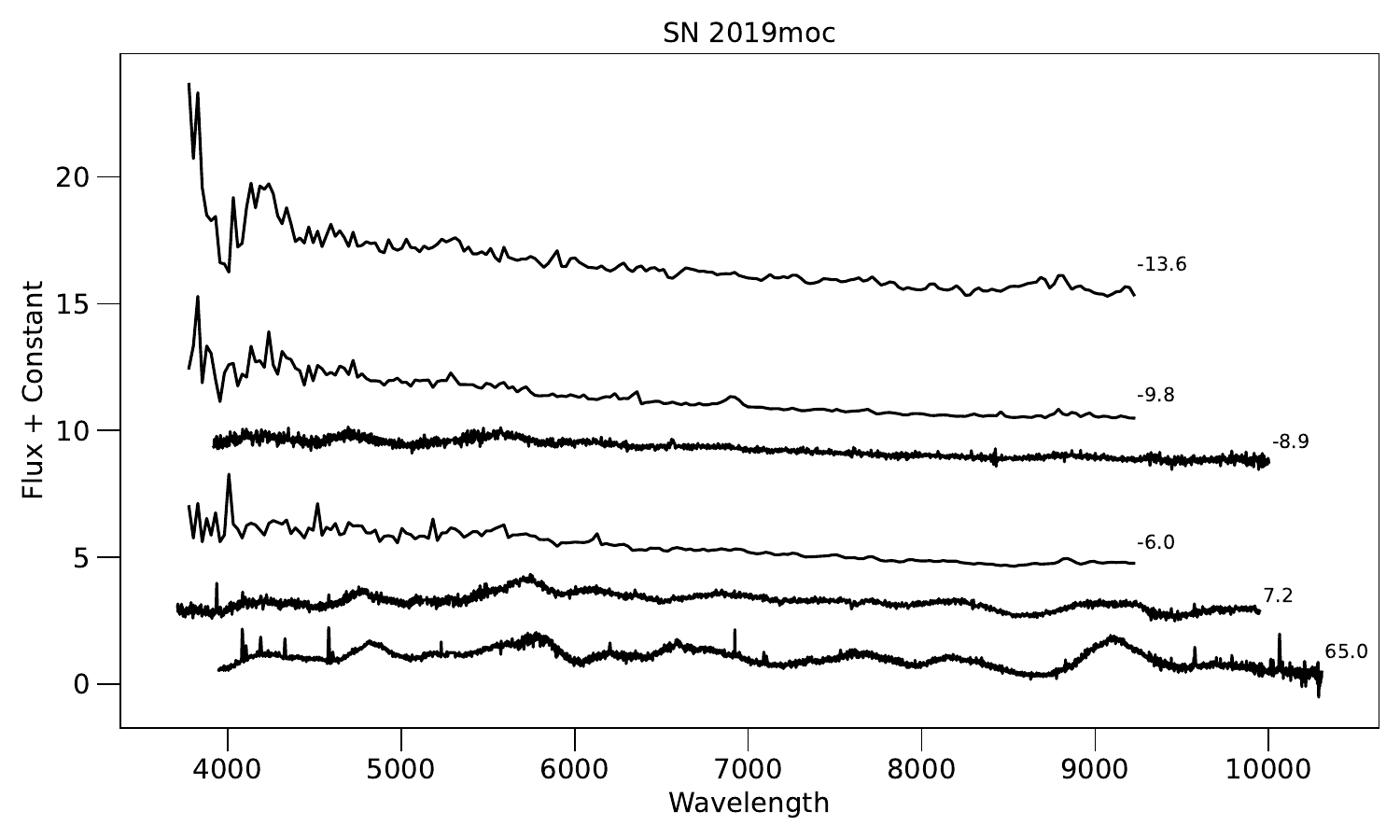} \\
  \includegraphics[width=12.5cm]{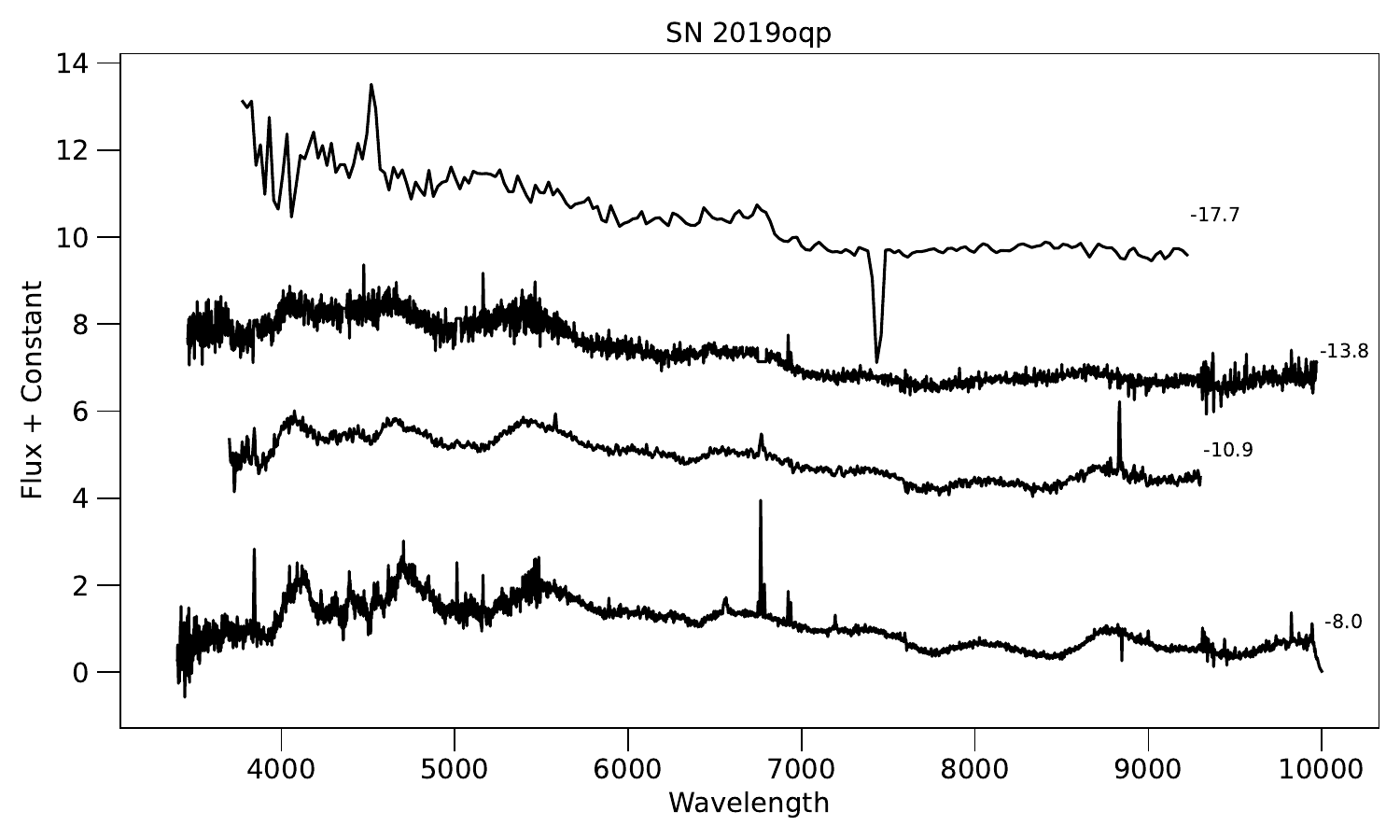} \\
  \includegraphics[width=12.5cm]{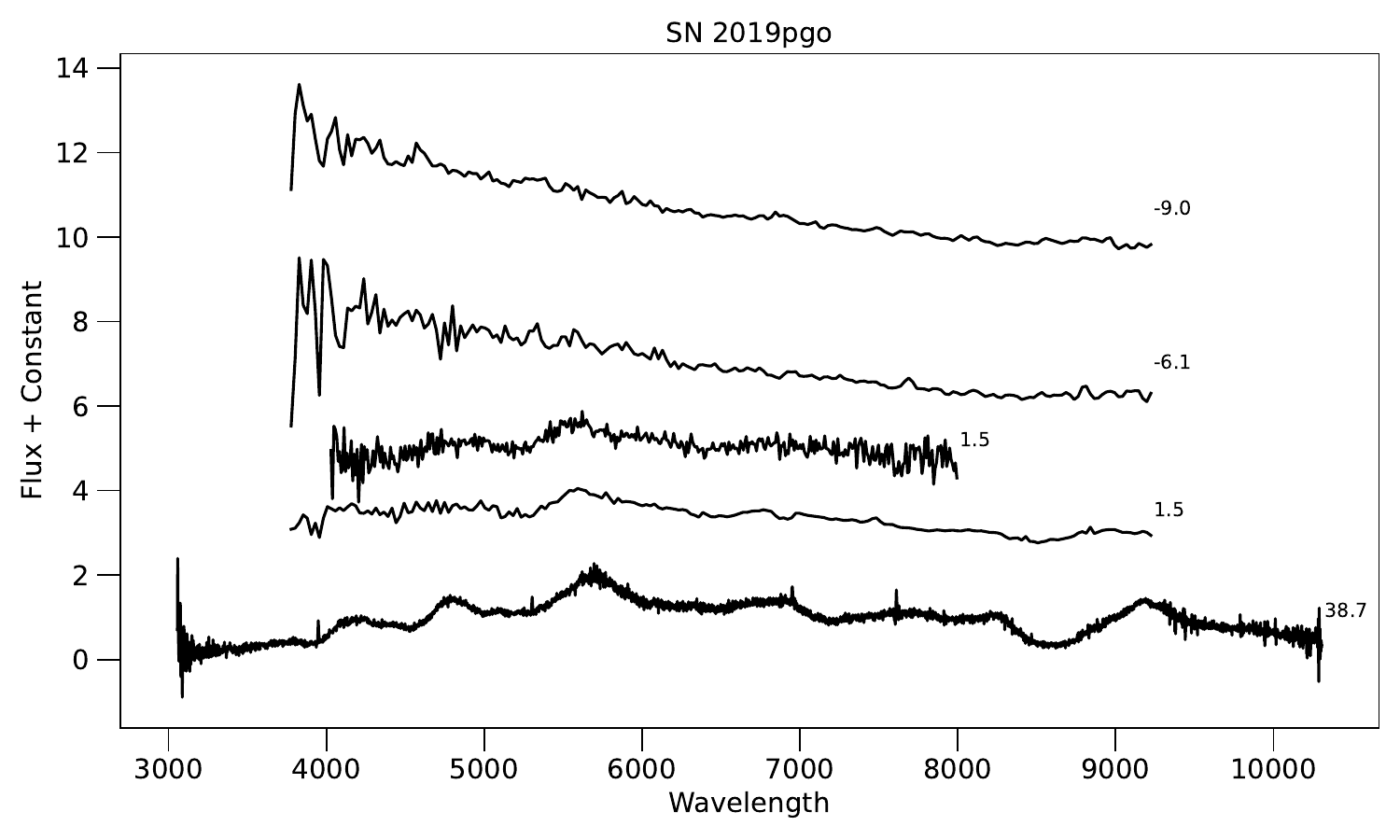} 
    \end{array}$
  \caption{\label{spec_seq2}Spectral sequences of SN 2019moc, SN 2019oqp, SN 2019pgo.}
 \end{figure}

 \begin{figure}[h!]
 \centering
 $\begin{array}{c}
      \includegraphics[width=12.5cm]{ 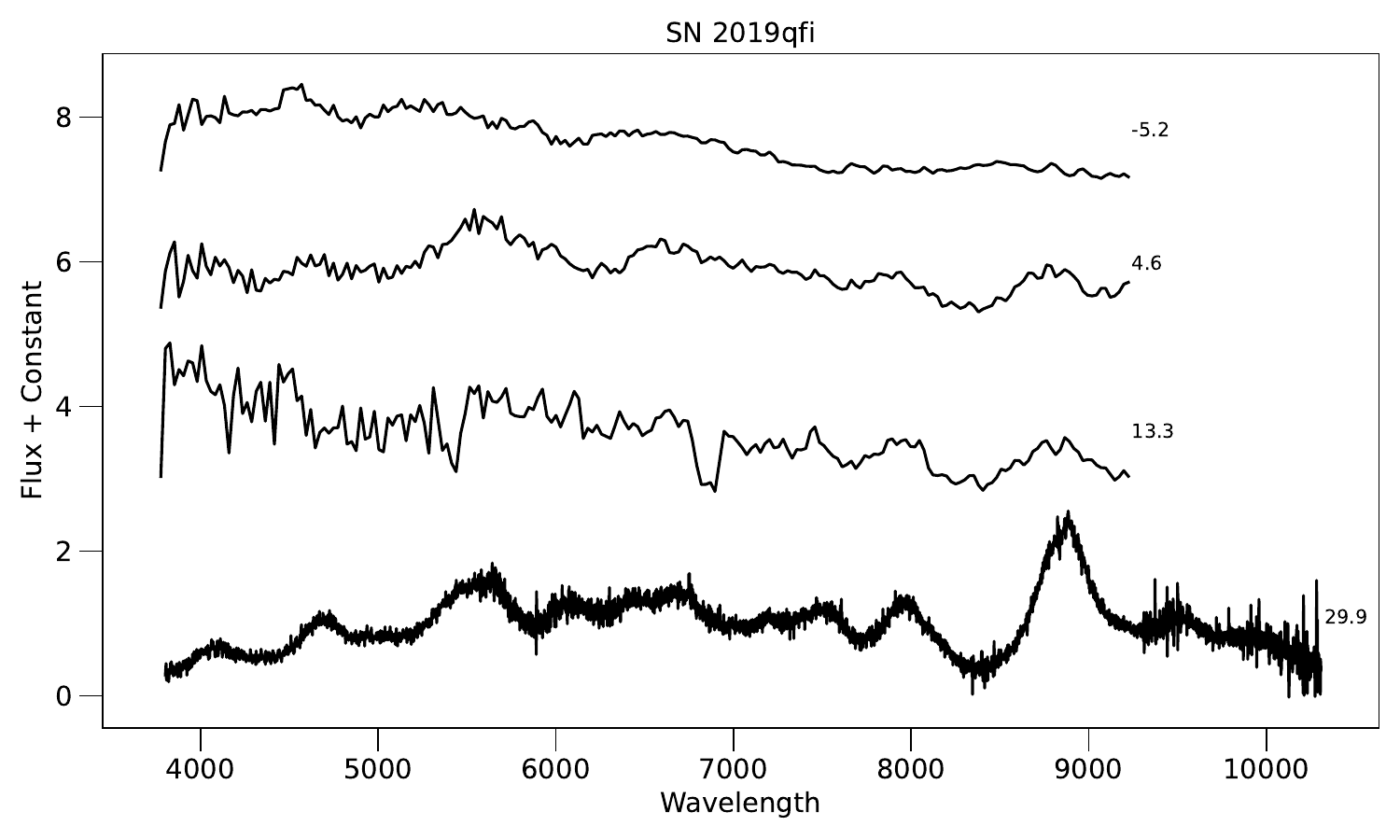} \\
  \includegraphics[width=12.5cm]{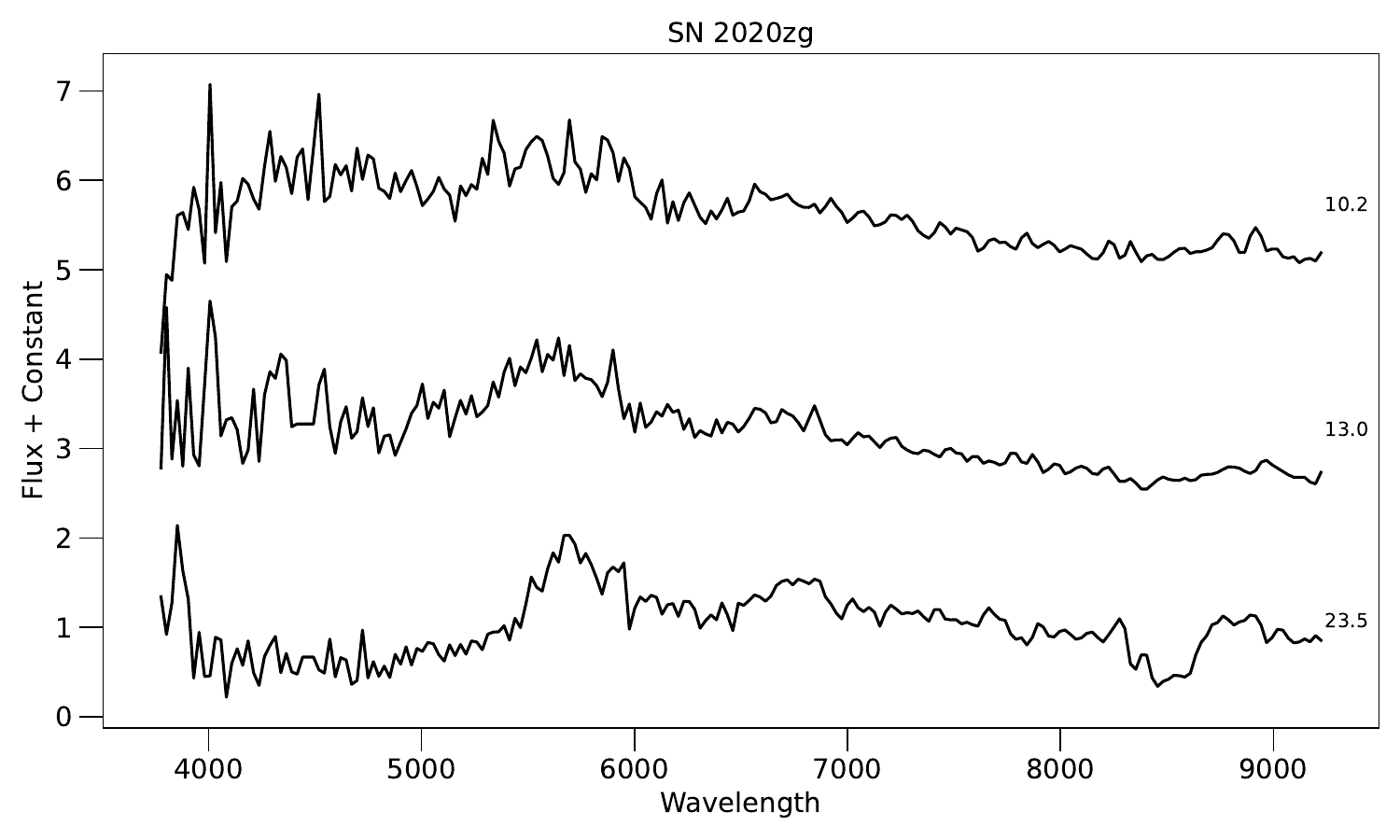} \\
  \includegraphics[width=12.5cm]{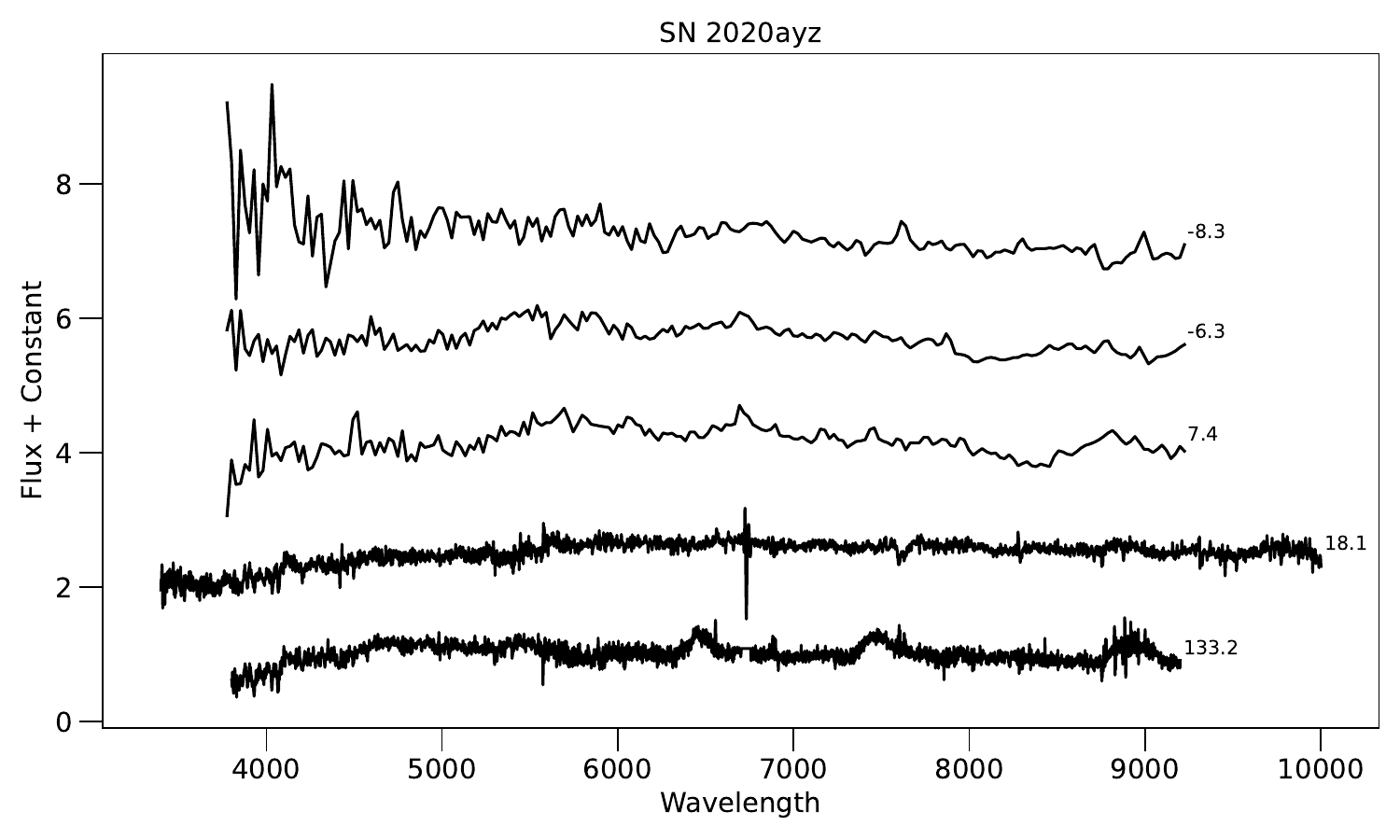} 
    \end{array}$
  \caption{\label{spec_seq2}Spectral sequences of SN 2019qfi, SN 2020zg, SN 2020ayz.}
 \end{figure}

 \begin{figure}[h!]
 \centering
 $\begin{array}{c}
      \includegraphics[width=12.5cm]{ 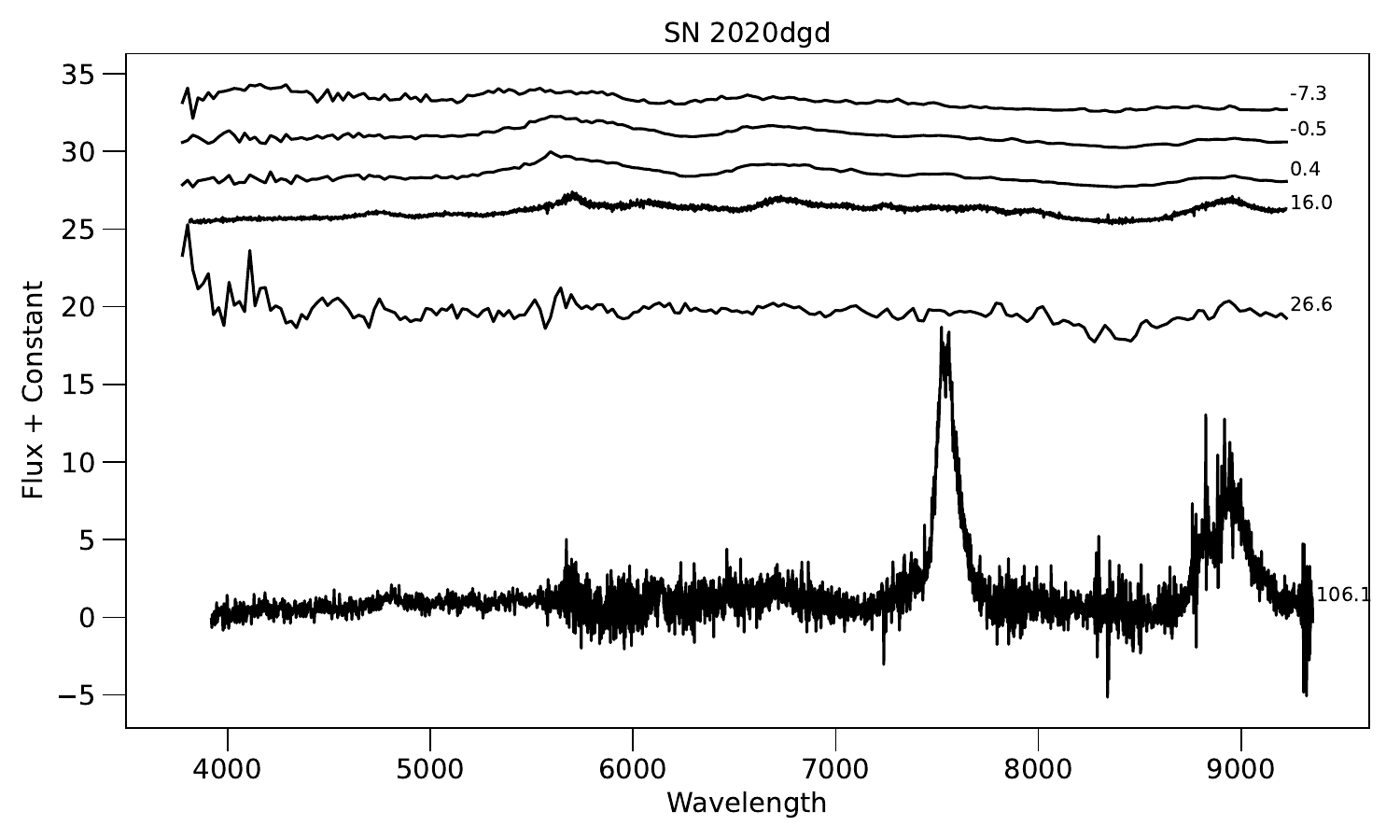} \\
  \includegraphics[width=12.5cm]{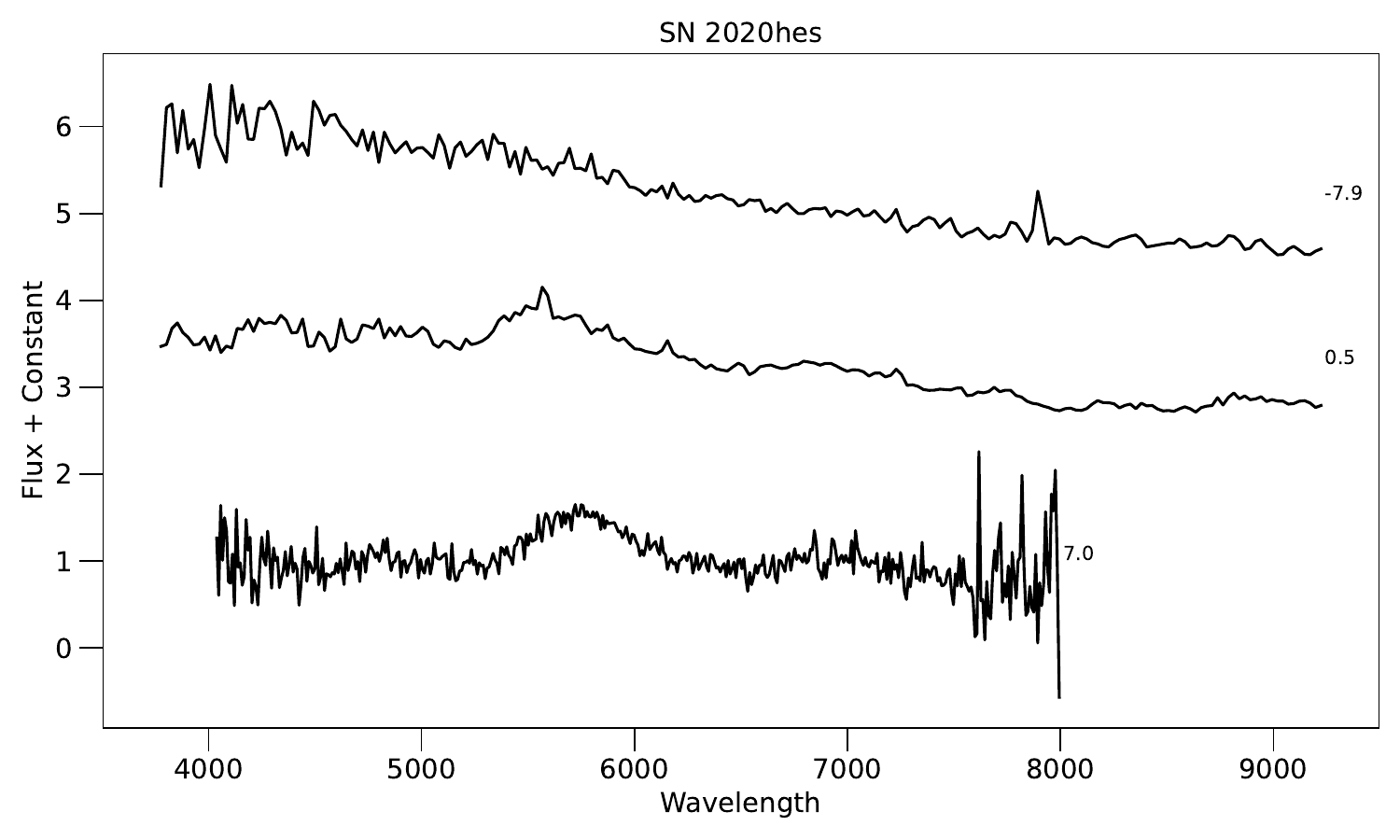} \\
  \includegraphics[width=12.5cm]{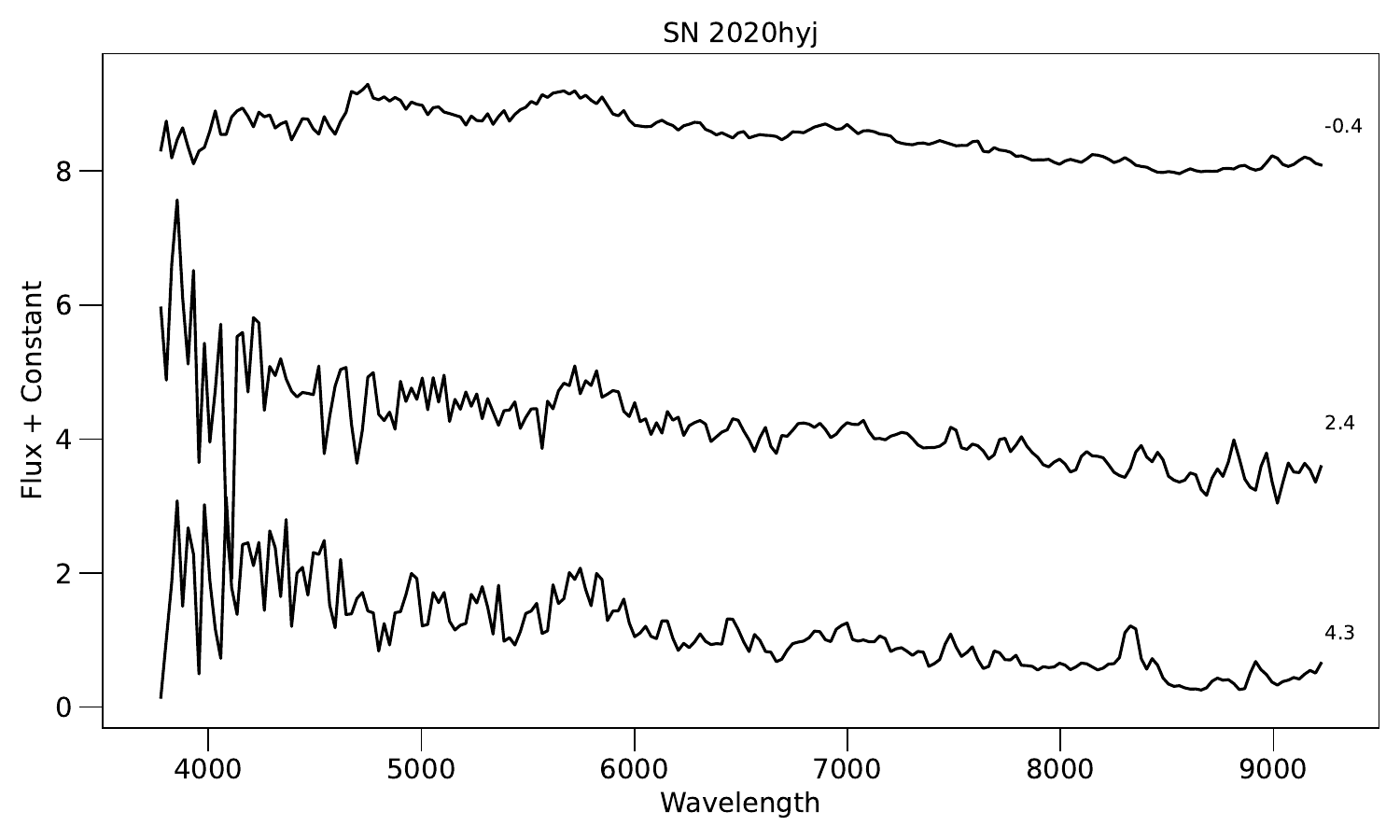} 
    \end{array}$
  \caption{\label{spec_seq2}Spectral sequences of SN 2020dgd, SN 2020hes, SN 2020hyj.}
 \end{figure}

 \begin{figure}[h!]
 \centering
 $\begin{array}{c}
  \includegraphics[width=12.5cm]{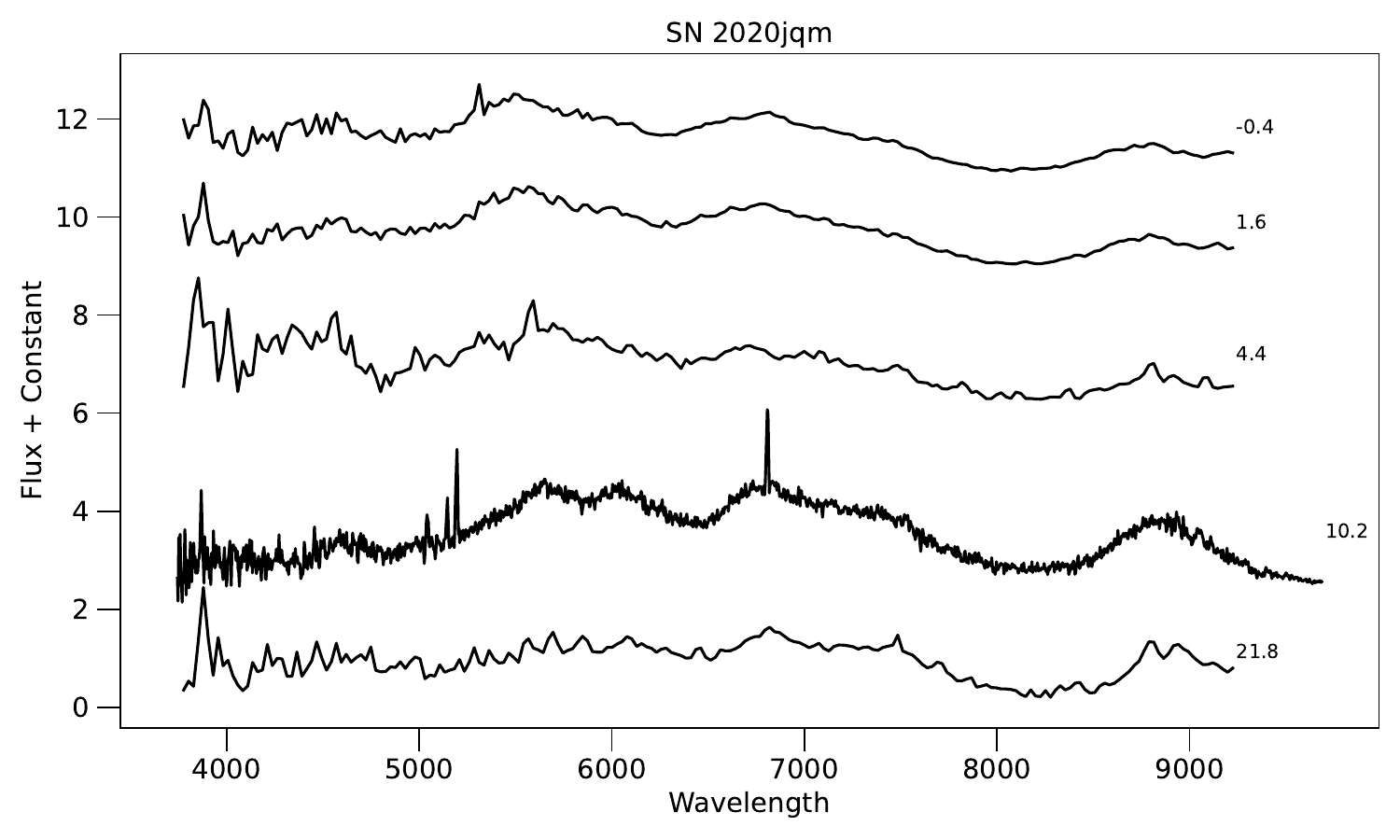} \\
  \includegraphics[width=12.5cm]{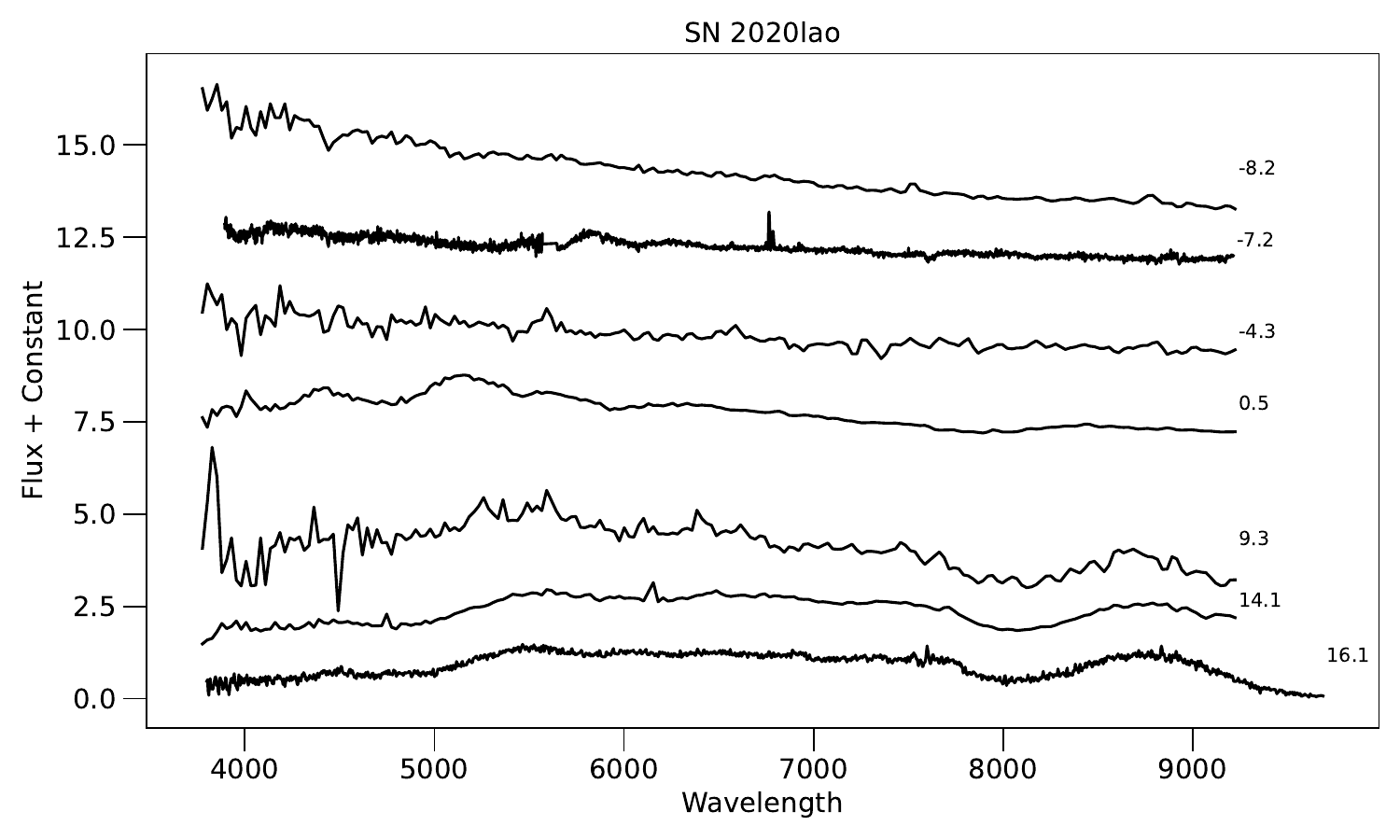} \\
  \includegraphics[width=12.5cm]{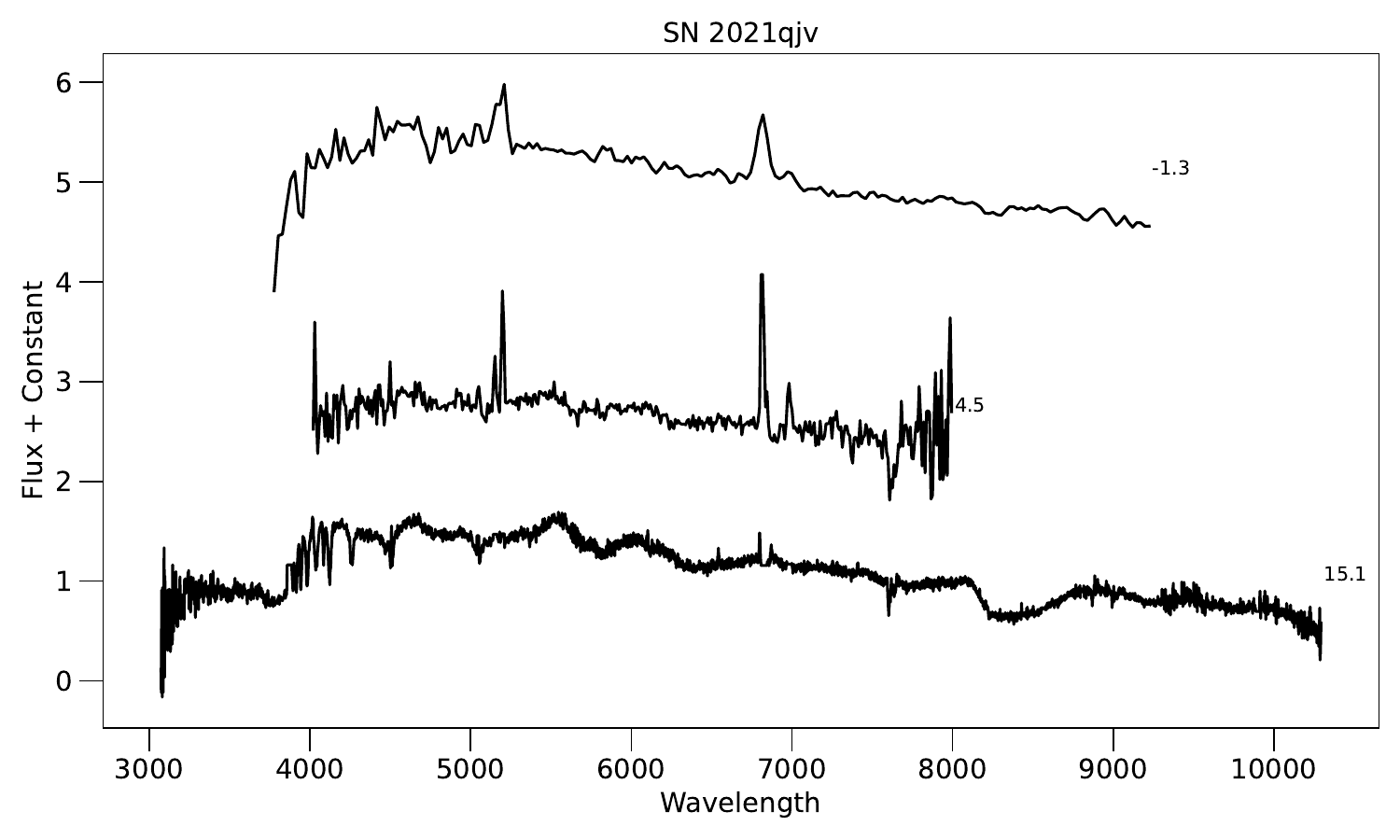}
    \end{array}$
  \caption{\label{spec_seq2}Spectral sequences of SN 2020jqm, SN 2020lao, SN 2021qjv.}
 \end{figure}

 \begin{figure}[h!]
 \centering
 $\begin{array}{c}
      \includegraphics[width=12.5cm]{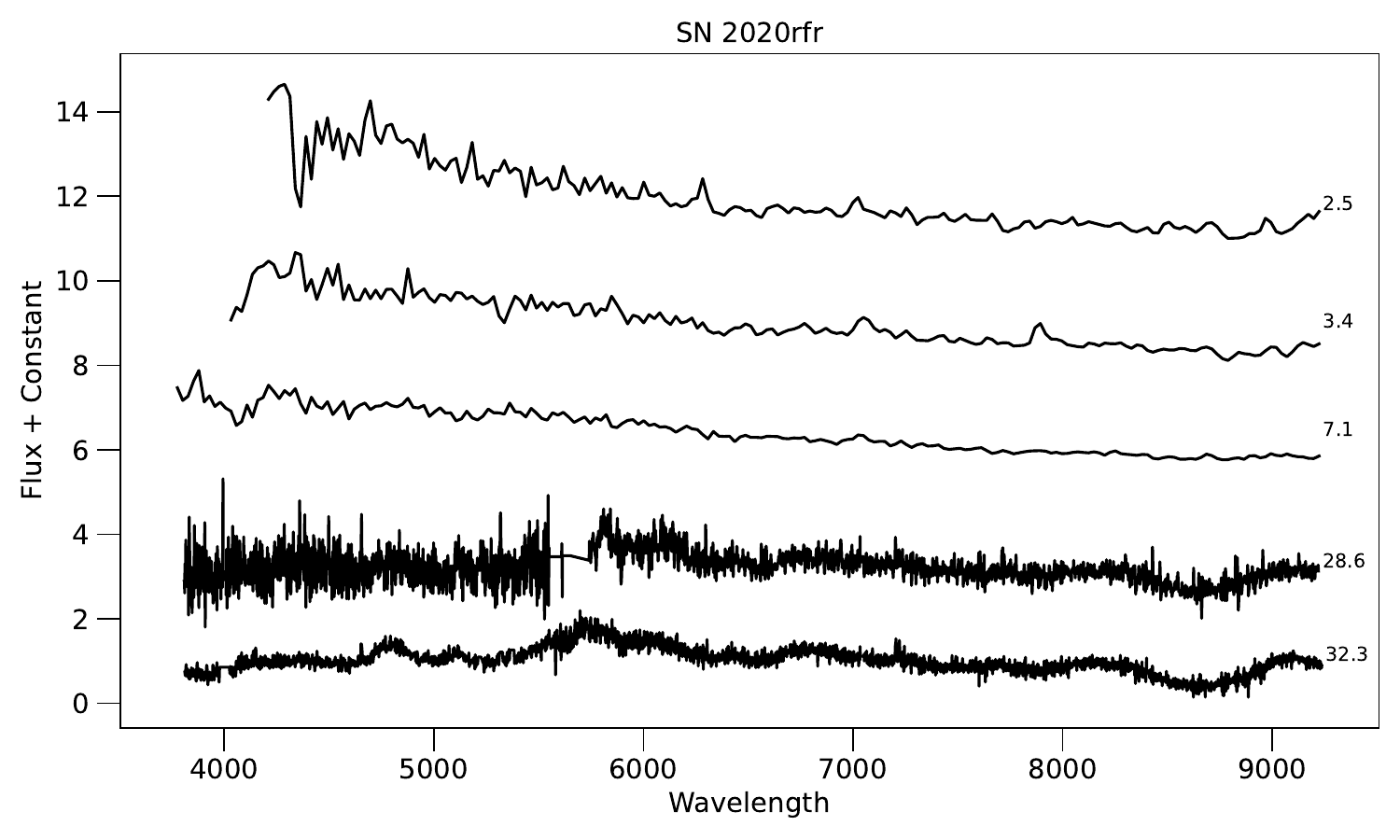} \\
  \includegraphics[width=12.5cm]{ 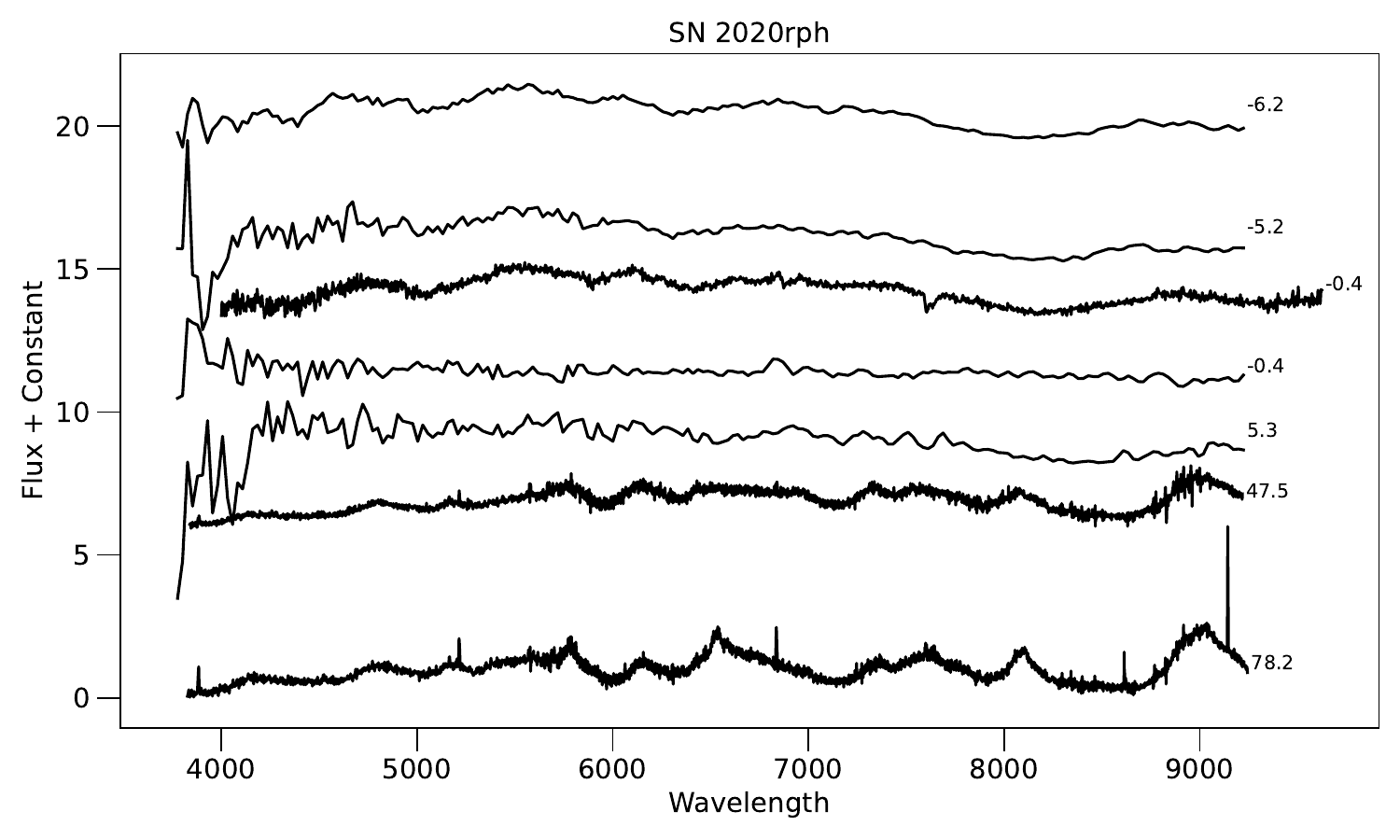} \\
  \includegraphics[width=12.5cm]{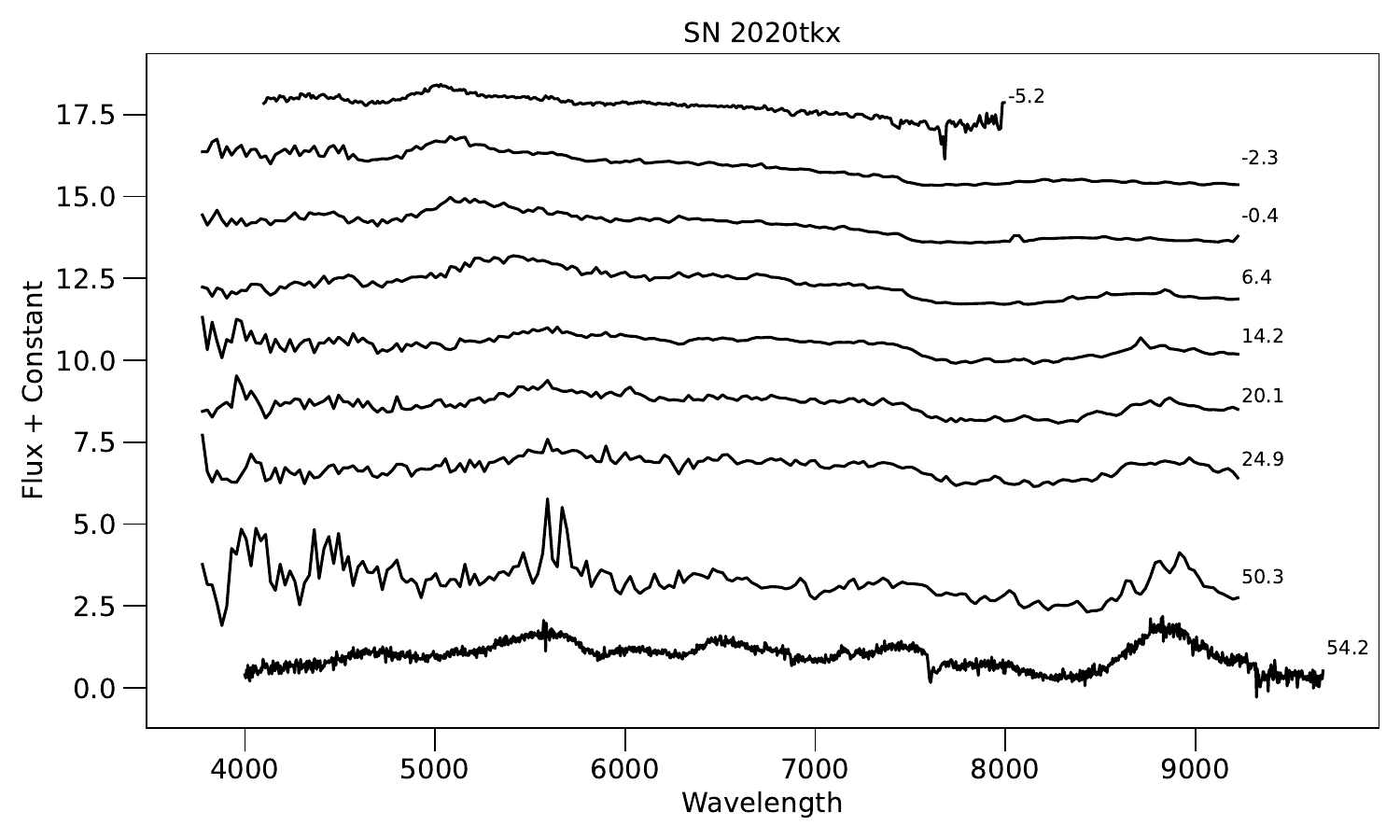}
    \end{array}$
  \caption{\label{spec_seq2}Spectral sequences of SN 2020rfr, SN 2020rph, SN 2020tkx.}
 \end{figure}

 %ZTF20achvlbs_spectra.pdf,ZTF20acvcxkz_spectra.pdf,

  \begin{figure}[h!]
 \centering
 $\begin{array}{c}
      \includegraphics[width=12.5cm]{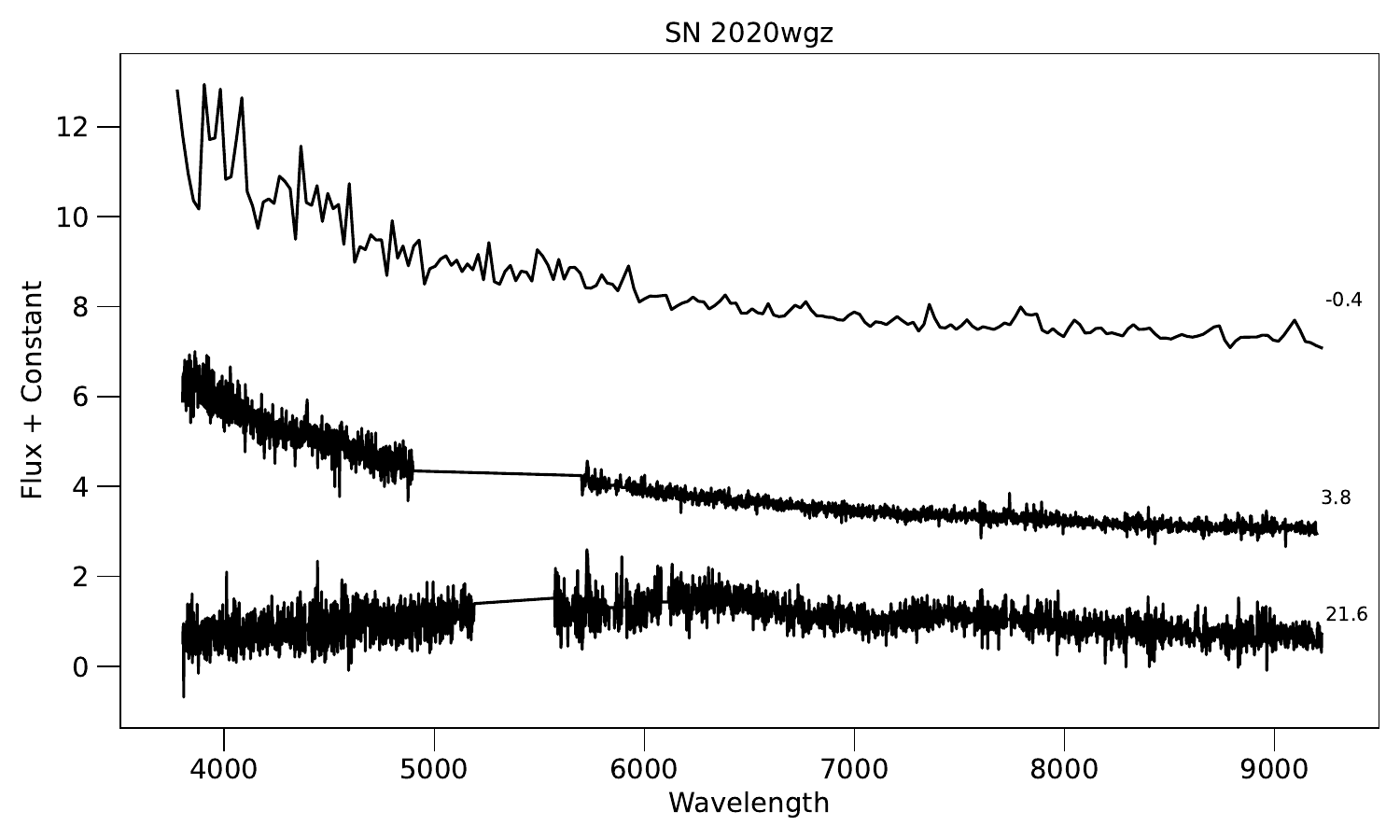}  \\
  \includegraphics[width=12.5cm]{ 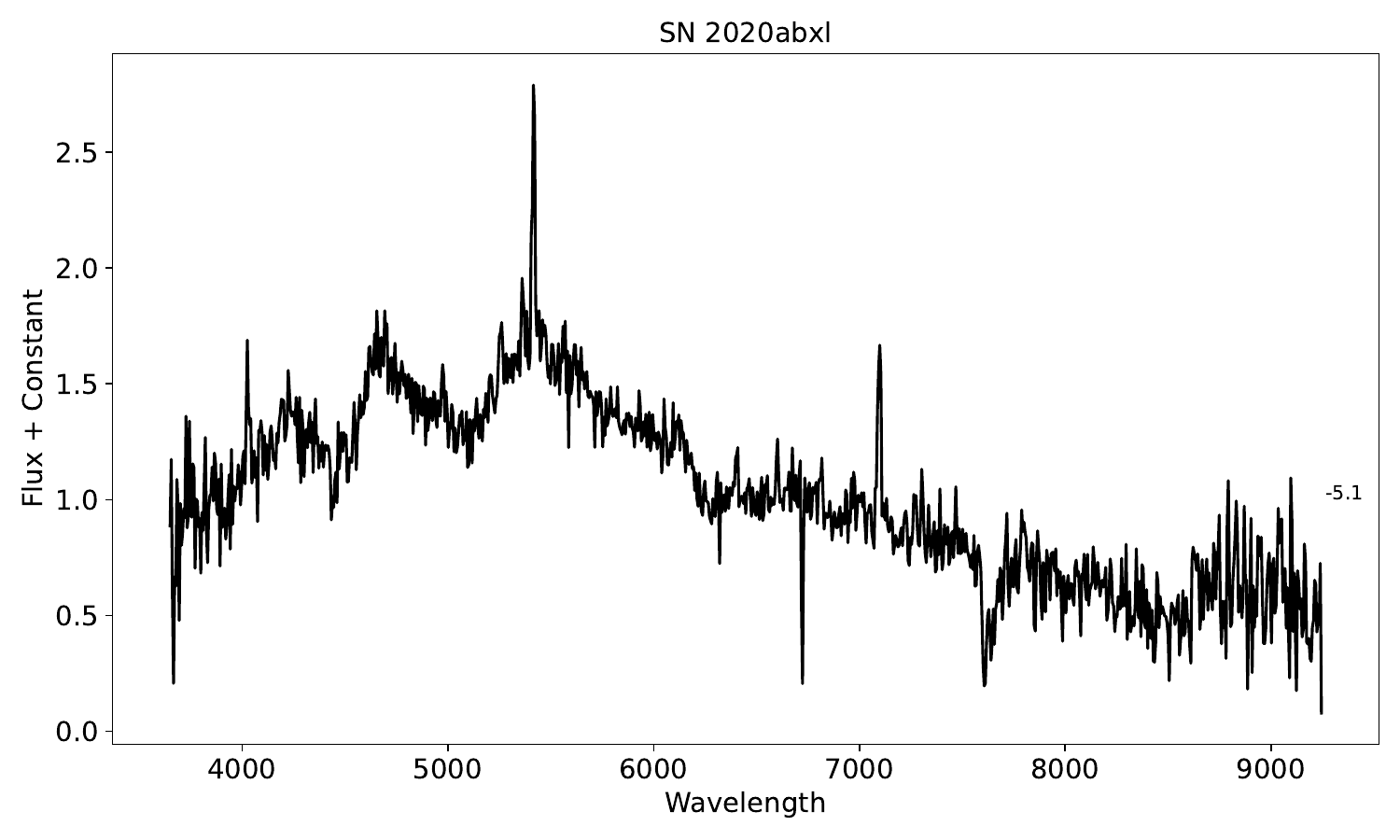} \\
  \includegraphics[width=12.5cm]{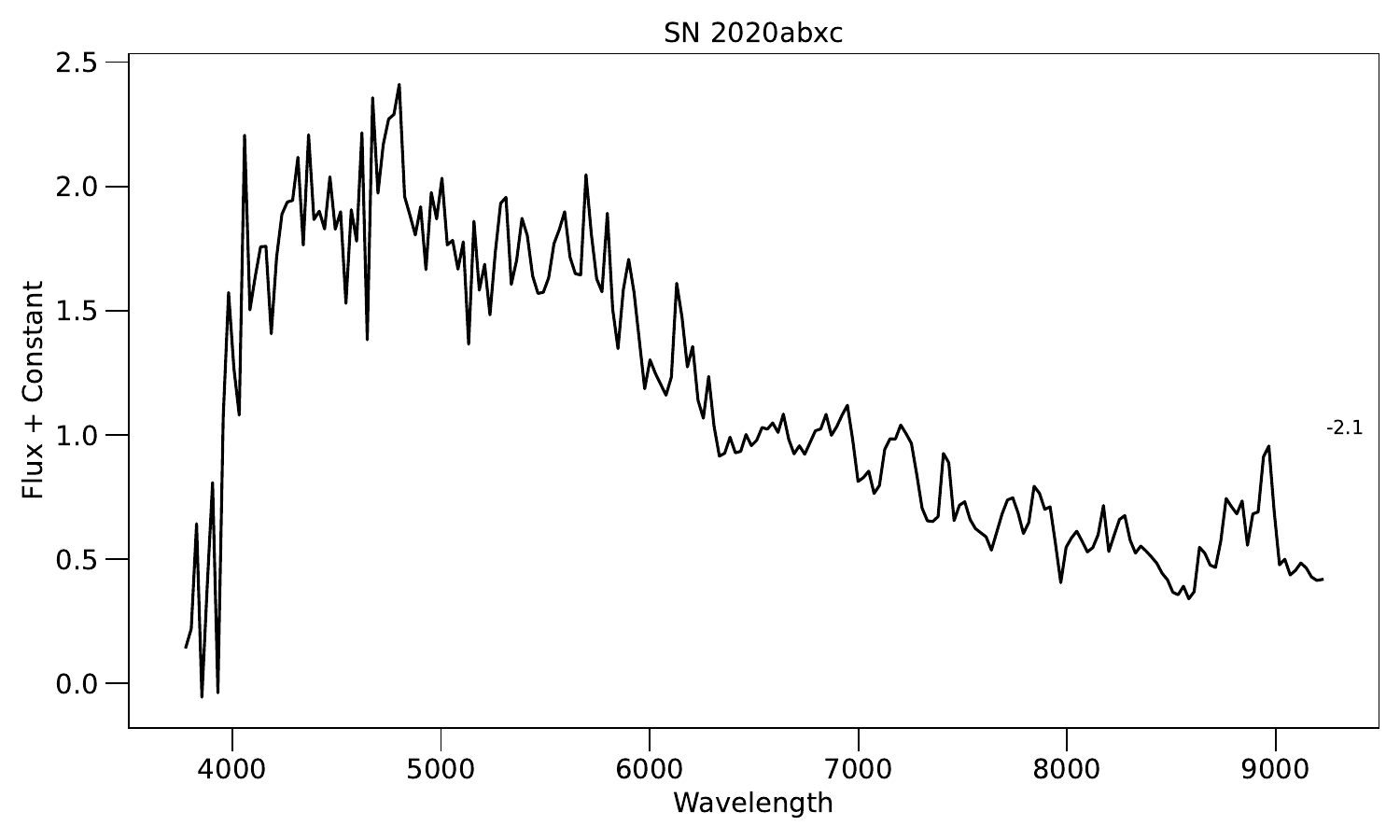}
    \end{array}$
  \caption{\label{spec_seq2}Spectral sequences of SN 2020wgz, SN 2020abxl, SN 2020abxc.}
 \end{figure} 

\begin{figure}[h!]
 \centering
 $\begin{array}{c}
      \includegraphics[width=12.5cm]{ 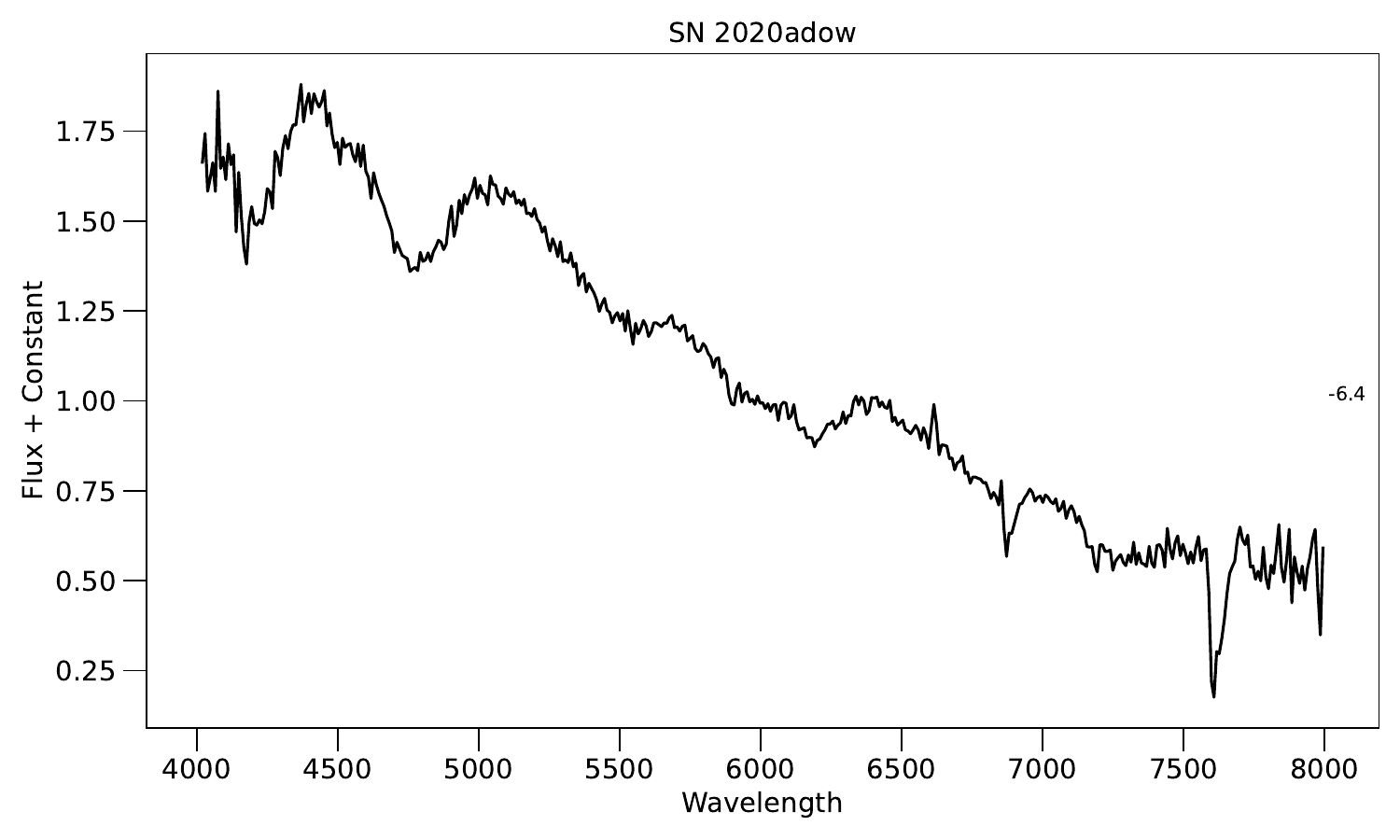}  \\
  \includegraphics[width=12.5cm]
  {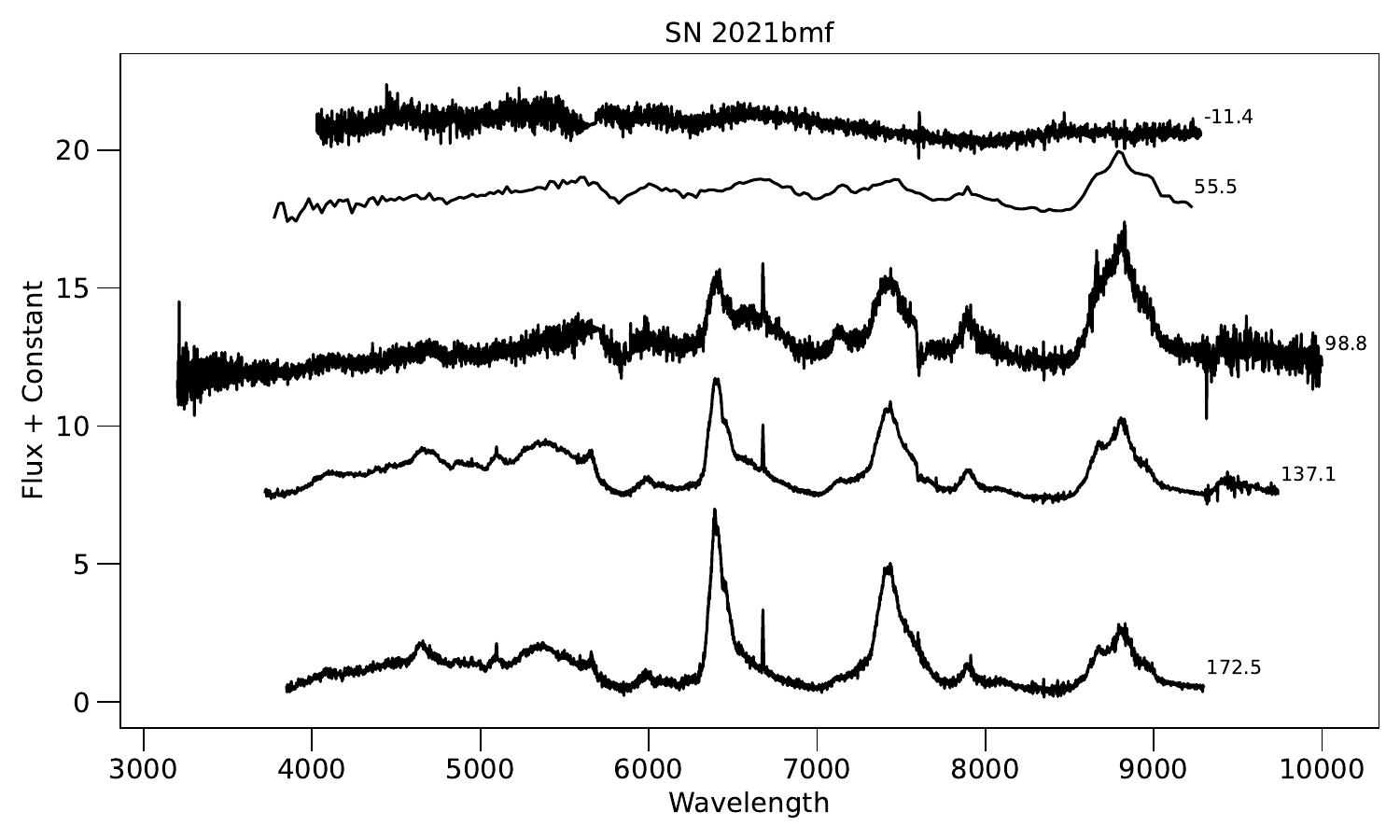}\\
  \includegraphics[width=12.5cm]{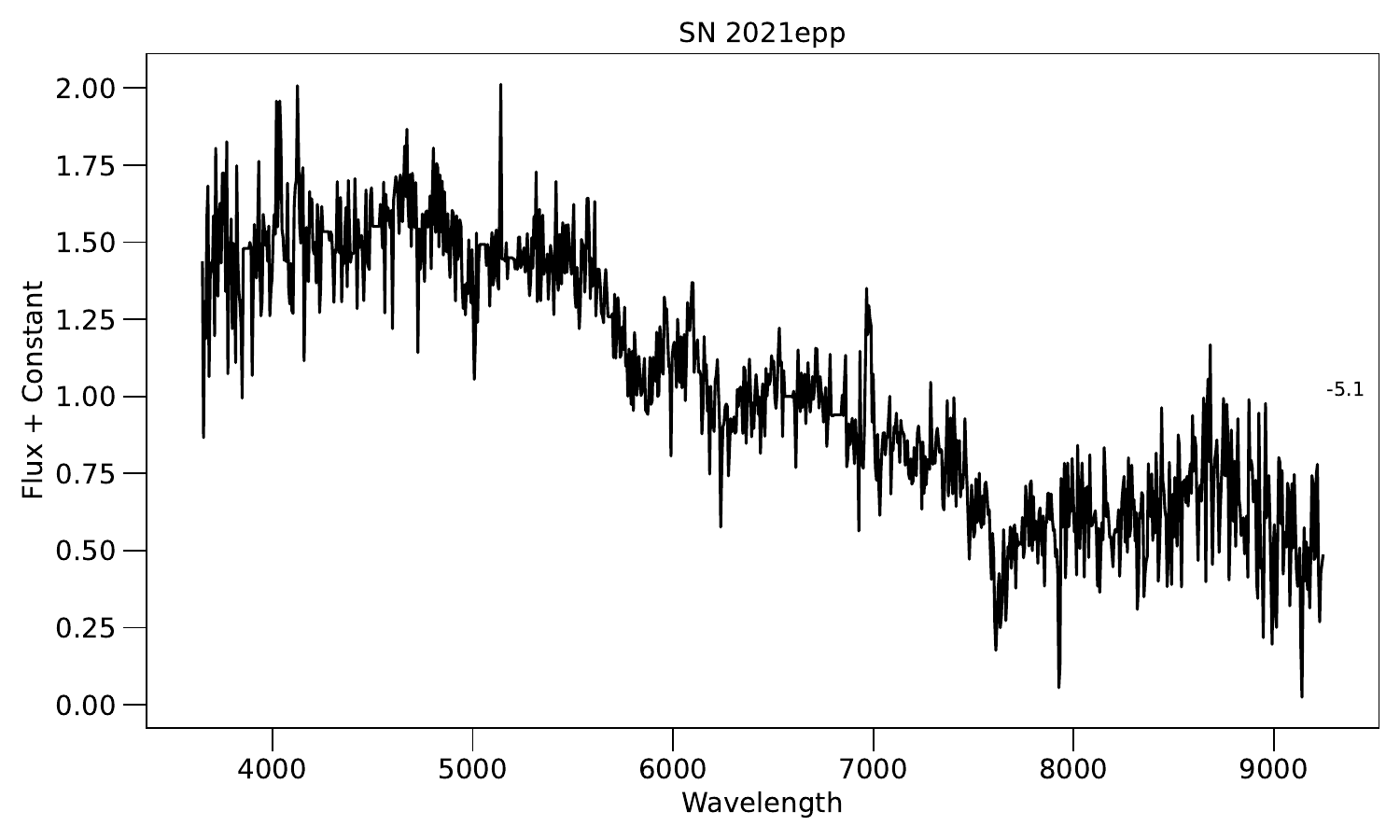} 
    \end{array}$
  \caption{\label{spec_seq2}Spectral sequences of SN 2020adow, SN 2021bmf, SN 2021epp.}
 \end{figure}

 \begin{figure}[h!]
 \centering
 $\begin{array}{c}
      \includegraphics[width=12.5cm] {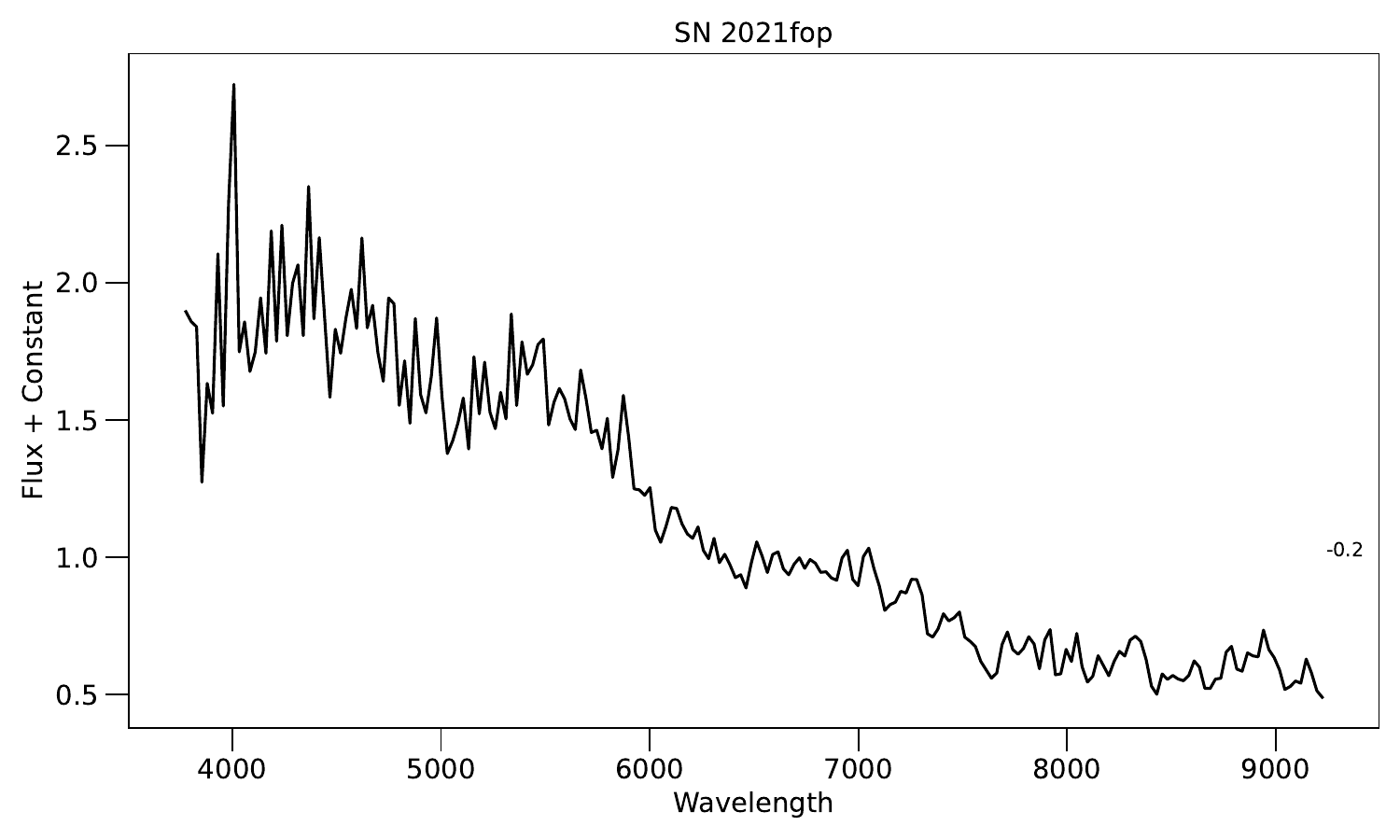}\\
  \includegraphics[width=12.5cm]{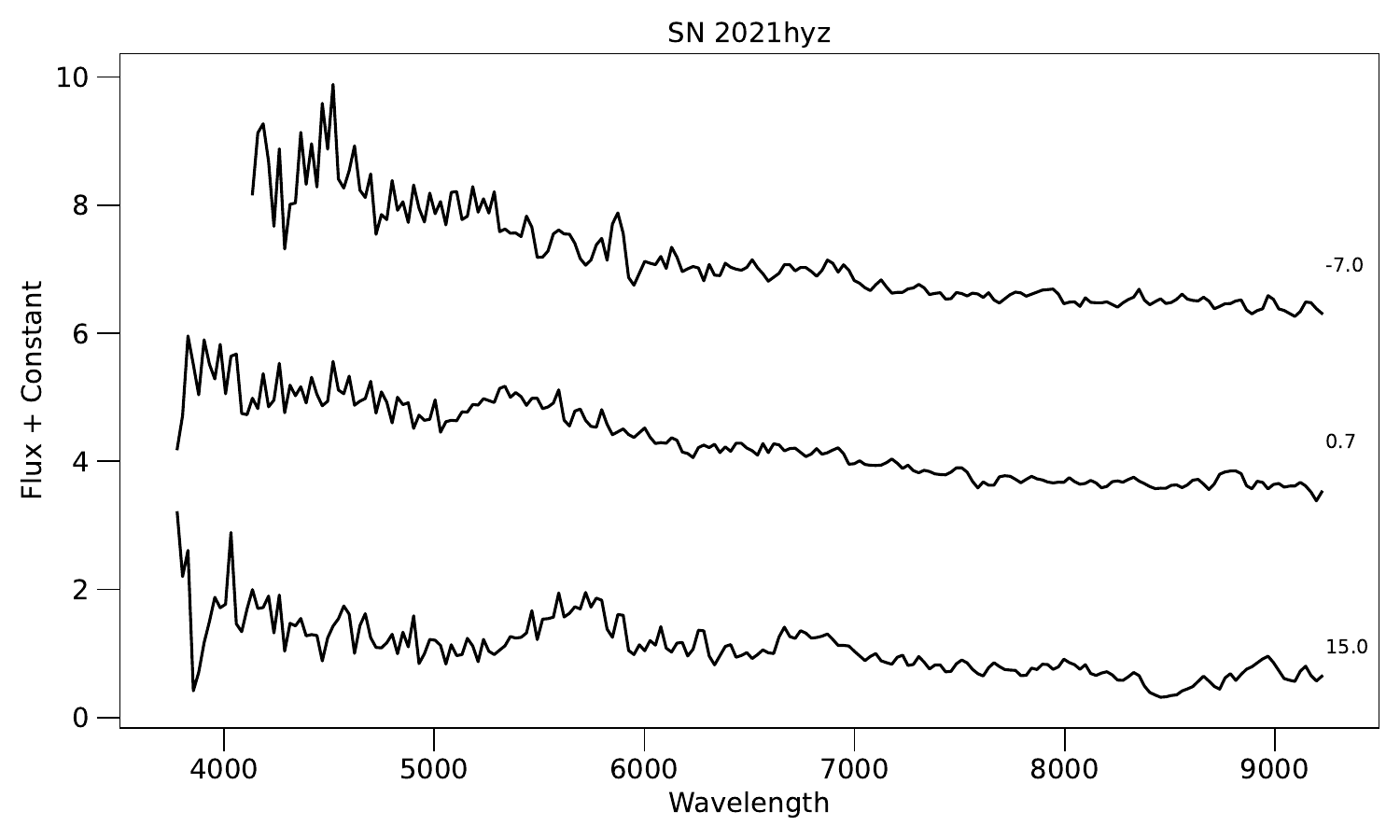}  \\
  \includegraphics[width=12.5cm]{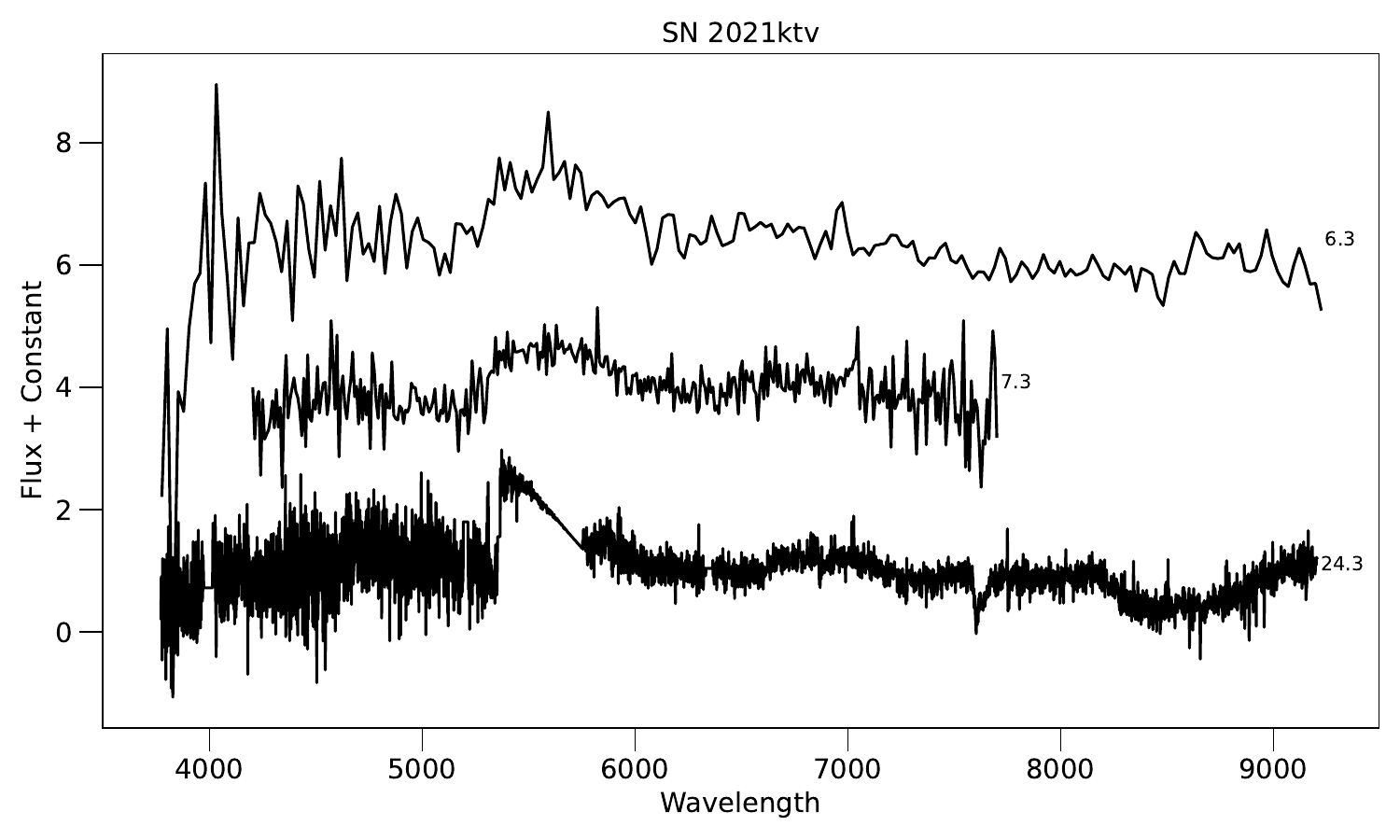}
    \end{array}$
  \caption{\label{spec_seq2}Spectral sequences of SN 2021fop, SN 2021hyz, SN2021ktv.}
 \end{figure} 

  \begin{figure}[h!]
 \centering
 $\begin{array}{c}
      \includegraphics[width=12.5cm]{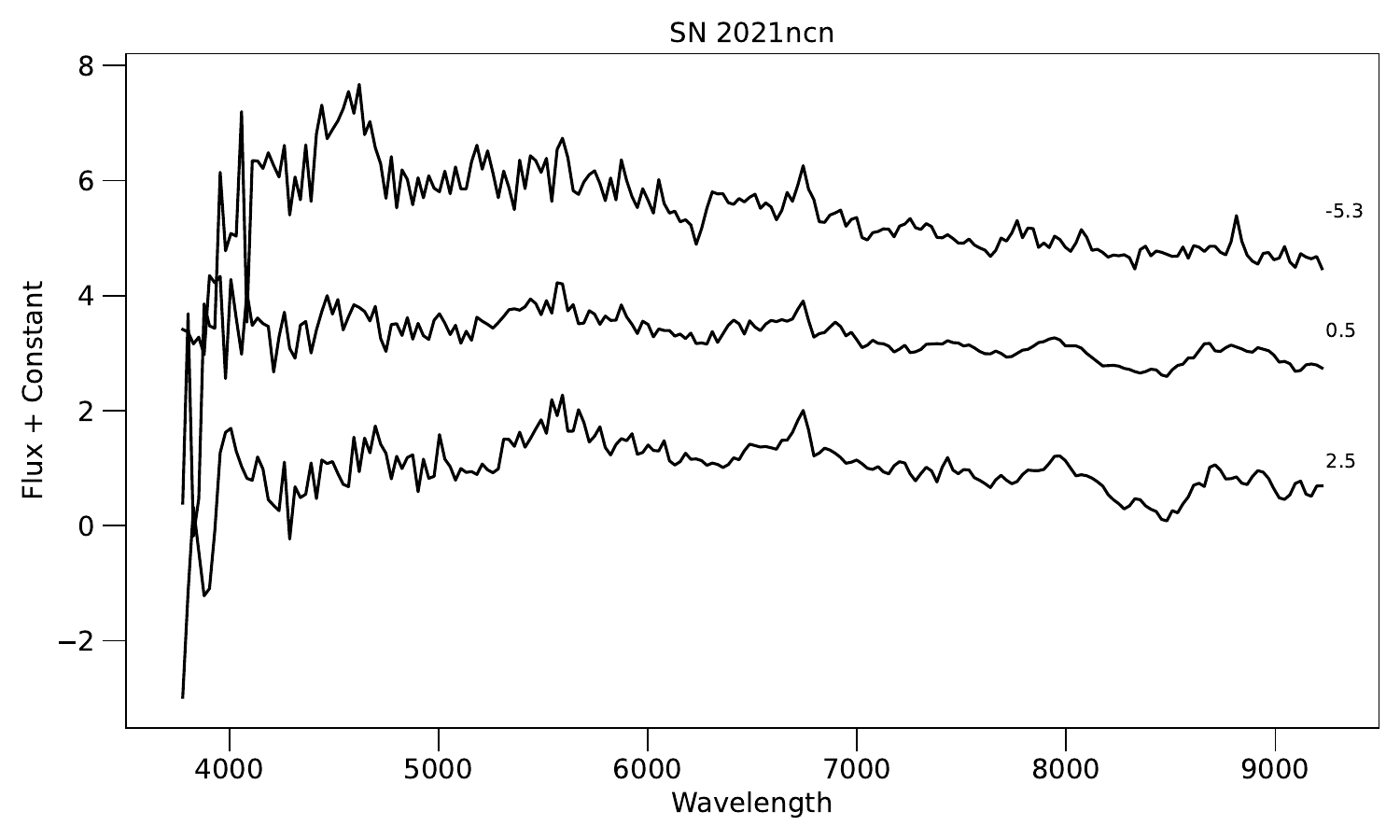}\\
  \includegraphics[width=12.5cm]{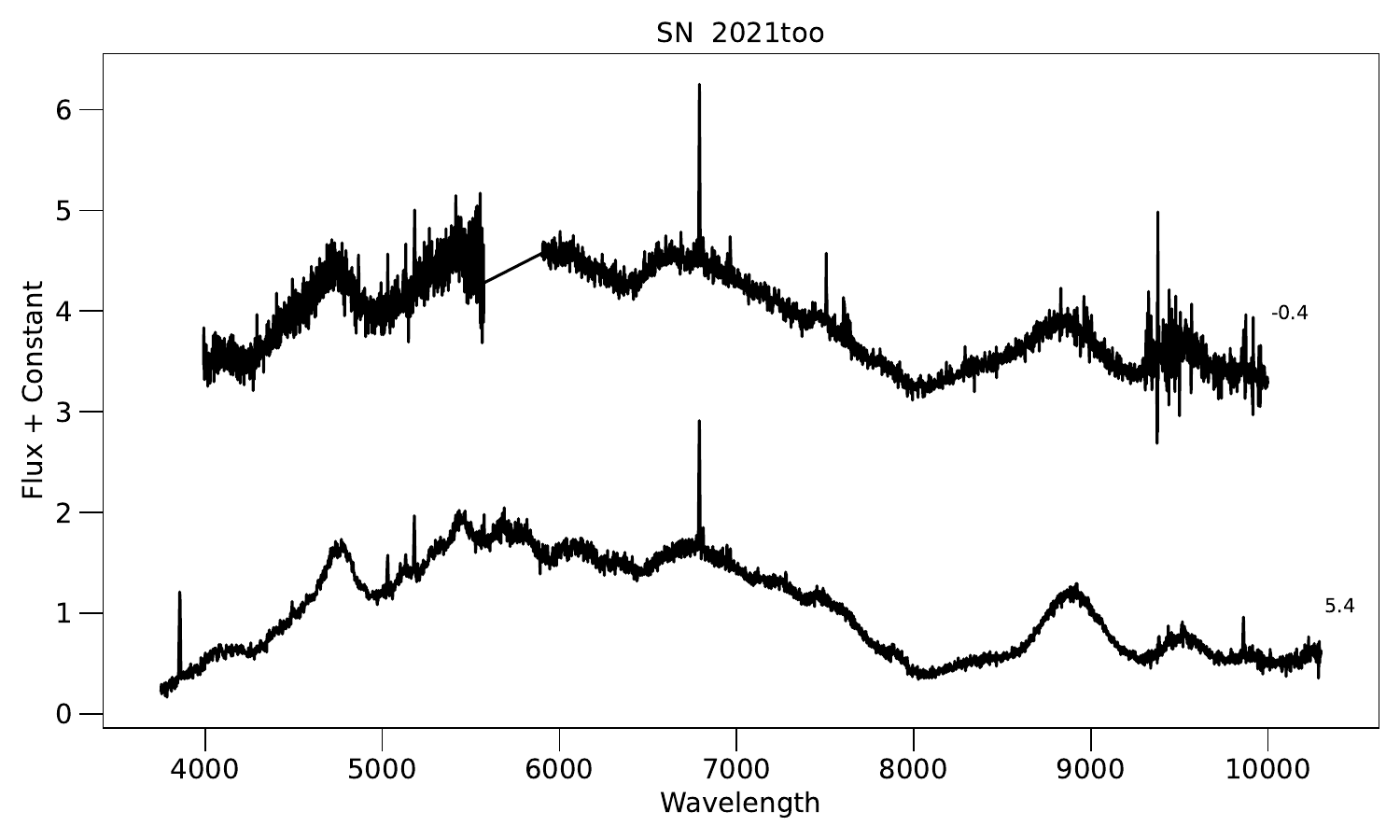}  \\
  \includegraphics[width=12.5cm]{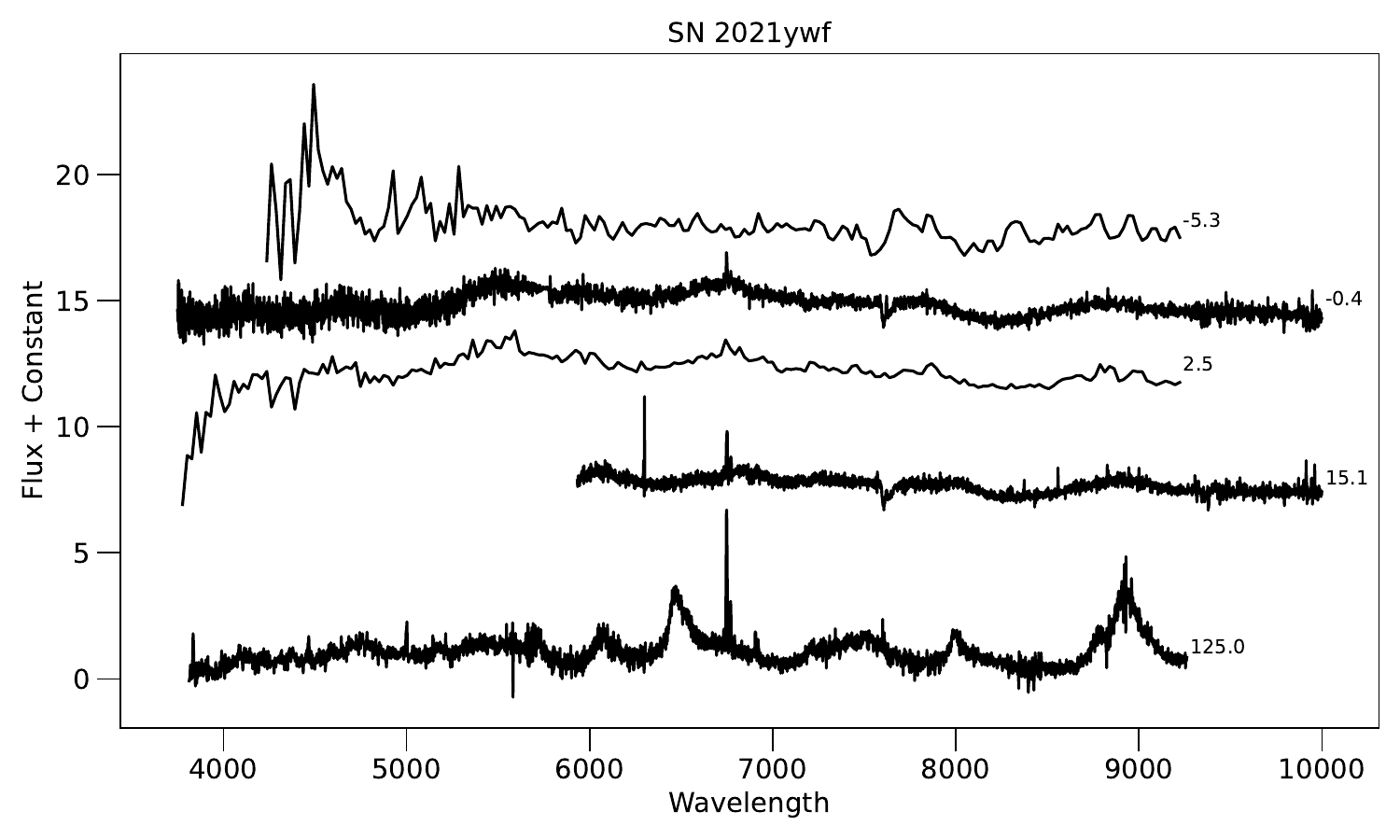}
    \end{array}$
  \caption{\label{spec_seq2}Spectral sequences of SN 2021ncn, SN 2021too, SN 2021ywf.}
 \end{figure} 

\section{Multiwavelength Modeling}
\label{physmech}

We use the open-source electromagnetic transient Bayesian fitting software package \texttt{redback} \citep{redback} to perform multiwavelength modeling of three events to determine if it is possible that these events had an associated relativistic jet, through modeling the optical, radio, and X-ray data simultaneously in flux density space. We use the \texttt{arnett + tophat} model in \texttt{redback}, which utilizes a combination of radioactive decay of Nickel from \citet{arnett1982}, along with a synchrotron component from an accompanying top-hat jet generated from the software package \texttt{afterglowpy} \citep{afterglowpy}, to model SN emission with an off-axis accompanying relativistic jet with a jet half-opening angle of 0.1 radian. 

We compare the Bayesian evidences for this model to those of the synchrotron self-absorption
(SSA) model for radio SNe (Eq. 4 from \citealt{Chevalier1998}), to compute a Bayes factor and determine which model can better explain the multiwavelength observations. We report the free fitting parameters for each of the models and the priors used in Tables \ref{arnetttable} and \ref{SSAtable}. Below we report the results for each of the events. Because there are only a few radio and X-ray detections/non-detections available for each event, we note that these results must be taken with a grain of salt, and therefore present them in the Appendix.

\subsubsection{SN 2020tkx}
SN 2020tkx has two radio detections and one X-ray non-detection reported in \citet{Corsi2024}. Because $\rm{v_{ph}}$ for this event was measured from a spectrum 53 days after peak, we allow it to vary as a free parameter with uniform priors. Furthermore, the radio LC shown in Figure 11 in \citet{Corsi2024} shows a rise in time, and there are no significant constraints that can be put on the jet's energy or the ISM number density, so we allow both those parameters to vary with uniform priors in logarithmic space. The corner plots for the fitting are shown in the Appendix, and we find that the \texttt{Arnett + tophat} model is favored with a Bayes factor of $10^{23.0}$.

\subsubsection{SN 2020adow}
SN 2020adow has two radio detections and one X-ray detection, and was not reported in \citet{Corsi2024}. There were two radio measurements obtained, at 5.3 days and 16.3 days after explosion, with flux densities of $28.5 \pm 7.1$ and $17.1 \pm 7.6 $ $\mu$Jy \citep{2020adowradio}. The X-ray measurement was obtained 42 days after explosion, with a 0.3 -- 10 keV flux of $4.9^{+2.8}_{-2.1} \times 10^{-14}$ erg cm$^{-2}$ s$^{-1}$. We convert the 0.3 -- 10 keV flux to a flux density at 5 keV for the fitting, through assuming a power-law spectrum of photon index $\Gamma = 2$. We constrain $\rm{v_{ph}}$ as a prior to the value obtained in \S \ref{spectra} (19,500 km s$^{-1}$), which utilized a spectrum taken 7 days before peak. We have no constraints on the jet energy or ISM number density for this object from previous works, so we allow both those parameters to vary with uniform priors in logarithmic space. The corner plots for the fitting are shown in the Appendix, and we find that the Radio + SSA model is favored qualitatively, though a conclusive Bayes factor was not able to be found for this event.

\subsubsection{SN 2021ywf}
SN 2021ywf has two radio detections and one X-ray detection reported in \citet{Corsi2024}. We convert the 0.3 -- 10 keV flux reported to a flux density at 5 keV for the fitting, through assuming a power-law spectrum of photon index $\Gamma = 2$. We constrain $\rm{v_{ph}}$ as a prior to the value obtained in \S \ref{spectra} (12,000 km s$^{-1}$), which utilized a spectrum taken 0.5 days after peak. \citet{Corsi2024} also was able to rule out top-hat jets with energies greater than $10^{49}$ erg and interstellar medium (ISM) number densities ($n_0$) greater than 0.1 cm$^{-3}$ for this event, so we utilize these priors in the fitting procedure as well. The corner plots for the fitting are shown in the Appendix, and we find that the Radio SSA model is favored with a Bayes factor of $K_{\rm{Bayes}} = 10^{28.5}$.

\begin{table*}[h]
\centering
\begin{tabular}{l|l|l}
\hline\hline
        Parameters & Description & Prior Boundaries\\ 
        \hline
         z & redshift & Set to values from Table \ref{discoverytable}\\ 
         $\rm{v_{ph}}$ & Peak photospheric expansion velocity & Set to values from Table \ref{explosiontable}\\
        $f_{\rm{Ni}}$ & Nickel fraction & [0.001, 1] (in $\rm{Log_{10}}$ space)\\
        $\rm{M_{ej}}$ & Ejecta mass & [0.0001, 100] $\rm{M_{\odot}}$ (in $\rm{Log_{10}}$ space)\\ 
        $A_v$ & MW extinction & 0 mag \\ 
        $\kappa$ & Optical opacity & 0.07 $\rm{cm^{2} \, g^{-1}}$ \\ 
        $\kappa_\gamma$ & Gamma-ray Opacity & [0.0001, 10000] $\rm{cm^{2} \, g^{-1}}$ (in $\rm{Log_{10}}$ space)   \\ 
        $T_{\rm{floor}}$ & Minimum temperature reached in early explosion & [1000, 100000] K (in $\rm{Log_{10}}$ space)  \\ 
        $\theta_{\rm{observer}}$ & Observing angle of jet & Sine[0.2, 1.57] \\ 
        $\theta_{\rm{core}}$ & Jet opening angle & 0.1  \\ 
        $\epsilon_e$ & $\rm{Log_{10}}$ fraction of thermal energy to electrons & 0.1  \\ 
        $\epsilon_b$ & $\rm{Log_{10}}$ fraction of thermal energy to magnetic field & 0.1  \\
        $\rm{E_0}$ & $\rm{Log_{10}}$ on axis isotropic equivalent energy & See individual event subsections  \\ 
        $n_0$ & $\rm{Log_{10}}$ ISM number density & See individual event subsections  \\ 
        $p$ & Electron distribution power-law index & [2, 3]\\ 
        $\xi_{N}$ & Fraction of electrons that get accelerated &  [0,1] \\ 
        $\Gamma_0$ & Initial Lorentz Factor & [100, 2000] \\ 
    \end{tabular}
    \caption{Model parameters, description, and priors used for the \texttt{arnett + tophat} model in the \texttt{redback} fitting.}
    \label{arnetttable}
\end{table*}

\begin{table*}[h]
\centering
\begin{tabular}{l|l|l}
\hline\hline
        Parameters & Description & Prior Boundaries \\ 
        \hline
        z & redshift & Set to values from Table \ref{discoverytable}\\
        $\rm{v_{ph}}$ & Peak photospheric expansion velocity & Set to values from Table \ref{explosiontable}\\
       $f_{\rm{Ni}}$ & Nickel fraction & [0.001, 1] (in $\rm{Log_{10}}$ space)\\
        $\rm{M_{ej}}$ & Ejecta mass & [0.0001, 100] $\rm{M_{\odot}}$ (in $\rm{Log_{10}}$ space)\\ 
        $A_v$ & MW extinction & 0 mag \\ 
        $\kappa$ & Optical opacity & 0.07 $\rm{cm^{2} \, g^{-1}}$ \\ 
        $\kappa_\gamma$ & Gamma-ray Opacity & [0.0001, 10000] $\rm{cm^{2} \, g^{-1}}$ (in $\rm{Log_{10}}$ space)   \\ 
        $T_{\rm{floor}}$ & Minimum temperature reached in early explosion & [1000, 100000] K (in $\rm{Log_{10}}$ space)  \\  
        $t_c$ & Time where emission has $\tau_{\rm{opt}} \sim 1$  at $\nu_c$  & [0, 1000] days\\ 
        $\nu_c$ & Characteristic frequency to scale $t_c$ & [$10^7$, $10^9$] Hz (in $\rm{Log_{10}}$ space)   \\ 
        $m$ & Power law index of emitting region expansion ($R^m$) & [$\frac{2}{3}$, 1]  \\ 
        $\gamma$ & electron spectral index & [1, 4] \\ 
        $k$ & Flux-scaling Factor & [0.0001, 100] (in $\rm{Log_{10}}$ space)   \\ 
        $\beta$ & Spectral power-law index & [0.5, 2]\\
        $\beta_{\rm{time}}$ & Temporal power-law index & [0.5, 3]\\
         \end{tabular}
    \caption{Model parameters, description, and priors used for the Radio SSA model \citep{Chevalier1998}  in the \texttt{redback} fitting.}
    \label{SSAtable}
\end{table*}

\begin{figure}[h!]
    \centering
    \includegraphics[width = 0.65\linewidth]{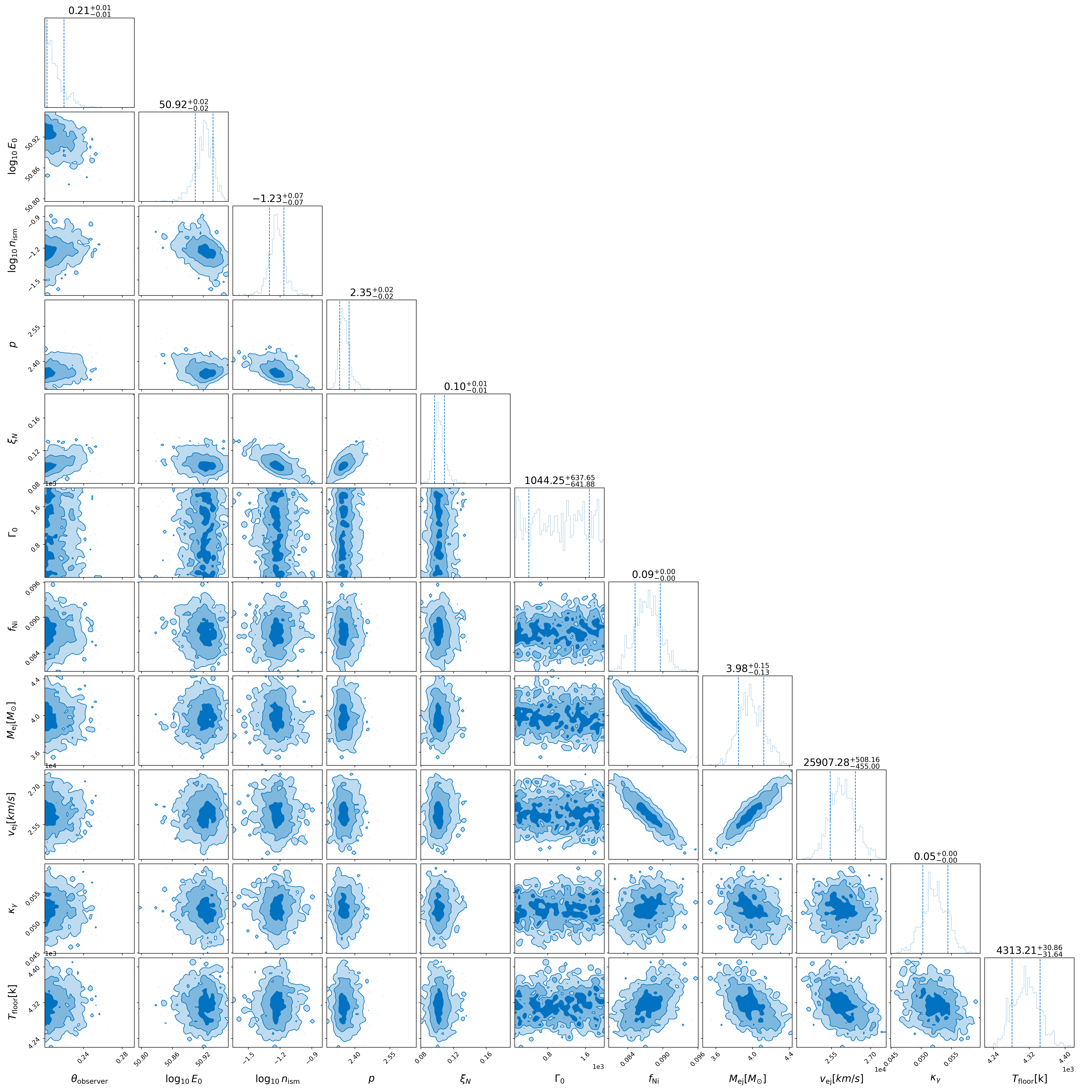}
     \includegraphics[width = 0.65\linewidth]{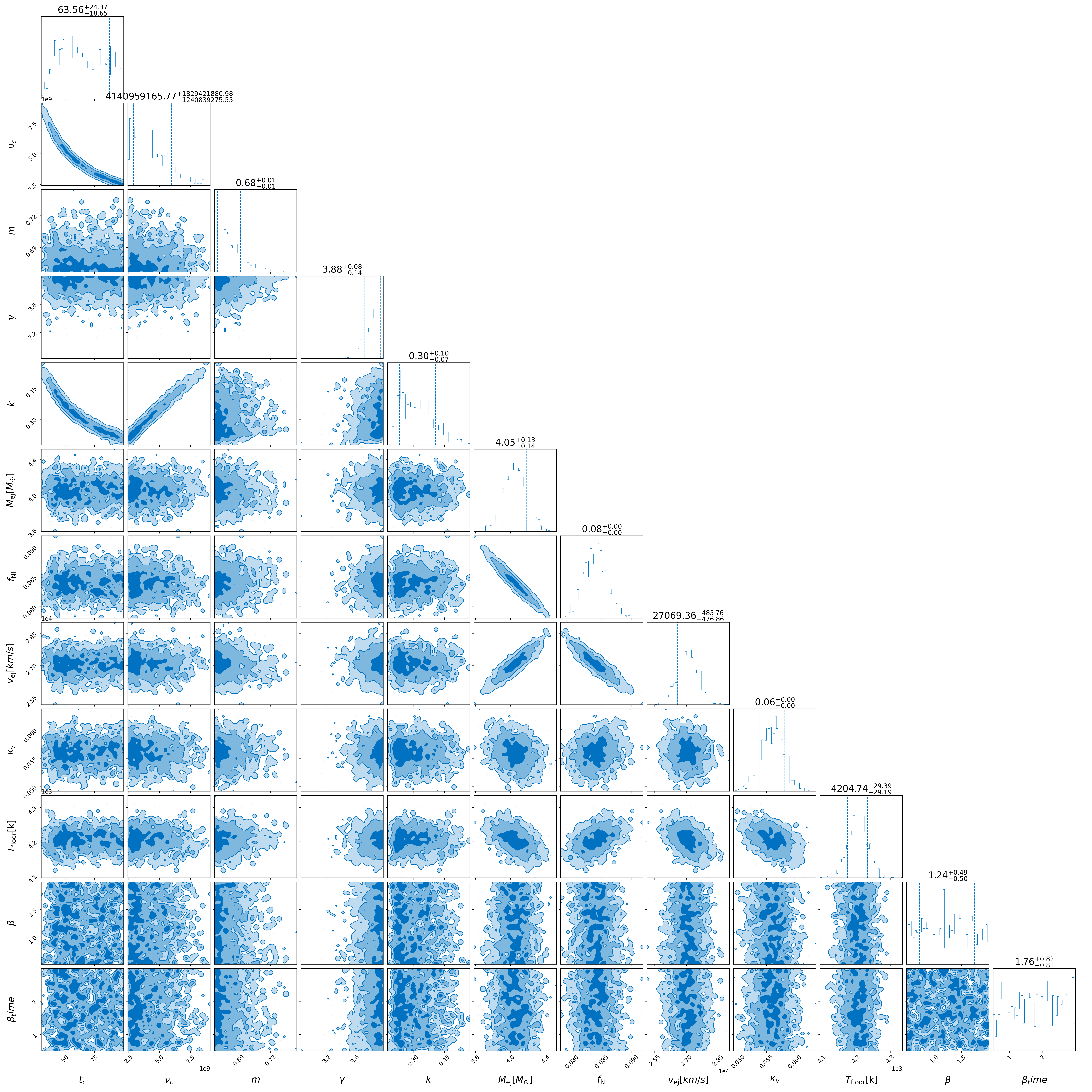}
    \caption{\texttt{Redback} modeling results for SN 2020tkx, with the \texttt{arnett + tophat} model above and the Radio SSA model below. }
    \label{fig:enter-label}
\end{figure}

\begin{figure}[h!]
    \centering
    \includegraphics[width = 0.65\linewidth]{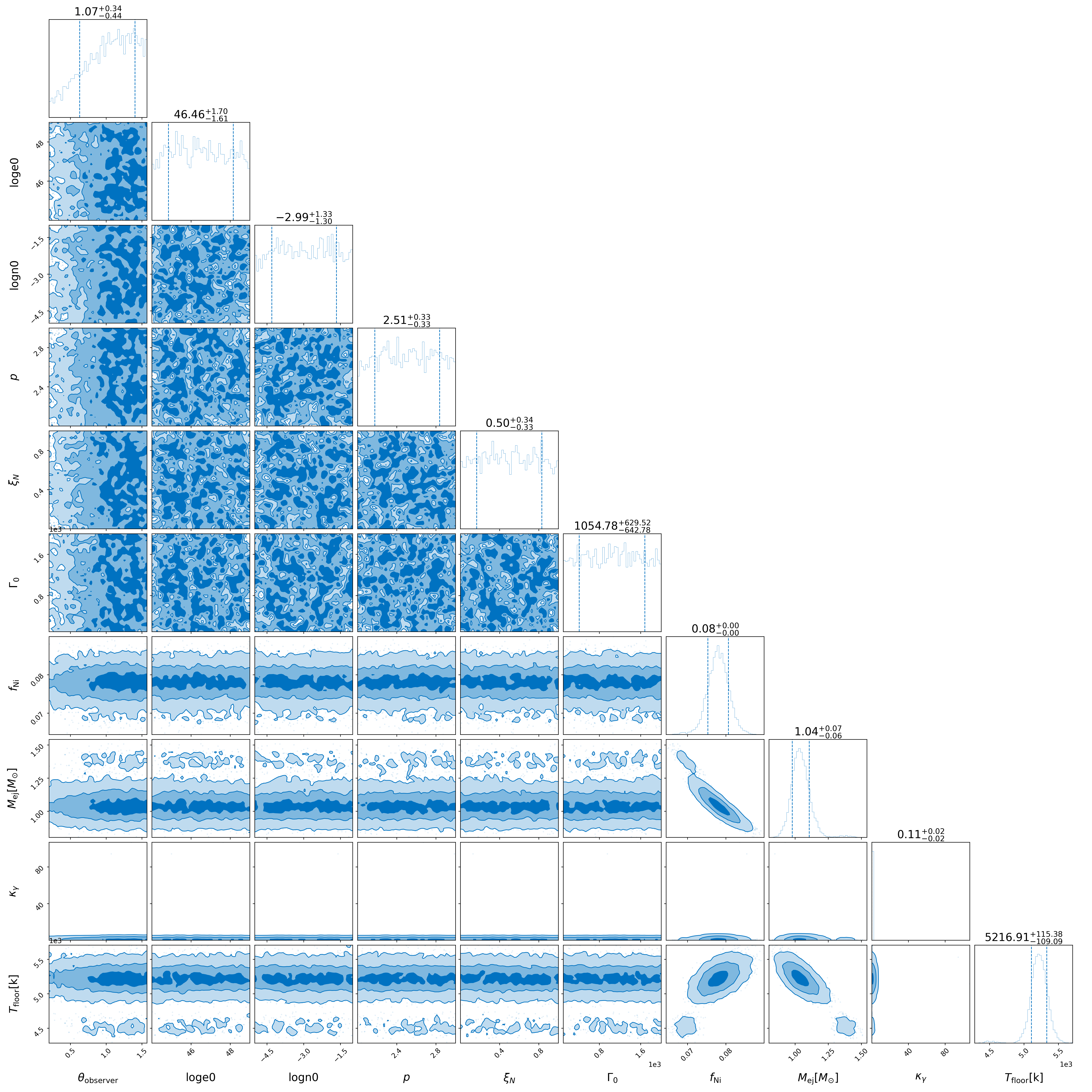}
     \includegraphics[width = 0.65\linewidth]{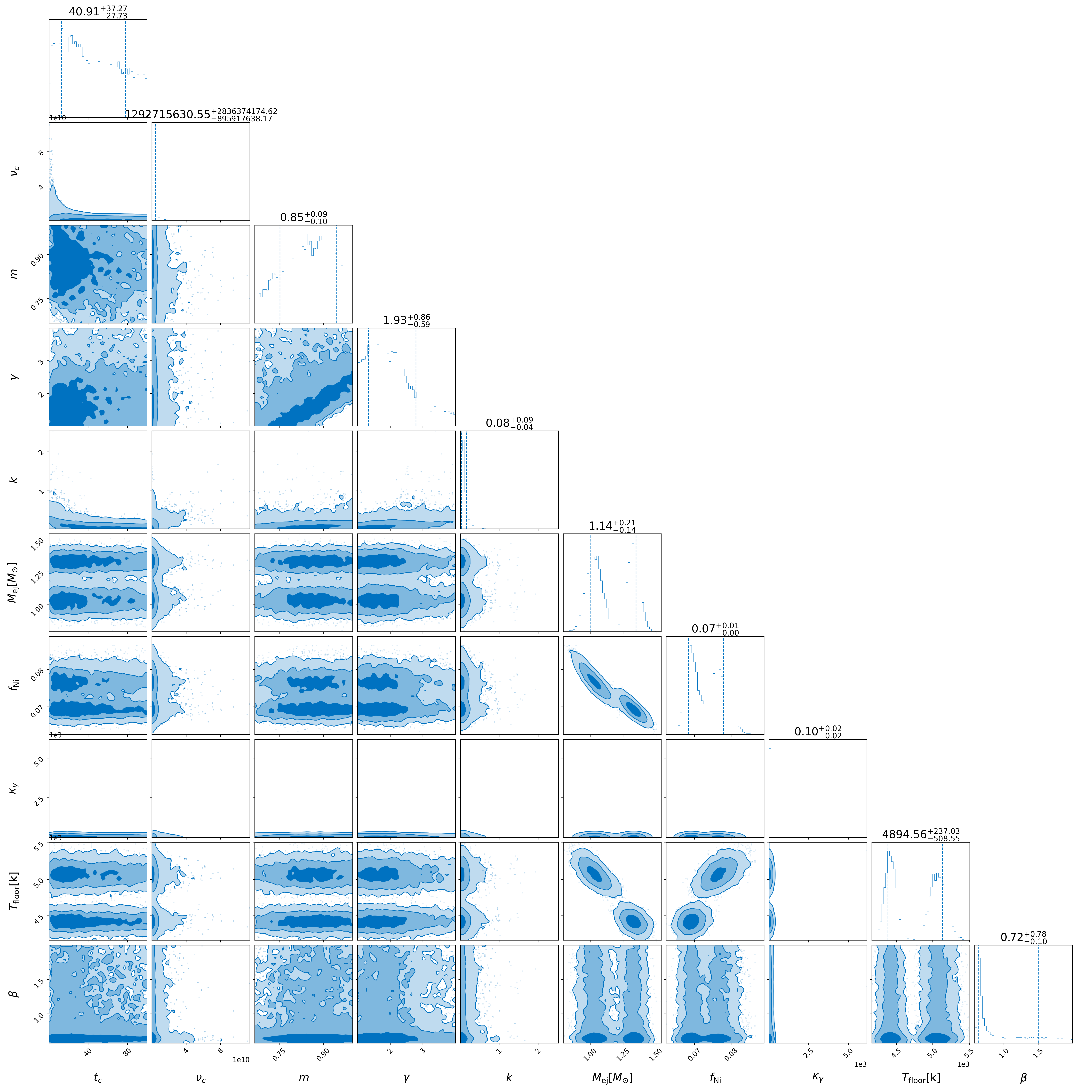}
    \caption{\texttt{Redback} modeling results for SN 2021ywf, with the \texttt{arnett + tophat} model above and the Radio SSA model below. }
    \label{fig:enter-label}
\end{figure}

\begin{figure}[h!]
    \centering
    \includegraphics[width = 0.65\linewidth]{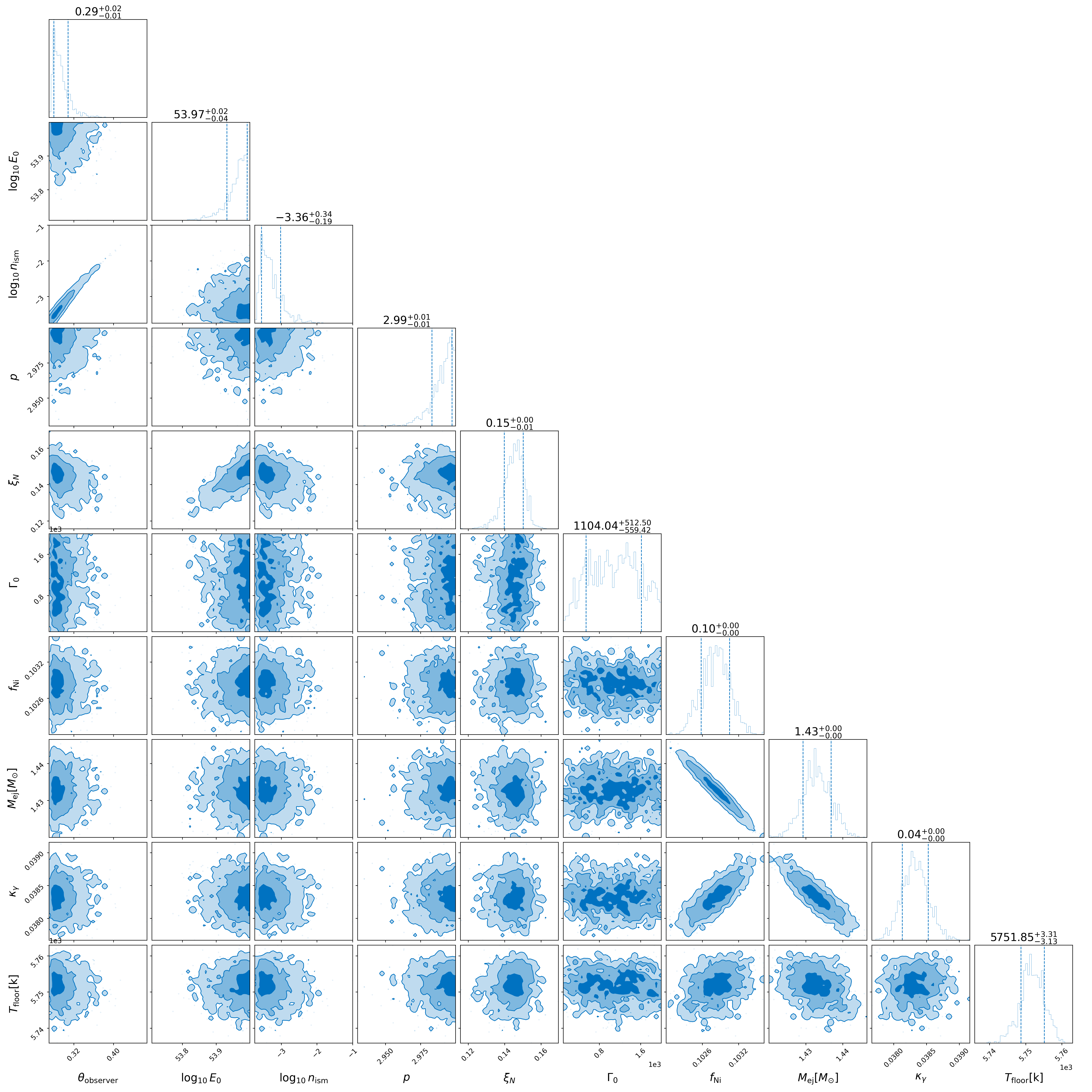}
     \includegraphics[width = 0.65\linewidth]{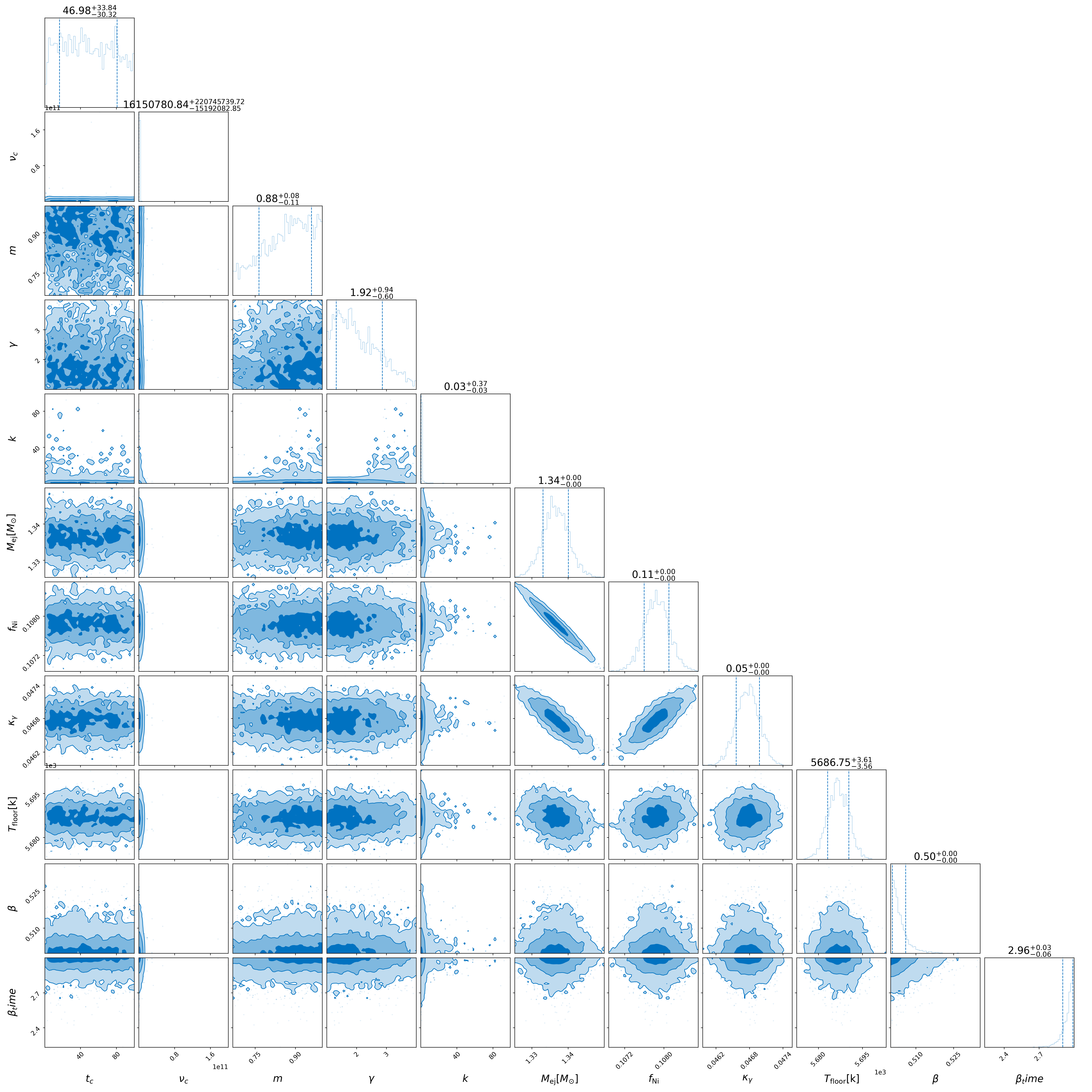}
    \caption{\texttt{Redback} modeling results for SN 2020adow, with the \texttt{arnett + tophat} model above and the Radio SSA model below. }
    \label{fig:enter-label}
\end{figure}

\begin{figure}[h!]
    \centering
    \includegraphics[width = 0.9\linewidth]{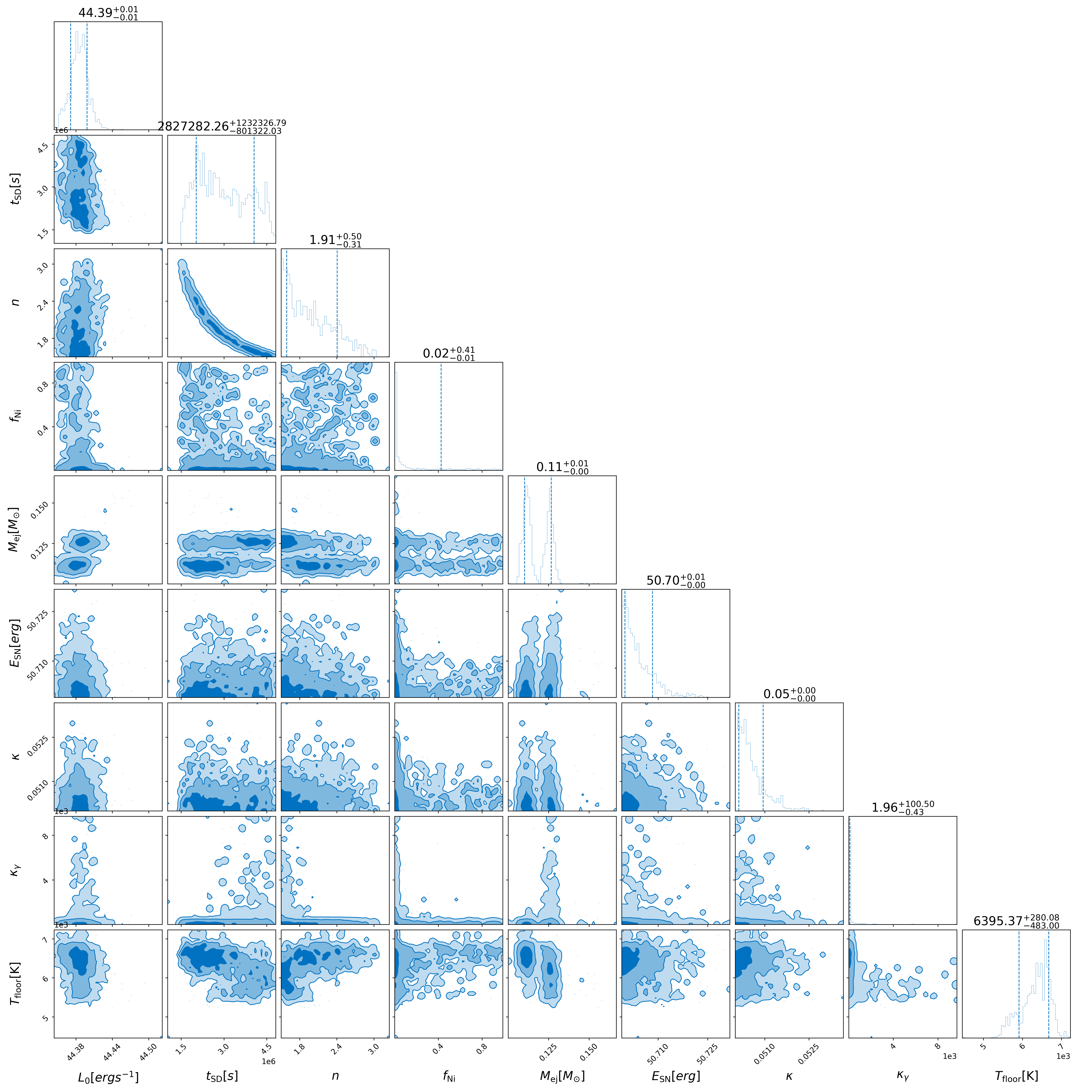}
    \caption{\texttt{Redback} modeling results for SN 2020wgz, for the magnetar and nickel powered model. }
    \label{fig:enter-label}
\end{figure}

\bibliography{main}{}
\bibliographystyle{aasjournal}
\end{document}